\documentclass[letterpaper,11pt,twoside]{article}

\usepackage{amsmath,amsthm,amssymb}
\usepackage[utf8]{inputenc}
\usepackage{enumitem}
\usepackage{fancyhdr}
\usepackage{lastpage}
\usepackage{hyperref}
\usepackage{tikz-cd}
\usepackage{verbatim}
\usepackage{leftidx}
\usepackage{footmisc}
\usepackage{subcaption}

\newcommand{\moy}[1]{\langle #1 \rangle}

\newcommand{\abs}[1]{\left|#1\right|}

\newcommand{\p}{\partial}

\newcommand{\Hom}{\operatorname{Hom}}
\newcommand{\Aut}{\operatorname{Aut}}

\newcommand{\Map}{\operatorname{Map}}
\newcommand{\coker}{\operatorname{coker}}
\newcommand{\inv}{\operatorname{inv}}
\newcommand{\id}{\operatorname{Id}}
\newcommand{\Cat}{\operatorname{\bf Cat}}
\newcommand{\Grp}{\operatorname{\bf Grp}}

\DeclareMathOperator*{\bigto}{\rightarrow}
\DeclareMathOperator*{\biglongto}{\longrightarrow}

\newcommand\rC {{\mathbf K}}

\newcommand\bG {{\mathbb G}}

\newcommand\bN {{\mathbb N}}
\newcommand\bP {{\mathbb P}}
\newcommand\bR {{\mathbb R}}

\newcommand\bZ {{\mathbb Z}}

\renewcommand\( {\left(}
\renewcommand\) {\right)}
\newcommand\nn {\nonumber}
\newcommand\ds {\displaystyle}
\newcommand\beq {\begin{equation}}
\newcommand\eeq {\end{equation}}
\newcommand\beqa {\begin{equation}\begin{array}}
\newcommand\eeqa {\end{array}\end{equation}}
\newcommand\bal {\begin{align}}
\newcommand\eal {\end{align}}

\theoremstyle{plain}
\newtheorem{thm}{Theorem}
\newtheorem{prop}[thm]{Proposition}

\theoremstyle{definition}
\newtheorem{defn}[thm]{Definition}
\newtheorem{eg}[thm]{Example}

\theoremstyle{remark}

\fancyhfoffset{0pt}

\title{\Huge Higher categorical groups and the classification of topological defects and textures}
\author{
J.~P.~Ang$^{a,b,}$\thanks{jianpeng.ang@gmail.com (corresponding author)}~~ and Abhishodh~Prakash$^{a,b,c,}$\thanks{abhishodh.prakash@icts.res.in} \\[10pt]
$^{a}${\it\small Department of Physics and Astronomy,} \\[-1ex]
	{\it\small State University of New York at Stony Brook,} \\[-1ex]
	{\it\small Stony Brook, NY 11794-3840, USA} \\[5pt]
$^{b}${\it\small C.~N.~Yang Institute for Theoretical Physics,} \\[-1ex]
	{\it\small Stony Brook, NY 11794-3840, USA} \\[5pt]
$^{c}${\it\small International Centre for Theoretical Sciences (ICTS-TIFR),} \\[-1ex]
	{\it\small Tata Institute of Fundamental Research,} \\[-1ex]
	{\it\small Shivakote, Hesaraghatta, Bangalore 560089, India}
}
\date{}

\begin{document}
\thispagestyle{plain}
\begin{flushright}
YITP-SB-18-29
\end{flushright}
{\let\newpage\relax\maketitle}

\begin{abstract}
Sigma models effectively describe ordered phases of systems with spontaneously broken symmetries. At low energies, field configurations fall into solitonic sectors, which are homotopically distinct classes of maps. Depending on context, these solitons are known as textures or defect sectors. In this paper, we address the problem of enumerating and describing the solitonic sectors of sigma models. We approach this problem via an algebraic topological method -- combinatorial homotopy, in which one models both spacetime and the target space with algebraic objects which are higher categorical generalizations of  fundamental groups, and then counts the homomorphisms between them. We give a self-contained discussion with plenty of examples and a discussion on how our work fits in with the existing literature on higher groups in physics.
\end{abstract}

\newpage
\section{Introduction}
Methods of algebraic topology have now been successfully used in physics for over half a century. One of the earliest uses of homotopy theory in physics was by Skyrme~\cite{Skyrme1962}, who attempted to model the nucleon using solitons, or topologically stable field configurations, which are now called skyrmions. Skyrmions have found a life in condensed matter systems and are reported to be observed in Bose-Einstein condensates~\cite{skyrmions2001} and liquid crystals~\cite{skyrmions2011}. Homotopy theory techniques were also employed in the classification of topological defects in ordered media~\cite{Mermin:1979zz} like vortices in superfluids and superconductors. Skyrmions and vortices are experimentally accessible examples of topological textures and defects in ordered media. Ordering refers to the phenomenon of spontaneous symmetry breaking whereby certain local order parameters dynamically pick up non-zero expectation values. In such a situation, the low-energy dynamics can be approximated by a non-linear sigma model of maps, known as order parameters, taking values in the space of cosets of the residual symmetry subgroup in the global symmetry group \cite{Nambu1960,Goldstone1961}. Textures and defects are topological sectors of the sigma model, i.e. homotopically distinct classes of these maps. It should also be noted that sigma models are not just useful in describing phases of spontaneously broken symmetry. For instance, the sigma model in two dimensions describes the  motion of a relativistic string in some background geometry, and topological sectors in this case correspond to different winding modes of the string. Gauge theories can be described as a sigma model with target space being the classifying space of the gauge group. In this setting, topological sectors are the different principal bundles (i.e. instanton sectors) of the gauge theory. In these examples, the target space is not restricted to be a coset space. In this paper, we are interested in the classification of topological sectors of sigma models; in other words, the classification of homotopy classes of maps.

The problem of homotopy classification of maps is an old problem in algebraic topology. One approach, which we adopt in this paper, is to model the spaces algebraically, and count the homomorphisms between the algebraic models. This is known as combinatorial homotopy. When the space is two dimensional, the relevant algebraic objects are known as crossed modules \cite{whitehead1949,whitehead1949b}. Later, it was realized that crossed modules are equivalent to categorical groups \cite{brown1976} and to strict 2-groups \cite{sinh1975}, which have received much attention in recent years from the physics community. While it was understood early on that physical theories often have higher categorical symmetries, interest in this topic has intensified in the last five years or so when its ubiquity in various areas in physics was appreciated. To give some examples, gapped phases of gauge theories can sometimes be understood as spontaneously broken phases of higher group symmetries \cite{Gukov:2013zka,Kapustin:2013uxa,Gaiotto:2014kfa}; anomalies of higher group symmetries can be used to constrain IR behaviors of theories and test conjectured infrared dualities \cite{Benini:2017dus,Benini:2018reh,Cordova:2017kue}; Green-Schwarz mechanisms have been understood as a manifestation of gauged higher group symmetries \cite{Cordova:2018cvg}; the classification of symmetry enriched topological phases can be understood in terms of coupling an ordinary group to a higher group \cite{Barkeshli:2014cna}. We would like to add to this wealth of literature by pointing out yet another physical phenomenon, that of topological sectors of defects and textures, which falls under this paradigm. Indeed, this was the original motivation of this paper.

This paper is written with two audiences in mind. First, for condensed matter physicists, we would like to present a systematic technique to classify textures and defect sectors. The configurations are represented by homomorphisms between algebraic models of the spaces, so in principle it should be possible to explicitly construct each configuration. To illustrate the technique, we include many examples, ordered by complexity, at the end of each section. Second, for mathematical physicists, we would like to highlight the role played by higher categorical groups in this story, and relate it to other occurences of higher groups in physics. A self-contained introduction to category theory and higher groups can be found in appendix \ref{app.category}. We point out in section \ref{sec.2gsect} that the classification of two dimensional textures and defects can be understood in terms of 2-group homomorphisms, as can (for example) the classification of symmetry enriched topological phases in two spatial dimensions. In appendix \ref{app.category}, we briefly discuss the situation in higher dimensions.

The mathematical framework underpinning this paper is the combinatorial approach to the homotopy classification of maps, which began with Whitehead's result that homotopy 2-types were modeled by crossed modules \cite{whitehead1949,whitehead1949b}. It was later realized that crossed modules are equivalent to categorical groups \cite{brown1976} and to strict 2-groups \cite{sinh1975}. It was expected that higher homotopy types could be modeled by higher categorical generalizations of groups. Indeed, following the seminal work of Brown \cite{brown1987,brown1987b} generalizing van Kampen-like theorems to higher categorical groups, several subtly different models of homotopy 3-types were introduced \cite{conduche1984,baues1991,Ellis1993}. For completeness, we should mention that there is another approach to the homotopy classification of maps, via obstruction theory, where the obstruction to the existence of a homotopy between two maps is studied by iteratively extending the homotopy on skeletons. This approach traces its origins to the work of Steenrod and Postnikov \cite{steenrod1947,postnikov1949} in the 40s; for an overview see the book \cite{baues2006}.

The rest of the paper is organized as follows. In section \ref{sec.sector}, we define topological textures and defect sectors and state the problem. The dependence of these sectors on a choice of base point is also explained. In section \ref{sec.one} we briefly recall the situation in one dimension, before quickly moving on to two dimensions in section \ref{sec.two}. The principal object of study in dimension two, the fundamental crossed module, is introduced in section \ref{sec.fundxm}. In section \ref{sec.2gsect}, we explain that crossed modules, categorical groups and (strict) 2-groups are equivalent, and point out other areas in physics where such structures are manifest. In section \ref{sec.three}, we turn to three dimensions, where the principal object is the fundamental crossed square, which is introduced in section \ref{sec.fundxs}. We conclude and list directions for further study in section \ref{sec.conc}. In appendix \ref{app.category}, we give a self-contained introduction to category theory and state our results in that language. In appendix \ref{app.freexm}, we define the notion of freeness for crossed modules. In appendix \ref{app.grpext}, we recall the relation between group cohomology and extensions of groups and crossed modules. In appendix \ref{app.fundxs}, we provide a detailed construction of the fundamental crossed square of a canonical CW complex. In appendix \ref{app.calc}, more details of a tedious calculation from example \ref{sec.t3s2} are presented.

\section{Topological sectors} \label{sec.sector}
The configuration space of a sigma model is the space\footnote{with the compact-open topology} of maps $\Map(M,X)$ from some connected topological space $M$ to some other connected topological space $X$ (the target space). For instance, in the low energy effective theory of a symmetry broken phase, $X$ is the coset space of Goldstone modes. $M$ can be thought of as either space or spacetime, depending on context, in which case the configurations are known as textures, or as the complement of a subspace $D$ of positive codimension in space or spacetime, in which case the configurations are known as defects (more precisely, the defect is supported on $D$).

\begin{figure}
\centering
\parbox{0.4\textwidth}{\begin{subfigure}{0.4\textwidth}
\centering
{\includegraphics[width=0.5\textwidth]{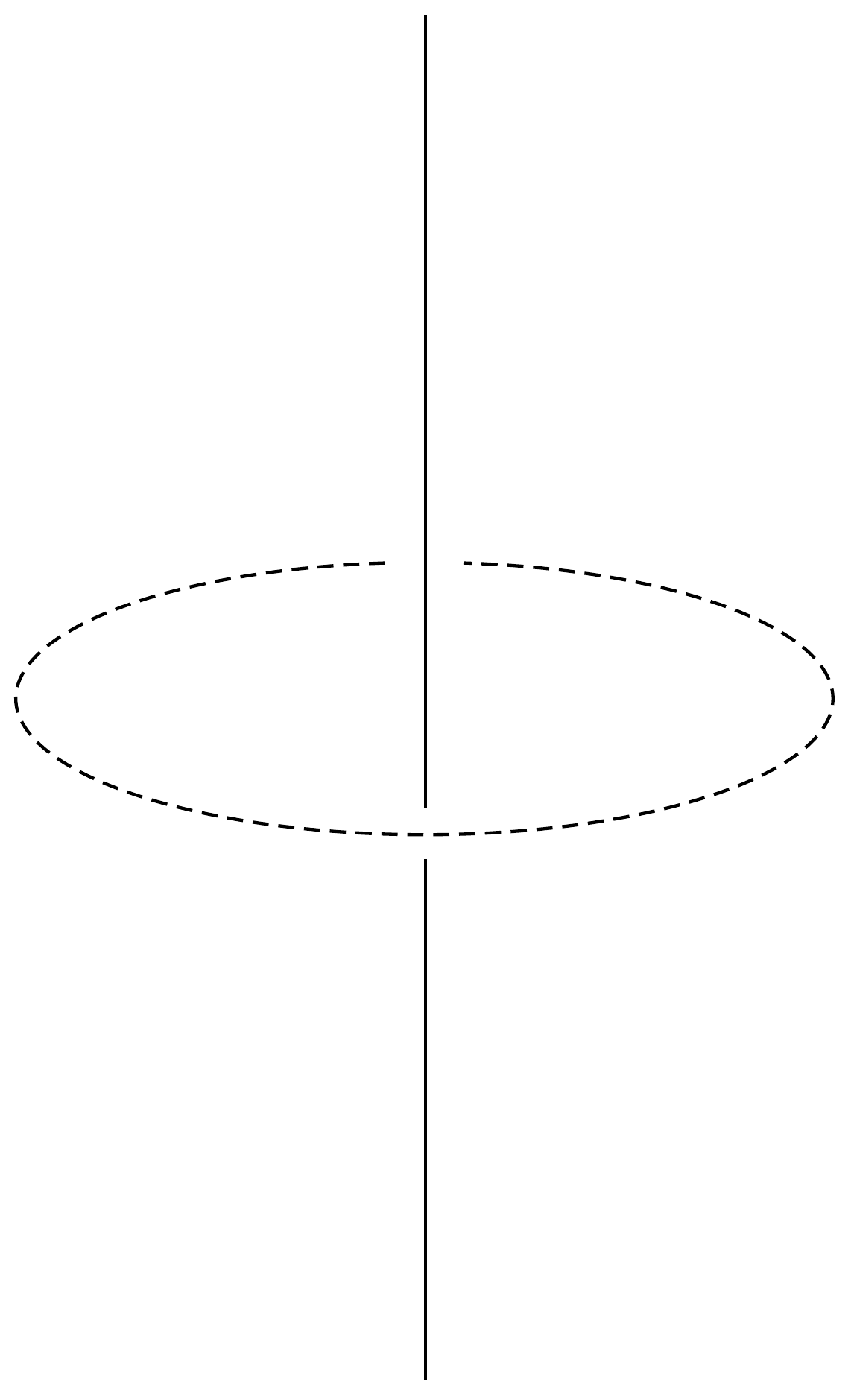}}
\caption{The line defect in three dimensions, $M=\bR^3\setminus\bR$, retracts onto the dashed circle $S^1$.}
\label{fig.linedefect}
\end{subfigure}}
\parbox{0.4\textwidth}{\begin{subfigure}{0.4\textwidth}
\centering
{\includegraphics[width=0.6\textwidth]{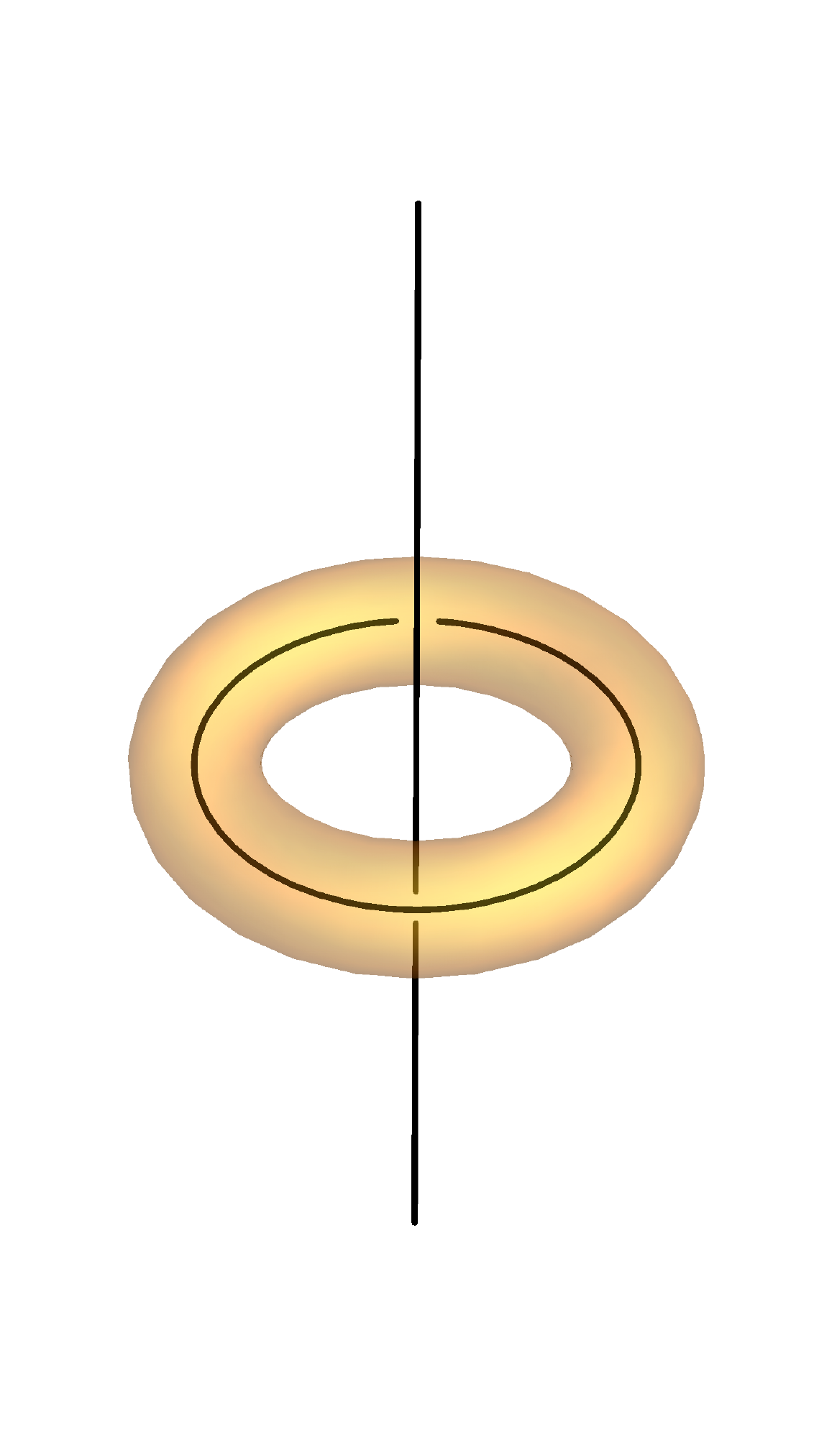}}
\caption{A line defect linked to a loop defect in three dimensions retracts onto the torus $T^2$.}
\label{fig.torusdefect}
\end{subfigure}}
\caption{Examples of spaces with defects.}
\end{figure}

Topological theta terms in the action of the sigma model assign phases in $U(1)$ to connected components of the configuration space $\pi_0\Map(M,X)=[M,X]$ which are homotopy classes of maps. We call these homotopy classes the topological sectors of the model. In this paper, we address the problem of classifying the topological sectors; i.e. describing the set $[M,X]$ for fixed $M$ and $X$. Clearly, $[M,X]$ depends only on the homotopy types of $M$ and $X$. Therefore, we can often replace $M$ and $X$ with homotopy equivalent spaces (such as a deformation retract), which is easier to deal with. For example, codimension $n+1$ defects in $\bR^d$ retracts onto a sphere linking the defect for $n\geq 1$; see figure \ref{fig.linedefect}. For another less trivial example, the Hopf link (two circles linking once) defect in $S^3$ is homotopy equivalent to a loop defect linking a line defect once in $\bR^3$ (by stereographic projection with respect to a point on the defect). Both of these configurations retract onto the 2-torus (see figure \ref{fig.torusdefect}). This procedure of finding ``simple'' homotopy equivalent spaces of defect configurations is a procedure often used in knot theory when computing knot or link groups (which are the fundamental groups of the knot/link complements).

\subsection{Base point dependence}
It is often easier to compute \emph{base pointed} homotopy classes $[M,X]_0$. This is defined by fixing base points\footnote{We abuse notation by denoting all base points by $\ast$.} $\ast\in M$ and $\ast\in X$, and considering only those maps from $M$ to $X$ carrying base point to base point. Such base pointed maps form a subspace $\Map_0(M,X)\subset\Map(M,X)$. The base pointed homotopy classes $[M,X]_0$ are the quotient of $\Map_0(M,X)$ by base point preserving homotopies (i.e. the base point of $M$ must be mapped to the base point of $X$ throughout the homotopy). Note that if we had quotiented $\Map_0(M,X)$ by all homotopies, we would have obtained $[M,X]$ instead. Base pointed homotopy classes often admit more structure than the unpointed versions; for instance, a natural group structure can be defined on $[S^n,X]_0=\pi_n(X,\ast)$, but not on the unpointed classes $[S^n,X]$. The following proposition allows one to recover the unbased, or free, homotopy classes from based homotopy classes.
\begin{prop} \label{thm.basept}
The inclusion of $\Map_0(M,X)$ into $\Map(M,X)$ leads to a surjection $[M,X]_0\to[M,X]$. That is, each free homotopy class contains a based representative. Furthermore, two based homotopy classes are free homotopic if and only if they differ by an action of the fundamental group $\pi_1(X,\ast)$. In other words, the free homotopy classes are the $\pi_1(X,\ast)$ orbits of the based homotopy classes
\beq [M,X]_0/\pi_1(X,\ast) \xrightarrow{\sim} [M,X]. \eeq
\end{prop}
The $\pi_1(X,\ast)$ action on $[M,X]_0$ can be described as follows: a loop $\gamma:[0,1]\to X$ acts on $[M,X]_0$ by sending a representative $f_0:M\to X$ to a representative $f_1:M\to X$ which is (free) homotopic via the homotopy $f_t$. The homotopy $f_t$ is defined by sending the base point $\ast\in M$ around the loop $\gamma$, in other words, $f_t(\ast)=\gamma(t)$. Such a homotopy always exists (due to the homotopy extension property of $(X,\ast)$ \cite{hatcher2003}). This action clearly depends on $\gamma$ only up to homotopy.

For example, when $M=S^1$, this action reduces to the conjugation action of $\pi_1(X,\ast)$ on $[S^1,X]_0=\pi_1(X,\ast)$. Applying proposition \ref{thm.basept}, this recovers the statement that $[S^1,X]$ is in bijection with the conjugacy classes of $\pi_1(X,\ast)$. When $M=S^n$, this action reduces to the usual action of $\pi_1(X,\ast)$ on $[S^n,X]_0=\pi_n(X,\ast)$, recovering the statement that $[S^n,X]$ is in bijection with the orbits of $\pi_n(X,\ast)$ under the action of $\pi_1(X,\ast)$. This recovers the classification of codimension $n+1$ defects (where $M=\bR^d\setminus\bR^{d-n-1}$) \cite{Mermin:1979zz}.

\subsection{Reduced CW complexes}
The remainder of this paper is devoted to computing the base pointed homotopy classes $[M,X]_0$ using combinatorial homotopy, as mentioned in the introduction. We shall assume that both $M$ and $X$ are CW complexes.\footnote{This is not a restrictive assumption, as CW structures can be defined on very general topological spaces, including for example all manifolds and even all simplicial complexes. Spaces for all cases of physical interest that we are aware of admit CW structures. See for example the appendix of \cite{hatcher2003} for an introduction to CW complexes.} Roughly speaking, a CW complex $M$ consists of sets of $d$-dimensional cells $\Sigma_d$ and gluing maps $\sigma_d$, which describe how $M$ can be ``built'' from its constituent cells. Cells of dimension 0 are just points; let the 0-skeleton $M^0$ be the collection of points in $\Sigma_0$. Inductively, for each $d$, $d\geq 1$, the gluing map $\sigma_d:\Sigma_d\to M^{d-1}$ specifies how the boundary of each $d$-cell is attached to the $d-1$-skeleton $M^{d-1}$. The space resulting after the $d$-cells are glued is the $d$-skeleton $M^d$.

The \emph{dimension} of a CW complex is the highest integer $n$ for which $\Sigma_n$ is non-empty, so $M=M^q$ for any $q\geq n$ (if $n$ exists).

\begin{figure}
\centering
\includegraphics[width=0.4\textwidth]{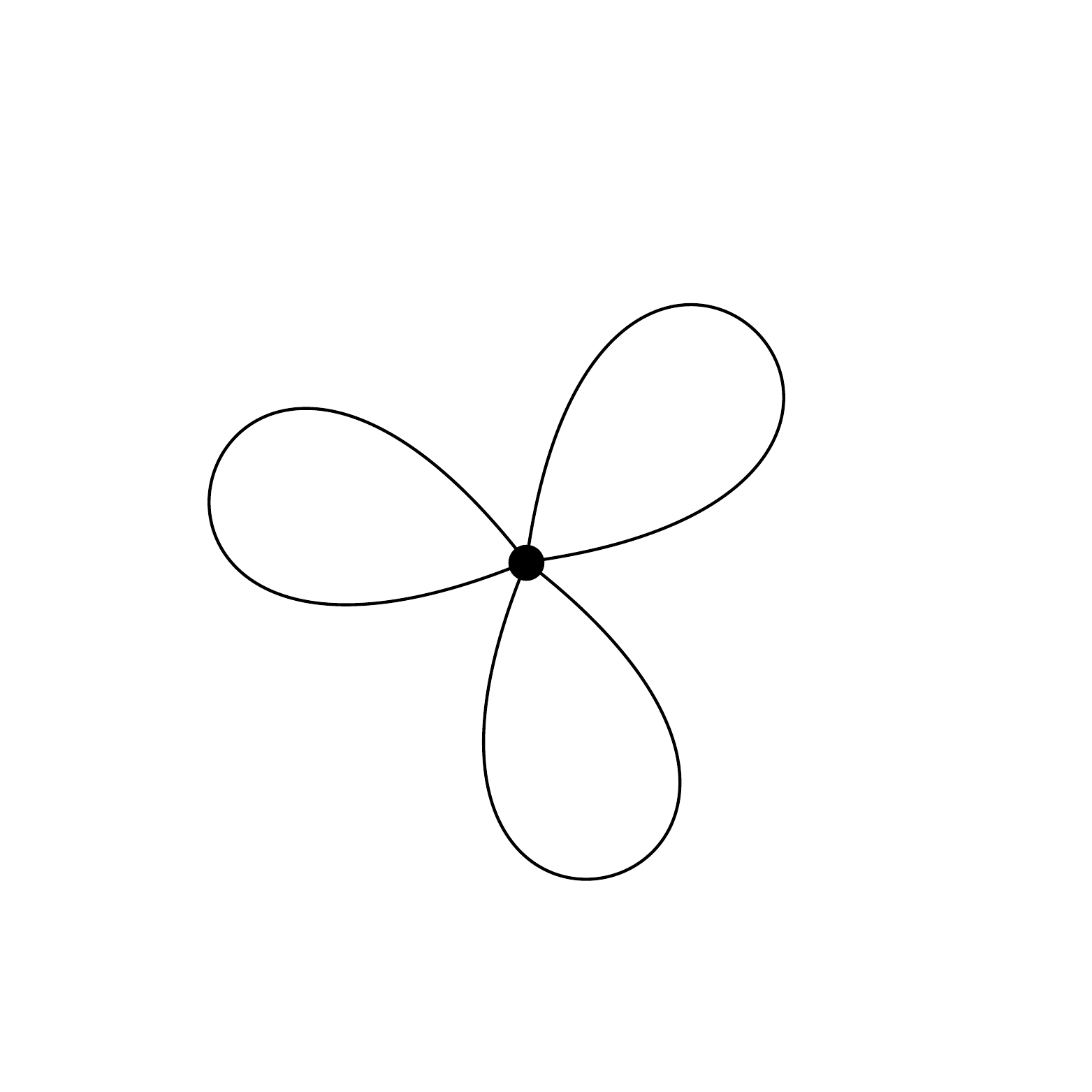}
\caption{An example of a one dimensional reduced CW complex with three 1-cells. The black blob is the base point and the unique 0-cell. There is no choice of gluing map $\sigma_1$ for a one dimensional reduced CW complex, since it must send the boundaries of every 1-cell to the base point. Indeed, this shows that all one dimensional reduced CW complexes are wedge sums\protect\footnotemark of circles, or ``bouquets''.}
\label{fig.cweg}
\end{figure}

\begin{figure}
\centering
\includegraphics[width=0.3\textwidth]{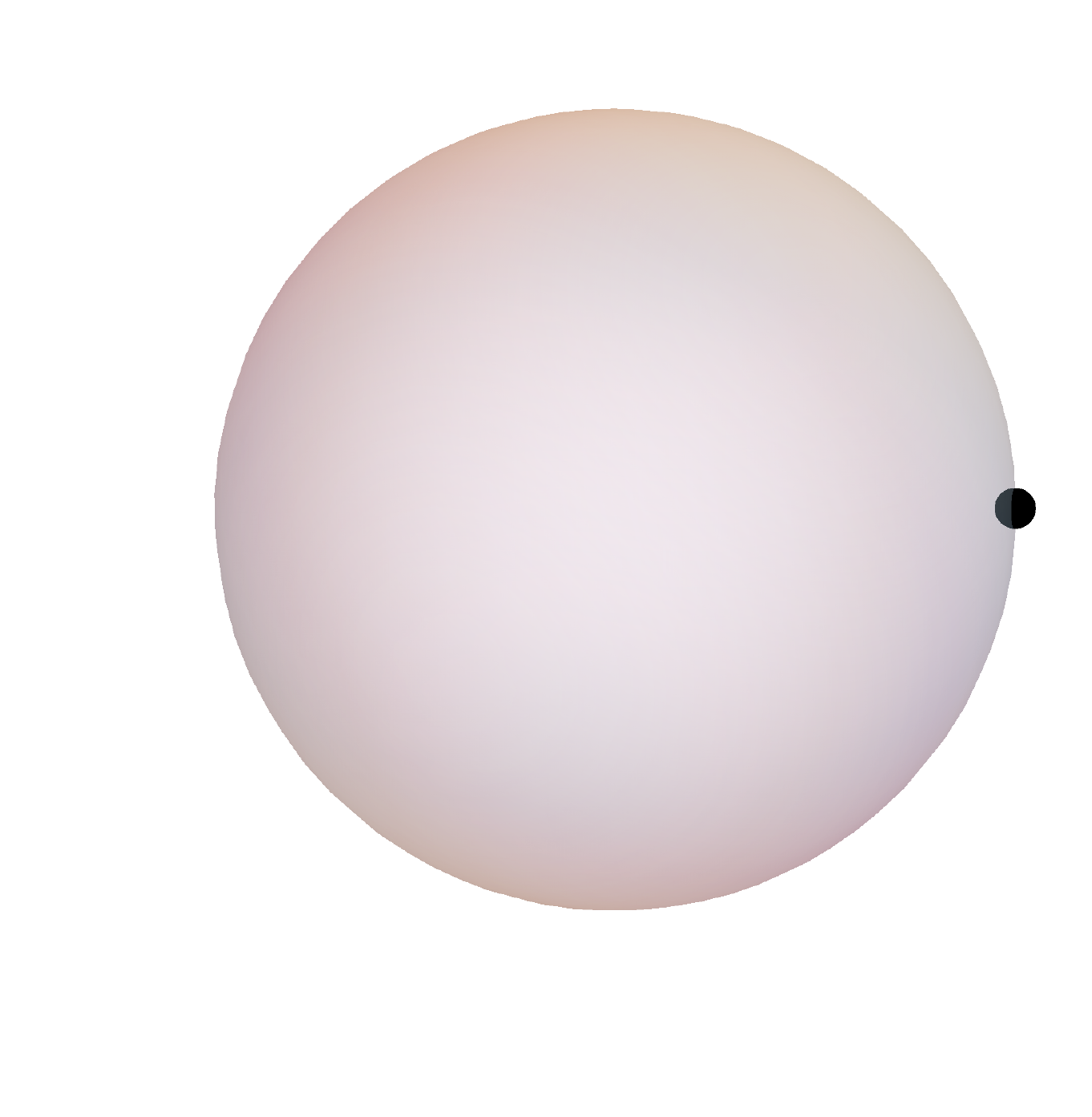}
\caption{A reduced CW structure on the 2-sphere, consisting of one 0-cell (the black blob), no 1-cells and one 2-cell attached trivially to the 0-cell.}
\label{fig.cwsphere}
\end{figure}

\begin{figure} 
\centering
\raisebox{-0.5\height}{\includegraphics[width=0.35\textwidth]{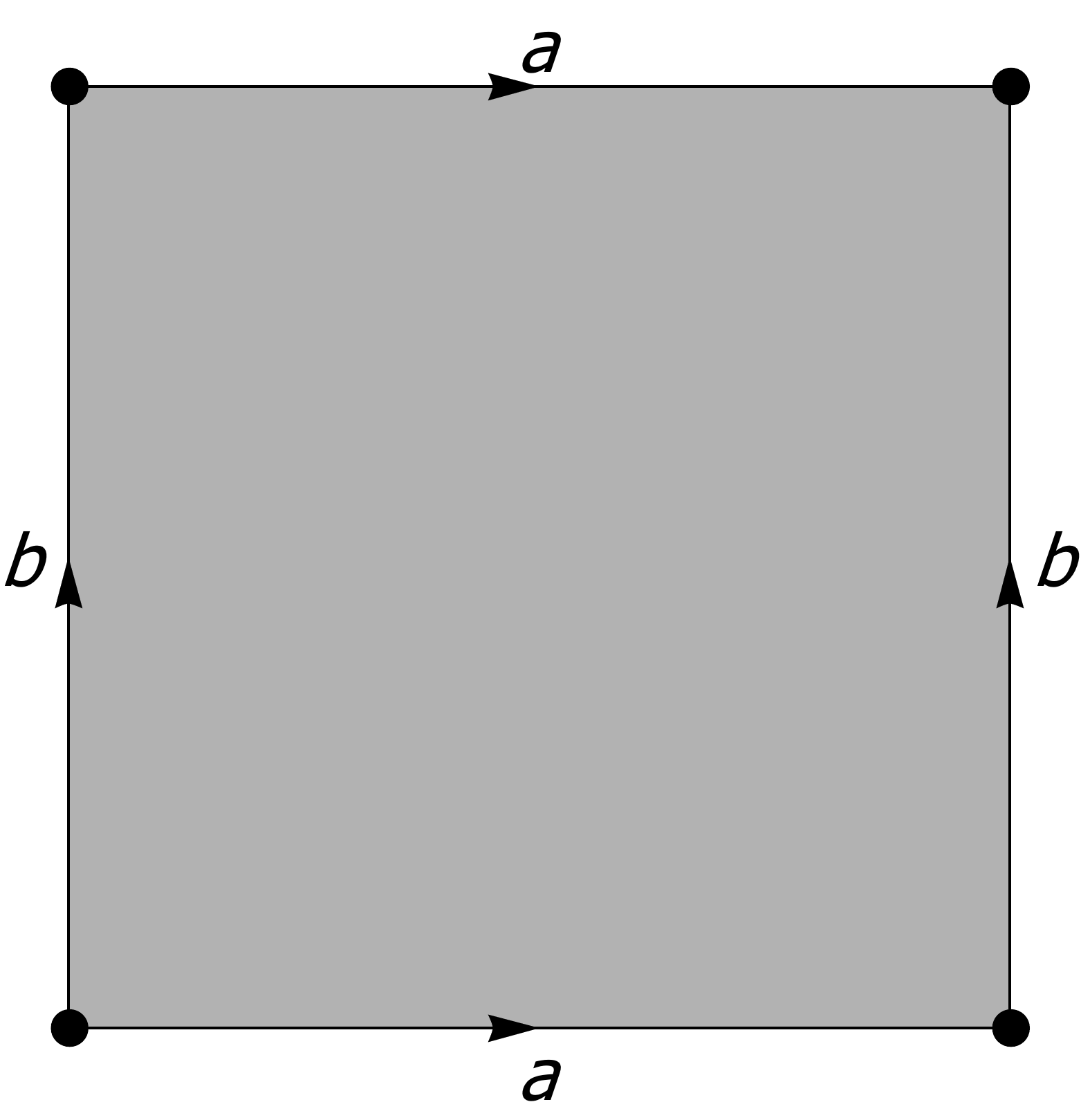}}
\raisebox{-0.5\height}{\includegraphics[width=0.6\textwidth]{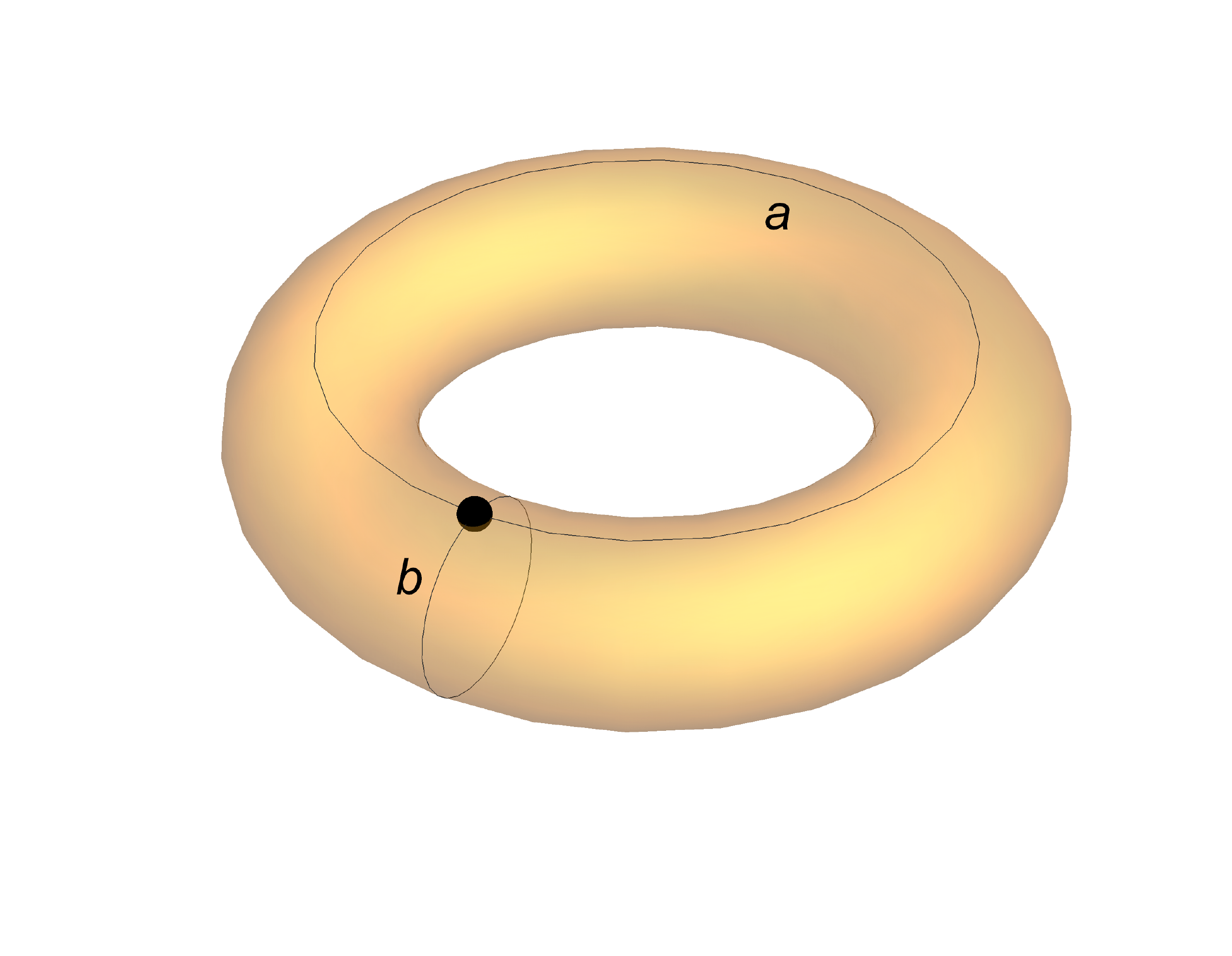}}
\caption{A reduced CW structure on the 2-torus, consisting of one 0-cell (the black blob), two 1-cells $\Sigma_1=\{a,b\}$ (opposite edges are identified in the figure on the left), and one 2-cell (shaded). The gluing map $\sigma_2$ maps the boundary of the 2-cell to the commutator $aba^{-1}b^{-1}$.}
\label{fig.cwtorus}
\end{figure}

We shall make the further assumption that $M$ and $X$ are \emph{reduced} CW complexes, which means that there is exactly one 0-cell, which also serves as the base point of the space.\footnote{Every CW complex is homotopy equivalent to a reduced version \cite{baues1998}.} See figures \ref{fig.cweg}, \ref{fig.cwsphere} and \ref{fig.cwtorus} for examples of reduced CW complexes.

\footnotetext{\label{foot.wedge}The wedge sum of a collection of pointed spaces is the space given by the union of the pointed spaces modulo identifying all the base points. This point serves as the base point of the wedge sum.}

Under the assumptions that $M$ and $X$ are reduced CW complexes, the homotopy information of the spaces can be presented in terms of an algebraic object, whose complexity increases as a function of dimension. In dimension one, it is the fundamental group, and therefore in higher dimensions the objects may be thought of as generalizations of the fundamental group. Homotopy classes of maps then correspond to classes of homomorphisms between these algebraic objects -- thus translating the problem of computing topological sectors into a combinatorial problem.

\section{Dimension one} \label{sec.one}
We warm up with the case where $M$ is a one dimensional CW complex. Homotopy 1-types are modeled by groups, so we have the following result.\footnote{From now on, to avoid cumbersome notation we suppress the base point dependence in homotopy groups $\pi_1M=\pi_1(M,\ast)$. The homotopy groups, and generalizations thereof, that we consider in this paper are always pointed.}
\begin{prop}
If $M$ is a reduced CW complex of dimension 1, and $X$ is a reduced CW complex, then
\beq [M,X]_0 \simeq \Hom\(\pi_1M,\pi_1X\). \eeq
\end{prop}
For instance, when $M=S^1$ we obtain $[S^1,X]_0\simeq\Hom(\bZ,\pi_1X)\simeq\pi_1X$. Under change of base point, $\pi_1X$ acts on $[S^1,X]_0$ by conjugation, so this recovers the classification of codimension 2 defects $M\equiv \bR^d\setminus\bR^{d-2}$ by conjugacy classes of $\pi_1X$.

\begin{figure}
\centering
\includegraphics[width=0.6\textwidth]{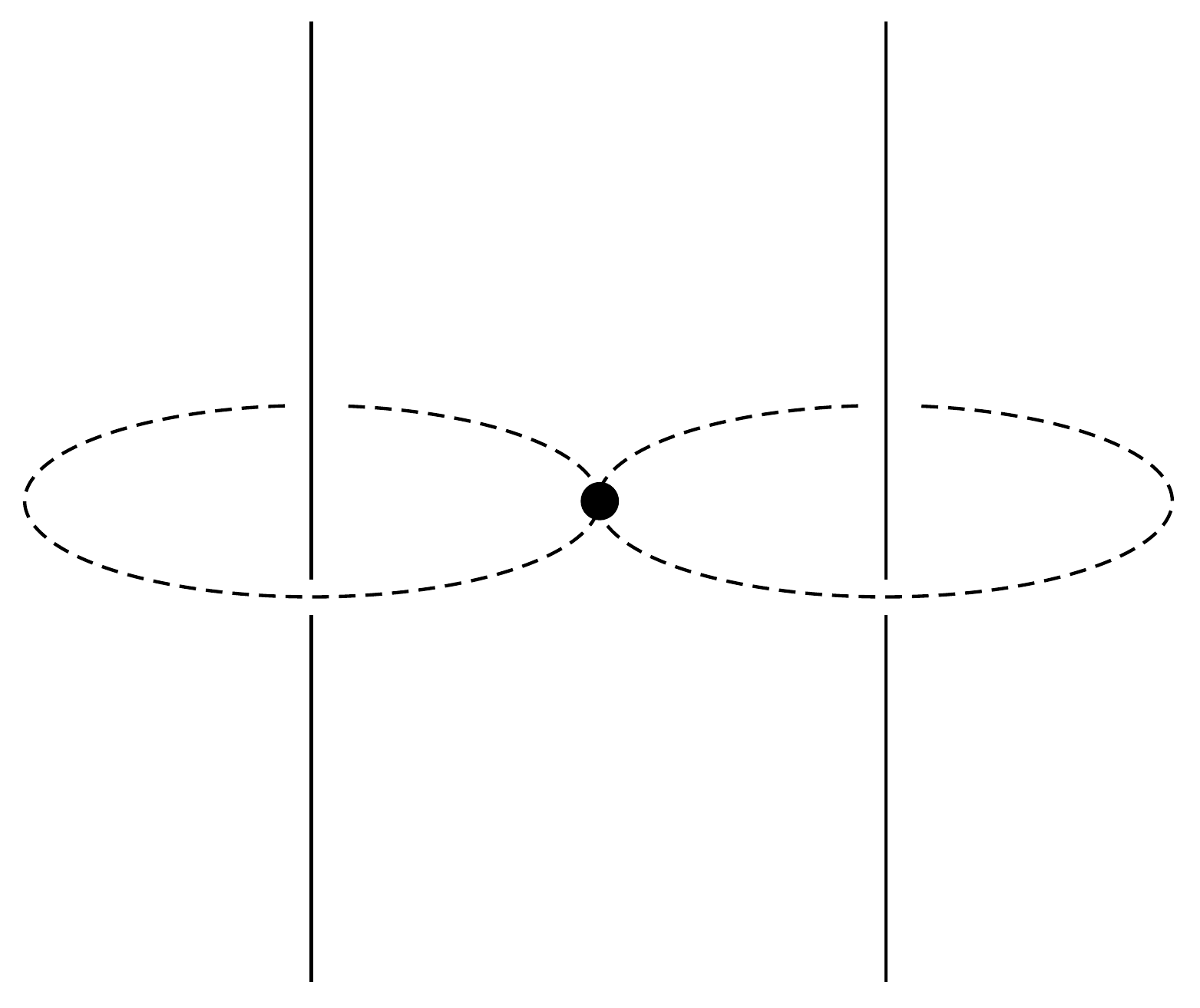}
\caption{Two isolated line defects in $\bR^3$ retracts onto the figure of eight (dashed).}
\label{fig.fig8}
\end{figure}

When $M=S^1\vee S^1$ is the wedge sum\footnote{See footnote \ref{foot.wedge} for the definition of the wedge sum, denoted by $\vee$.} of two circles, commonly known as the figure of eight, we obtain $[S^1\vee S^1,X]_0\simeq\Hom(\bZ\ast\bZ,\pi_1X)\simeq\pi_1X\times\pi_1X$. Upon taking the quotient by the conjugation action of $\pi_1X$, this recovers the classification of two isolated codimension 2 defects by two conjugacy classes of $\pi_1X$ (see figure \ref{fig.fig8}). (As we mentioned in figure \ref{fig.cweg}, all reduced CW complexes of dimension 1 are wedge sums of circles, so the most general one dimensional case is an easy generalization of this.)

\section{Dimension two} \label{sec.two}
Homotopy 2-types are modeled by algebraic objects known as crossed modules, which we shall define later in this section. The homotopy classification result in two dimensions is similar to that in one dimension, with two differences: one, the fundamental group replaced by the \emph{fundamental crossed module}, and two, there is a notion of homotopy between homomorphisms which we need to quotient by (this happened to be trivial in the one dimensional case).
\begin{thm}[\cite{whitehead1949b}] \label{thm.2d}
Let $M$ be a reduced CW complex of dimension 2, and $X$ be a reduced CW complex. Then
\beq [M,X]_0 \simeq [\Pi_{\leq 2}M,\Pi_{\leq 2}X]_0, \eeq
and
\beq [M,X] \simeq [\Pi_{\leq 2}M,\Pi_{\leq 2}X], \eeq
where $\Pi_{\leq 2}M$ is the fundamental crossed module of $M$, and $[\Pi_{\leq 2}M,\Pi_{\leq 2}X]_0$ (resp. $[\Pi_{\leq 2}M,\Pi_{\leq 2}X]$) denotes the equivalence classes of $\Hom(\Pi_{\leq 2}M,\Pi_{\leq 2}X)$ under the based (resp. free) homotopy equivalence relation.
\end{thm}
These algebraic concepts will be explained below, and a combinatorial method of computing the homotopy classes of homomorphisms will be presented. Before proceeding, we note that, for readers more familiar with algebraic topology, the equivalence classes of homomorphisms between two fundamental crossed modules can in fact be expressed in terms of cohomology groups.
\begin{prop} \label{thm.2dcoh}
The equivalence classes of homomorphisms between the fundamental crossed modules of $M$ and $X$ can be expressed as
\beq [\Pi_{\leq 2}M,\Pi_{\leq 2}X]_0 = \bigcup_{\phi_1}H^2_{\phi_1}(M,\pi_2X), \eeq
where the union is over homomorphisms $\phi_1\in\Hom(\pi_1M,\pi_1X)$ of fundamental groups. The cohomology groups have local coefficients ($\pi_2X$ is viewed as a module over $\pi_1M$ via the homomorphism $\phi_1$ and the action of $\pi_1X$).
\end{prop}
The cohomology group with local coefficients may be computed explicitly by constructing the cellular cochain complex $C^\ast(M,\pi_2X)$. Alternatively, it is isomorphic to the $\pi_1M$-equivariant cohomology group $H^2_{\phi_1(\pi_1M)}(\tilde M,\pi_2X)$ of the universal cover $\tilde M$, which may be computed using the cochain complex $\Hom^{\pi_1M}(C_\ast\tilde M,\pi_2X)$. Yet another alternative is to use the Cartan-LeRay spectral sequence to derive a K\"unneth-like formula for these cohomology groups in terms of ordinary group cohomology groups. A discussion of how to compute these cohomology groups is beyond the scope of this paper; we refer the interested reader to the textbook \cite{whitehead78} for more details about cohomology with local coefficients, and to \cite{mccleary1985} for more details about spectral sequences.

We would like to stress that knowledge of these cohomology theories is \emph{not} required, as we will present below an explicit algorithm for computing the homotopy classes of homomorphisms. Readers who are interested in examples without getting too involved in the details are invited to skip ahead to section \ref{sec.2deg} where examples are worked out.

\subsection{Fundamental crossed module} \label{sec.fundxm}
\begin{defn} \label{def.xm}
A crossed module is a homomorphism $\p:H\to G$ of groups together with an action of $G$ on $H$, denoted by $(g,h)\mapsto \leftidx{^g}{h}$, where $g\in G$ and $h\in H$, satisfying the properties
\begin{enumerate}
\item $\p(\leftidx{^g}{h})=g\p(h)g^{-1}$ and
\item $\leftidx{^{\p(h)}}{h'}=hh'h^{-1}$, for all $g\in G$ and $h,h'\in H$.
\end{enumerate}
If condition 1 is satisfied but not condition 2, then the homomorphism is known as a pre-crossed module.
\end{defn}

\begin{defn}
A homomorphism $\phi$ of crossed modules from $\p:H\to G$ to $\p':H'\to G'$ consists of a group homomorphism $\phi_1:G\to G'$, and a $\phi_1$-equivariant group homomorphism $\phi_2:H\to H'$ (that is, $\phi_2(\leftidx{^g}{h})=\leftidx{^{\phi_1(g)}}{\phi_2(h)}$ for all $g\in G, h\in H$) such that the diagram
\beq \begin{tikzcd}
H \ar[r,"\p"] \ar[d,"\phi_2"]& G \ar[d,"\phi_1"] \\
H' \ar[r,"\p'"]& G'
\end{tikzcd} \eeq
commutes.
\end{defn}

\begin{defn}
Two homomorphisms $\phi,\psi$ of crossed modules from $\p:H\to G$ to $\p':H'\to G'$ are free homotopic if there exists a function $\theta:G\to H'$
\beq \begin{tikzcd}
H \ar[r,"\p"] \ar[d,"{\phi_2,\psi_2}"'] & G \ar[d,"{\phi_1,\psi_1}"] \ar[ld,"\theta"'] \\
H' \ar[r,"\p'"] & G'
\end{tikzcd} \eeq
and some element $\gamma\in G'$ satisfying
\beq \label{eqn.homotopy} \begin{cases}
\theta(gg') =& \theta(g)\ \leftidx{^{\gamma\psi_1(g)\gamma^{-1}}}{\theta(g')}, \\
\p'\theta(g) =& \phi_1(g)\gamma\psi_1(g)^{-1}\gamma^{-1}, \\
\theta(\p h) =& \phi_2(h)\ \leftidx{^\gamma}{\psi_2(h)}^{-1}.
\end{cases} \eeq
If $\gamma$ can be chosen to be the identity, then $\phi,\psi$ are based homotopic (or simply homotopic).
\end{defn}
Note that homotopy is symmetric: if $(\theta,\gamma)$ is a homotopy from $\phi$ to $\psi$, then $(\tilde\theta,\tilde\gamma)$ is a homotopy from $\psi$ to $\phi$, where $\tilde\theta(g)=\leftidx{^{\gamma^{-1}}}{\theta(g)}^{-1}$ and $\tilde\gamma=\gamma^{-1}$. This implies that both notions of homotopy are equivalence relations on the set $\Hom(\p,\p')$ of homomorphisms between two crossed modules.

Just as groups model homotopy 1-types, crossed modules model homotopy 2-types. We define a functor $\Pi_{\leq 2}$, known as the \emph{fundamental crossed module} \cite{whitehead1949b}, from reduced CW complexes to crossed modules, as follows. The fundamental crossed module $\Pi_{\leq 2}M$ of a reduced CW complex $M$ is the boundary homomorphism
\beq \p:\pi_2(M,M^1,\ast) \to \pi_1(M^1,\ast) \eeq
from the homotopy group of $M$ relative to the 1-skeleton $M^1$ to the fundamental group of the 1-skeleton, together with the standard action of $\pi_1M^1$ on $\pi_2(M,M^1)$ (see the discussion after \ref{thm.basept}).\footnote{In this paper, all homotopy groups, including relative ones, and later in section \ref{sec.three} triad ones, are pointed. To avoid cumbersome notation we suppress the base point dependence from now on.}

The fundamental crossed module $\Pi_{\leq 2}M$ contains all the homotopy information\footnote{
The fundamental crossed module $\Pi_{\leq 2}M$ does depend on the choice of CW structure on $M$, but only up to free product with the fundamental crossed module $\Pi_{\leq 2}(D^2,S^1)=\id:\bZ\to\bZ$ of the two dimensional disk with its boundary as its 1-skeleton \cite{Martins:2008uvd}.
} of $M$ as a 2-type -- in particular the fundamental group is
\beq \pi_1M = \coker\p = \pi_1M^1/\p\pi_2(M,M^1) \eeq
and the second homotopy group is
\beq \pi_2M = \ker\p. \eeq
A further piece of homotopy datum, known as the Postnikov class of $M$, can also be extracted from the fundamental crossed module. This will be discussed later in section \ref{sec.2gsect}.

The fundamental crossed module of a two dimensional complex has the property of being free (as a crossed module), so it is completely determined by the cells $\Sigma_1$, $\Sigma_2$ and the gluing map $\sigma_2:\Sigma_2\to\moy{\Sigma_1}$. The definition of freeness for a crossed module and an explicit construction of $\Pi_{\leq 2}M$ from its CW structure can be found in appendix \ref{app.freexm}.

\begin{eg}[2-sphere $S^2$]
The standard reduced CW structure on $S^2$ has no 1-cells and one 2-cell (see figure \ref{fig.cwsphere}), yielding the fundamental crossed module
\beq \Pi_{\leq 2}S^2 = \p:\bZ\to 0. \eeq
\end{eg}

\begin{eg}[2-torus $T^2$]
\label{eg.t2}
The standard reduced CW structure on $T^2$ has two 1-cells $\Sigma_1=\{a,b\}$ and one 2-cell $\Sigma_2=\{t\}$ attached via the commutator map $\sigma_2(t)=aba^{-1}b^{-1}$ (see figure \ref{fig.cwtorus}). The fundamental group of the 1-skeleton is the free group $G=\moy{a,b}$ on two generators. The fundamental crossed module is the inclusion of the derived subgroup $G'$ into $G$
\beq \Pi_{\leq 2}T^2 = \p:G' \hookrightarrow G=\moy{a,b}. \eeq
The derived subgroup $G'$ of $G$ is the subgroup generated by elements of the form $g_1g_2g_1^{-1}g_2^{-1}$ where $g_1,g_2\in G$. $G'$ is in fact a normal subgroup, so $G$ acts on it by conjugation.
\end{eg}

\begin{eg}[Real projective plane $\bR\bP^2$] \label{eg.rp2}
The standard CW structure for $\bR\bP^2$ has one 1-cell $\Sigma_1=\{a\}$ and one 2-cell $\Sigma_2\{t\}$ attached via $\sigma_2(t)=a^2$. Its fundamental crossed module is
\beq \Pi_{\leq 2}\bR\bP^2 = \p:\bZ[\bZ_2] \to \bZ. \eeq
Here, $\bZ[\bZ_2]$ denotes the free $\bZ$ module on two basis elements $\bZ_2=\{e_0,e_1\}$. The crossed module homomorphism sends a generic element $n_0e_0+n_1e_1$ of $\bZ[\bZ_2]$ to $\p(n_0e_0+n_1e_1)=2(n_0+n_1)$. $\bZ$ acts on $\bZ_2$ through parity: even elements act trivially while odd elements exchange $e_0$ with $e_1$.
\end{eg}

\subsection{Maps of CW complexes as maps of fundamental crossed modules} \label{sec.2dmap}
Just like each base pointed map between topological spaces induces a homomorphism between their fundamental groups, each base pointed cellular map between CW complexes induces a homomorphism between their fundamental crossed modules. Conversely, each homomorphism between the fundamental crossed modules can be realized as a base pointed cellular map \cite{whitehead1949b}.

In the previous section, we defined a notion of homotopy for homomorphisms of crossed modules, and this turns out to coincide with the notion of homotopy for maps of spaces. We provide some motivation for this equivalence below. Note that this justifies theorem \ref{thm.2d} -- that homotopy classes of maps between CW complexes are in bijection with homotopy classes of homomorphisms between their fundamental crossed modules.

\begin{figure}
\centering
\includegraphics[width=0.5\textwidth]{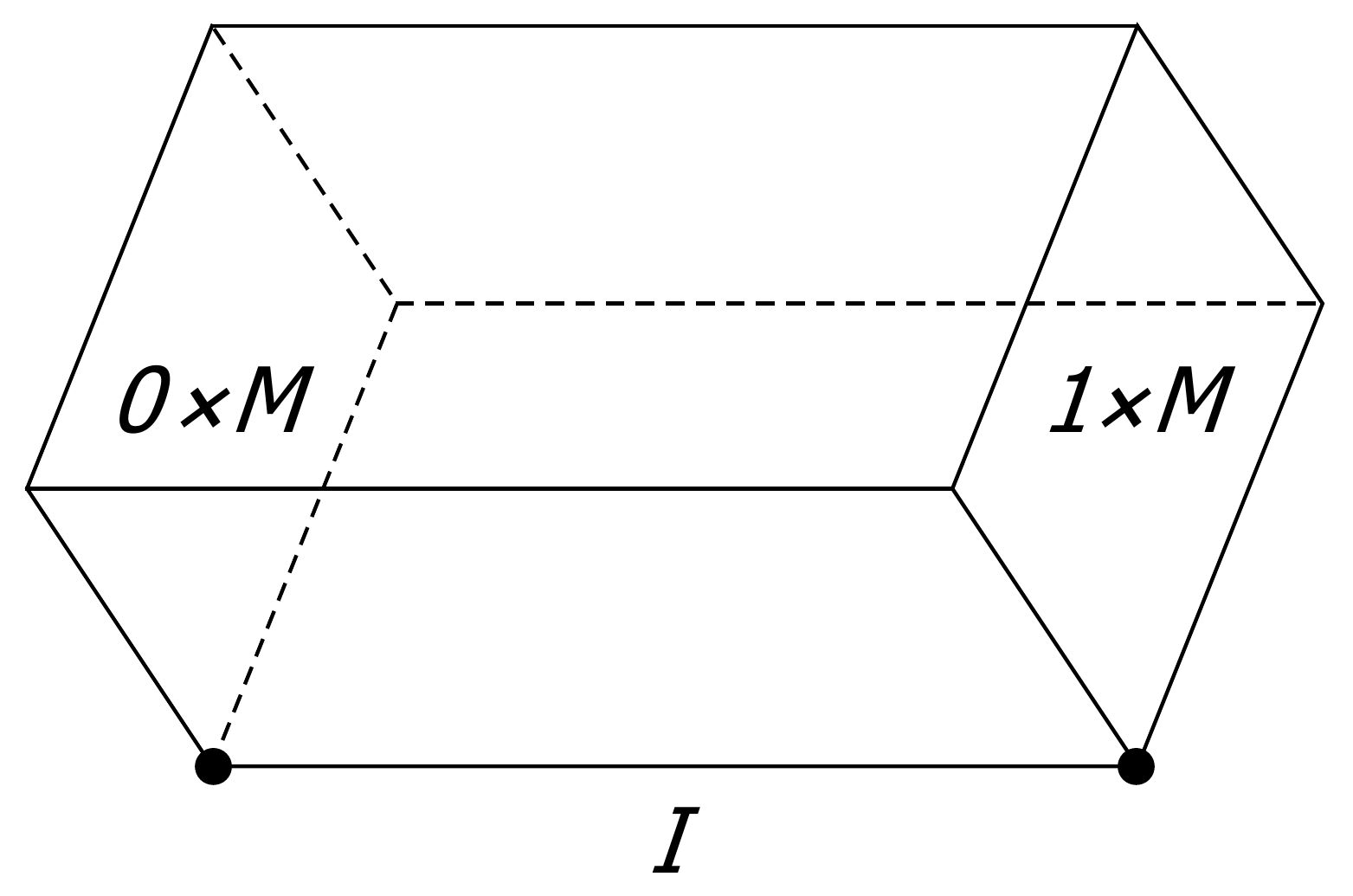}
\caption{The space $IM$ is a prism with base $M$. The black blobs are identified and serve as the base point. For $I_0M$, the entire interval labeled $I$ is identified. $IM$ and $I_0M$ contains two subcomplexes, $0\times M$ and $1\times M$, which are isomorphic to $M$.}
\label{fig.im}
\end{figure}

Given a pointed topological space $(M,\ast)$, let $IM$ denote the pointed space $[0,1]\times M$ with the two points $(0,\ast)$ and $(1,\ast)$ identified and also serving as the base point of $IM$ (see figure \ref{fig.im}). Similarly, let $I_0M$ denote the pointed space $[0,1]\times M$ with all points of the form $(t,\ast)$ identified, for $t\in[0,1]$. This also serves as the base point. A free homotopy (resp. based homotopy) between pointed maps $\phi,\psi:(M,\ast)\to (X,\ast)$ of topological spaces is a map $\Phi:(IM,\ast)\to (X,\ast)$ (resp. $\Phi:(I_0M,\ast)\to (X,\ast)$) such that $\Phi$ restricts to $\phi$ on $0\times M$ and to $\psi$ on $1\times M$. If $M$ is a CW complex, then $IM$ and $I_0M$ both admit natural CW structures admitting two copies, $0\times M$ and $1\times M$, of $M$ as CW subcomplexes. The homotopy $\Phi$ then induces a map, which we call $\Pi_{\leq 2}\Phi$, of fundamental crossed modules $\Pi_{\leq 2}IM$ (resp. $\Pi_{\leq 2}I_0M$) to $\Pi_{\leq 2}X$, coinciding with the induced maps $\Pi_{\leq 2}\phi$ and $\Pi_{\leq 2}\psi$ of fundamental crossed modules when restricted to the subcomplexes $0\times M$ and $1\times M$ respectively.

The 1-cells of the CW complex $IM$ comprise two copies of $\Sigma_1$ (the 1-cells of $M$), and the interval $I=[0,1]$; while the 2-cells of the CW complex comprise two copies of $\Sigma_2$ (the 2-cells of $M$) as well as one copy of $\Sigma_1$, corresponding to the cartesian products of 1-cells of $M$ with the interval $I$. The CW structure of $I_0M$ is the same, except that the interval $I$ omitted from the set of 1-cells. It is straightforward to show that a free or based homotopy between two crossed module homomorphisms $\phi,\psi:\Pi_{\leq 2}M\to\Pi_{\leq 2}X$ exists if and only if there exists a function $\theta:\pi_1M^1\to\pi_2(X^2,X^1)$ satisfying the properties
\beq \begin{cases}
\theta(gg') =& \theta(g)\ \leftidx{^{\gamma\psi_1(g)\gamma^{-1}}}{\theta(g')}, \\
\p_X\theta(g) =& \phi_1(g)\gamma\psi_1(g)^{-1}\gamma^{-1}, \\
\theta(\p_M h) =& \phi_2(h)\ \leftidx{^\gamma}{\psi_2(h)}^{-1},
\end{cases} \eeq
for all $g,g'\in\pi_1M^1$, $h\in\pi_2(M^2,M^1)$ and for some $\gamma\in\pi_1X^1$ in the case of a free homotopy, and for $\gamma=1$ in the case of a based homotopy. $\theta$ can be taken to be the map sending the 1-cell $g$ to the image of the 2-cell $I\times g$ under the cellular homotopy map $\Pi_{\leq 2}\Phi$. This is exactly the notion of free (resp. based) homotopy between crossed module homomorphisms (c.f. equation \eqref{eqn.homotopy}).

This prescription yields a combinatorial method of computing the topological sectors $[M,X]$ without knowledge of equivariant cohomology. Let us illustrate it in a few examples.

\subsection{Examples of two dimensional textures and defects} \label{sec.2deg}
\subsubsection{Ring defects in $\bR^3$} \label{eg.ringr3}
\begin{figure}
\centering
\includegraphics[width=0.5\textwidth]{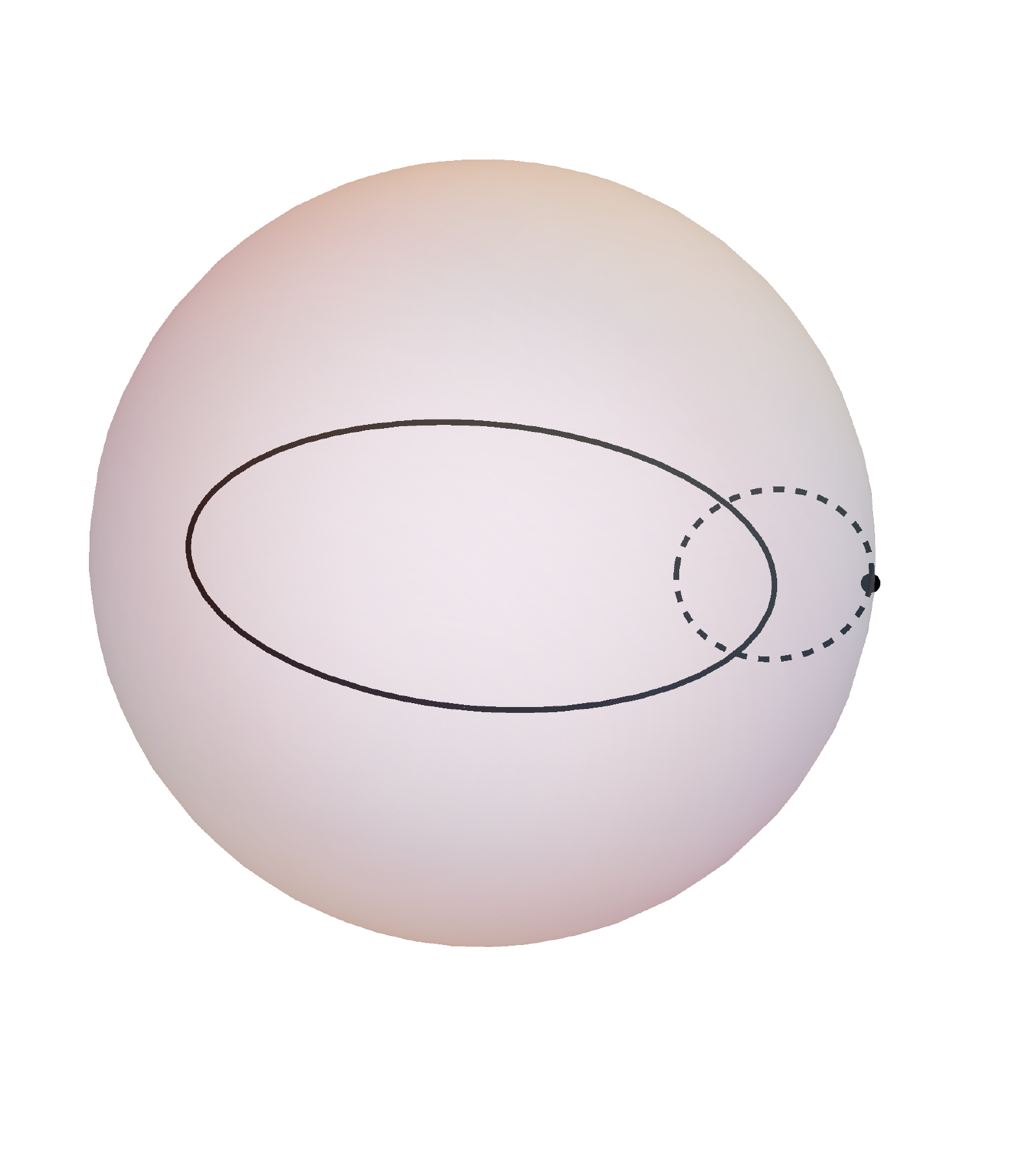}
\caption{A ring defect (solid line) in $\bR^3$ is homotopy equivalent to a circle (dashed) linking the ring defect, connected to a sphere encircling the whole ring defect at one point.}
\label{fig.ringdef}
\end{figure}

Consider one ring defect on $\bR^3$, or the combination of one line and one point defect on $\bR^3$. These two configurations are homotopy equivalent as both can be stereographically projected from one ring and one point defect on $S^3$. These configurations are all homotopy equivalent to the wedge sum $M=S^1\vee S^2$ (with the $S^1$ linking the ring and $S^2$ wrapping the point -- see figure \ref{fig.ringdef}). Thus
\beq \bR^3\setminus S^1\approx \bR^3\setminus(\bR\cup\textrm{pt}) \approx S^3\setminus(S^1\cup\textrm{pt}) \approx S^1\vee S^2. \eeq
The standard reduced CW structure on $M$ comprises one 1-cell and one 2-cell, each attached trivially to the base point.

Using our method for this relatively simple $M$ is perhaps using a sledgehammer to crack a nut, but it is instructive to see how the crossed module reduces to give the expected result. The fundamental crossed module of $M=S^1\vee S^2$ is the trivial map
\beq \p=0:\bZ[\bZ]\to\bZ, \eeq
where $\bZ[\bZ]$ denotes the free $\bZ$-module on $\bZ$, on which $\bZ$ acts by translating the basis elements. A crossed module homomorphism
\beq \begin{tikzcd}
\bZ[\bZ] \ar[r,"0"] \ar[d,"\phi_2"]& \bZ \ar[d,"\phi_1"] \\
\pi_2(X,X^1) \ar[r,"\p"]& \pi_1X^1
\end{tikzcd} \eeq
from $\Pi_{\leq 2}M$ to $\Pi_{\leq 2}X$ is determined by the image $\phi_1(1)$ of the generator $1\in\pi_1M^1$ and the image of any one basis element of $\pi_2(M^2,M^1)$, say, $e_0$. Indeed, the images of other basis elements are fixed in terms of the given one by equivariance: $\phi_2(e_n)=\phi_2(\leftidx{^n}{e_0})=\leftidx{^{\phi_1(n)}}{\phi_2(e_0)}$. Furthermore, the commutativity constraint says that $\phi_2(e_0)\in\ker\p$. Meanwhile, two homomorphisms $\phi,\psi$ are based homotopic if $\phi_1(1)\psi_1(1)^{-1}\in\p\pi_2(X,X^1)$ and $\phi_2(1)=\psi_2(1)$, so we recover the result that based homotopy classes are classified by\footnote{Note that we are considering $[S^1\vee S^2,X]_0$ as a set and not as a group. Indeed, it is possible to define a group structure on this set as a \emph{semi}direct product $\pi_2X\rtimes\pi_1X$.}
\beq [S^1\vee S^2,X]_0 \simeq \ker\p \times \coker\p = \pi_2X\times \pi_1X. \eeq
The free homotopy classes are the $\pi_1X$ orbits
\beq [S^1\vee S^2,X] \simeq (\pi_2X\times\pi_1X)/\pi_1X, \eeq
where the action is the usual action on $\pi_2X$ and conjugation on $\pi_1X$. This recovers the result of \cite{Nakanishi1988}.

\subsubsection{2d textures with periodic boundary conditions and Hopf link defects in nematic liquid crystals}
We now consider maps from the 2-torus $M=T^2=S^1\times S^1$. The most obvious example when this arises is the case of 2d textures with periodic boundary conditions. Another less obvious example which is homotopy equivalent to $T^2$ is that of the Hopf link, which consists of two ring defects linking once on $S^3$, or, equivalently, a ring defect encircling a line defect on $\bR^3$. This space retracts onto $T^2$ (see figure \ref{fig.torusdefect}).

We computed the fundamental crossed module of $T^2$ in example \ref{eg.t2}. It is the inclusion of the derived subgroup into the free group on two generators
\beq \p_{T^2}:G' \hookrightarrow G=\moy{a,b}, \eeq
with $G$ acting on its derived subgroup $G'$ via conjugation.

As target space we consider the low energy configurations of uniaxial nematic phase of liquid crystals. At low energies the liquid crystals form uniaxial molecules, which break the $SO(3)$ rotational symmetry to $O(2)$, so the coset space of Goldstone modes is $SO(3)/O(2)\simeq\bR\bP^2$ \cite{kleman1989}. The fundamental crossed module of $\bR\bP^2$ was computed in example \ref{eg.rp2}. Its fundamental crossed module is
\beq \p_{\bR\bP^2}:\bZ[\bZ_2]\to \bZ \eeq
where $\bZ_2=\{e_0,e_1\}$ and the crossed module map sends both $e_0$ and $e_1$ to $2\in\bZ$. The generator $1\in\bZ$ acts on $\bZ[\bZ_2]$ by exchanging $e_0$ and $e_1$.

The crossed module homomorphisms
\beq \begin{tikzcd}
{G'} \arrow[r, hook] \ar[d,"\phi_2"]& G \ar[d,"\phi_1"] \\
\bZ[\bZ_2] \ar[r,"\p_{\bR\bP^2}"]& \bZ
\end{tikzcd}. \eeq
are completely determined by three integers, $\phi_1(a),\phi_1(b)\in\bZ$ together with one component of $\phi_2(t)=\phi_2(t)_0e_0+\phi_2(t)_1e_1$, which we can take to be $\phi_2(t)_0\in\bZ$. (Indeed, the other component $\phi_2(t)_1$ is fixed by the commutativity of the diagram to be $\phi_2(t)_1=-\phi_2(t)_0$.) Conversely, we can check that any three integers determines a crossed module homomorphism as above. Hence, the crossed module homomorphisms are characterized by a triple of integers $(\phi_1(a),\phi_1(b),\phi_2(t)_0)\in\bZ^3$.

Two crossed module homomorphisms $\phi,\psi$ are based homotopic if there exists a function $\theta:G\to\bZ[\bZ_2]$, satisfying (see equation \ref{eqn.homotopy})
\beq \label{eqn.t2_1} \begin{cases}
\theta(gg') = \theta(g)\ \leftidx{^{\psi_1(g)}}{\theta(g')} \\
2(\theta(a)_0+\theta(a)_1) = \phi_1(a)-\psi_1(a) \\
\theta(aba^{-1}b^{-1})_0 = \phi_2(t)_0 - \psi_2(t)_0.
\end{cases} \eeq
From the first equation, we find
\begin{align} \theta(aba^{-1}b^{-1}) &= \theta(a)+\leftidx{^{\psi_1(a)}}{\theta(b)}-\leftidx{^{\psi_1(aba^{-1})}}{\theta(a)}^{-1}-\leftidx{^{\psi_1(abab^{-1})}}{\theta(b)} \nn \\
&= \theta(a)-\leftidx{^{\psi_1(b)}}{\theta(a)} -\theta(b)+\leftidx{^{\psi_1(a)}}{\theta(b)}.  \label{eqn.t2_2} \end{align}
From equation \eqref{eqn.t2_1} it is clear that a necessary condition for two homomorphisms $\phi,\psi$ to be homotopic is that the parities of $\phi_1(a)$ and $\psi_1(a)$, and also of $\phi_1(b)$ and $\psi_1(b)$, must coincide. We now split the analysis into four cases, corresponding to the parities of $\phi_1(a)$ and $\phi_1(b)$.

When $\phi_1(a)$ and $\phi_1(b)$ are both even, the second line of \eqref{eqn.t2_2} vanishes identically, and so each homomorphism with a different $\phi_2(t)_0$ belongs to its own homotopy class. When $\phi_1(a)$ is even and $\phi_1(b)$ is odd, the second line of \eqref{eqn.t2_2} is $(\theta(a)_0-\theta(a)_1)(e_0-e_1)$. Thus there are two homotopy classes, corresponding to the parity of $\frac{1}{2}\phi_1(a)+\phi_2(t)_0$. Similarly, when $\phi_1(a)$ is odd and $\phi_1(b)$ is even, there are two classes corresponding to the parity of $\frac{1}{2}\phi_1(b)+\phi_2(t)_0$, and when both $\phi_1(a)$ and $\phi_1(b)$ are odd, there are once again two classes corresponding to the parity of $\frac{1}{2}(\phi_1(a)+\phi_1(b))+\phi_2(t)_0$.

This brings us to
\beq [T^2,\bR\bP^2]_0 = \bZ \cup \bZ_2 \cup \bZ_2 \cup \bZ_2, \eeq
with the four factors corresponding respectively to the four homomorphisms $(0,0),(1,0),(0,1),(1,1)$ of fundamental groups $\bZ^2\to\bZ_2$. Notice that this is consistent with the cohomology groups
\beq H^2_{(0,0)}(T^2,\bZ) = \bZ,~~~ H^2_{(1,0)}(T^2,\bZ)=\bZ_2,~~~ H^2_{(0,1)}(T^2,\bZ)=\bZ_2,~~~ H^2_{(1,1)}(T^2,\bZ)=\bZ_2. \eeq
Explicit representatives of these homomorphisms, in the format $(\phi_1(a),\phi_1(b),\phi_2(t)_0)$, are respectively\footnote{We use $(n_1,n_2,n_3)$ to denote the group homomorphism, and $[n_1,n_2,n_3]$ to denote the homotopy class to which it belongs.}
\beq [T^2,\bR\bP^2]_0 = \{[0,0,n],n\in\bZ\} \cup \{[1,0,0],[1,0,1]\} \cup \{[0,1,0],[0,1,1]\} \cup \{[1,1,0],[1,1,1]\}. \eeq

To obtain free homotopy classes, note that $\pi_1\bR\bP^2$ acts trivially on $\phi_1$ but exchanges $\phi_2(t)_0$ and $\phi_2(t)_1$ (or, equivalently, it flips the sign of $\phi_2(t)_0$), so
\beq [T^2,\bR\bP^2] = \{[0,0,n],n\geq 0,n\in\bZ\} \cup \{[1,0,0],[1,0,1]\} \cup \{[0,1,0],[0,1,1]\} \cup \{[1,1,0],[1,1,1]\}. \eeq

\begin{figure}
\centering
\begin{tabular}{ccc}
\includegraphics[width=0.3\textwidth]{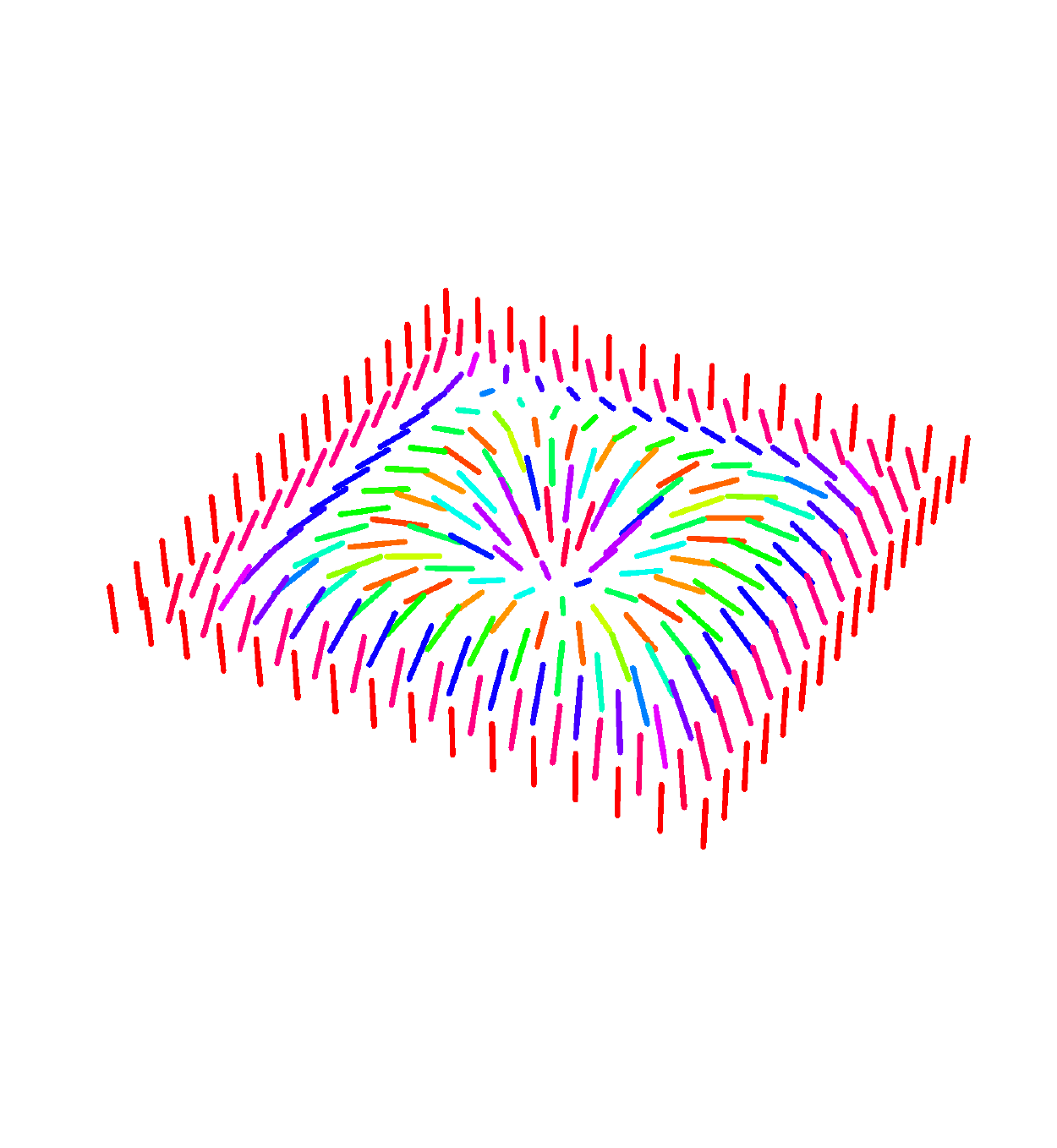}&
\includegraphics[width=0.3\textwidth]{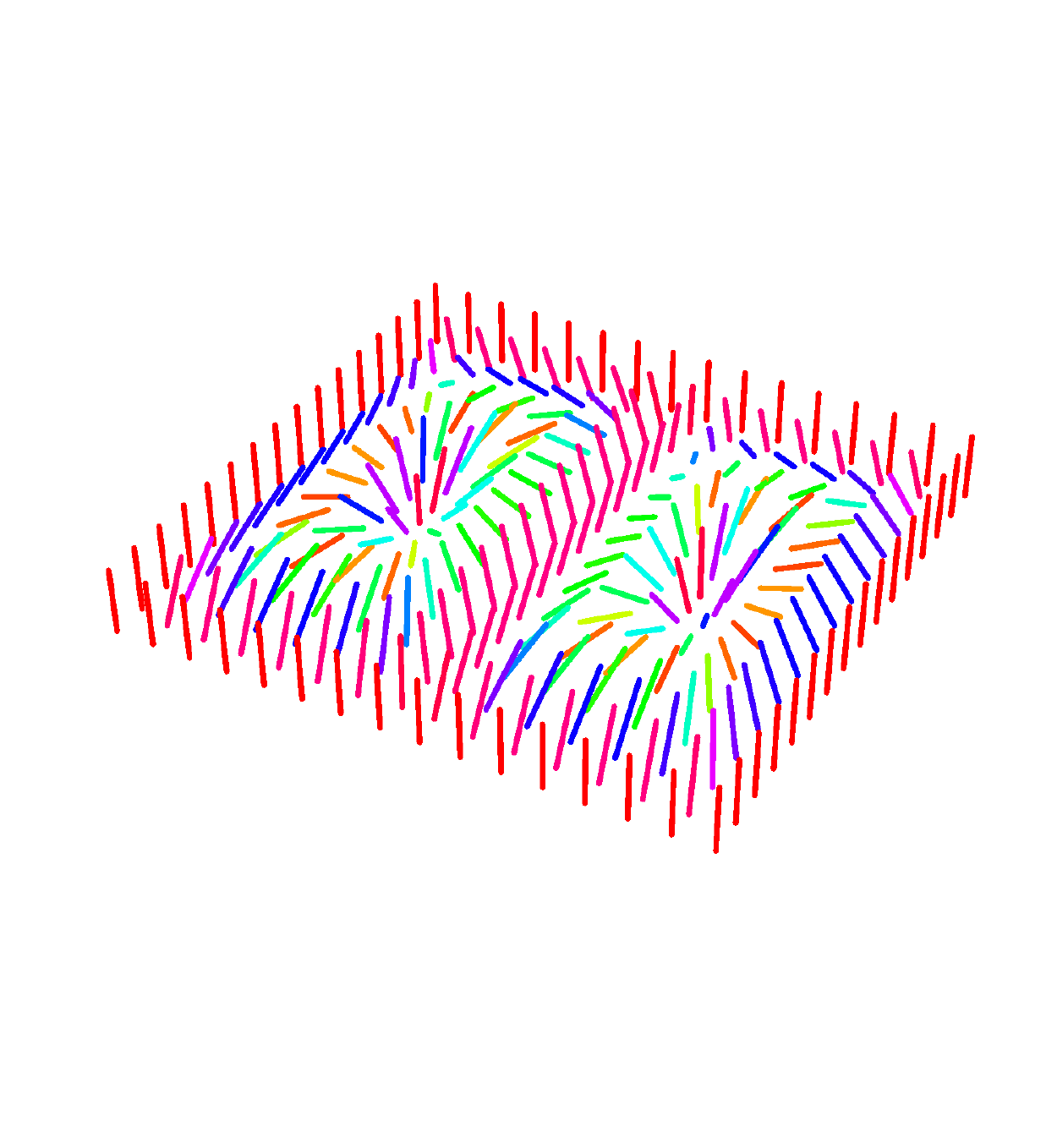}&
\includegraphics[width=0.3\textwidth]{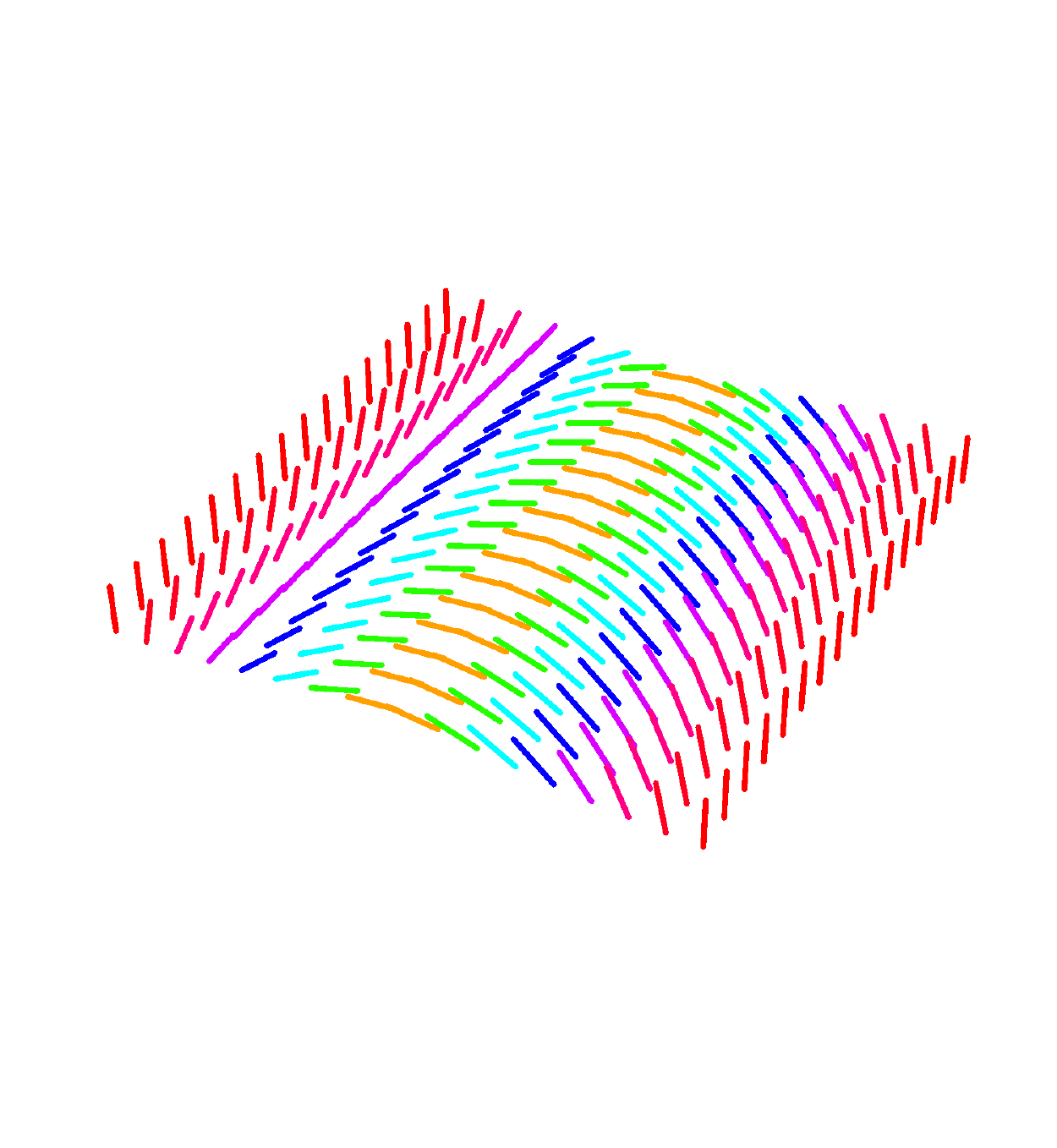}\\
\includegraphics[width=0.3\textwidth]{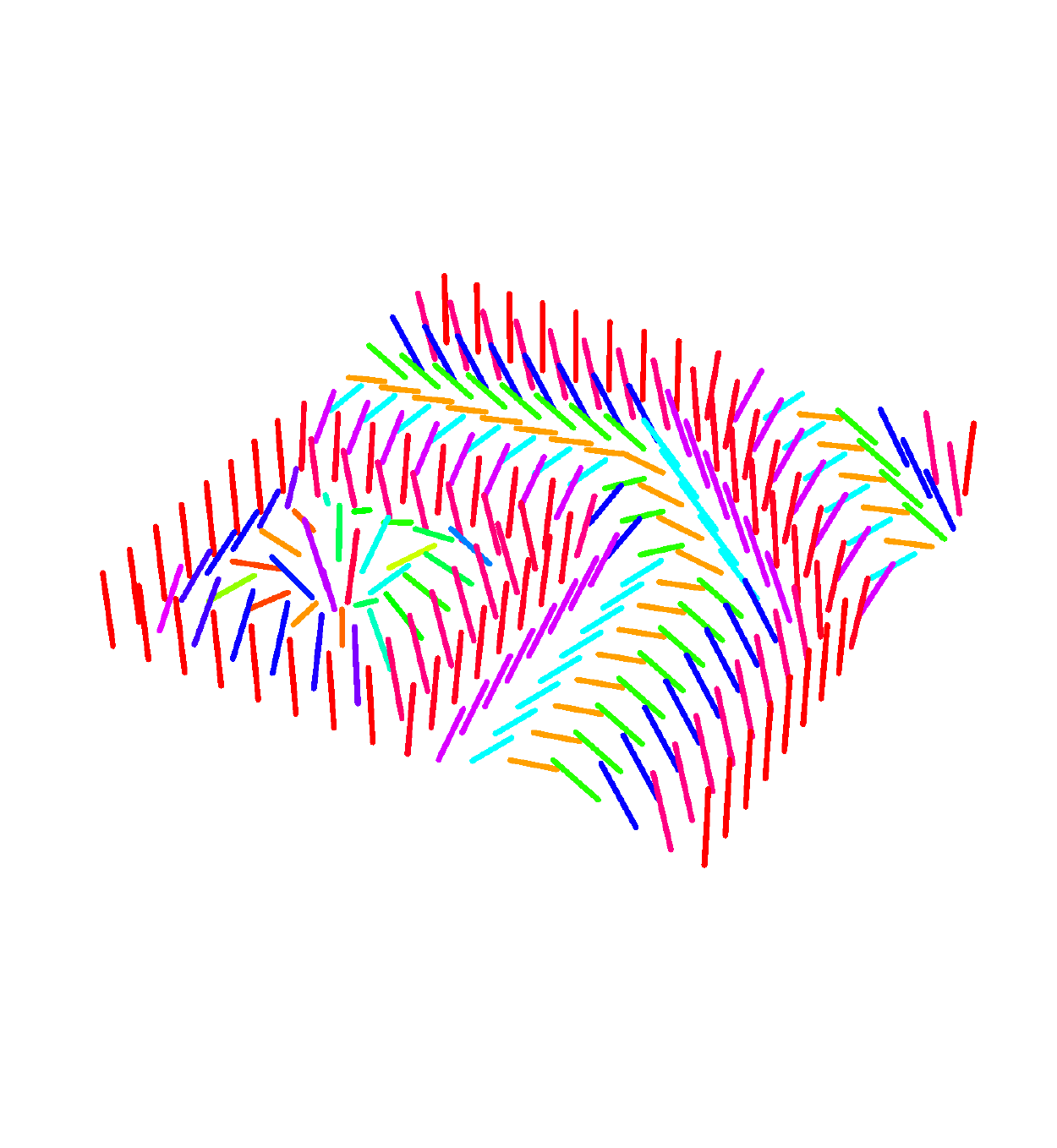}&
\includegraphics[width=0.3\textwidth]{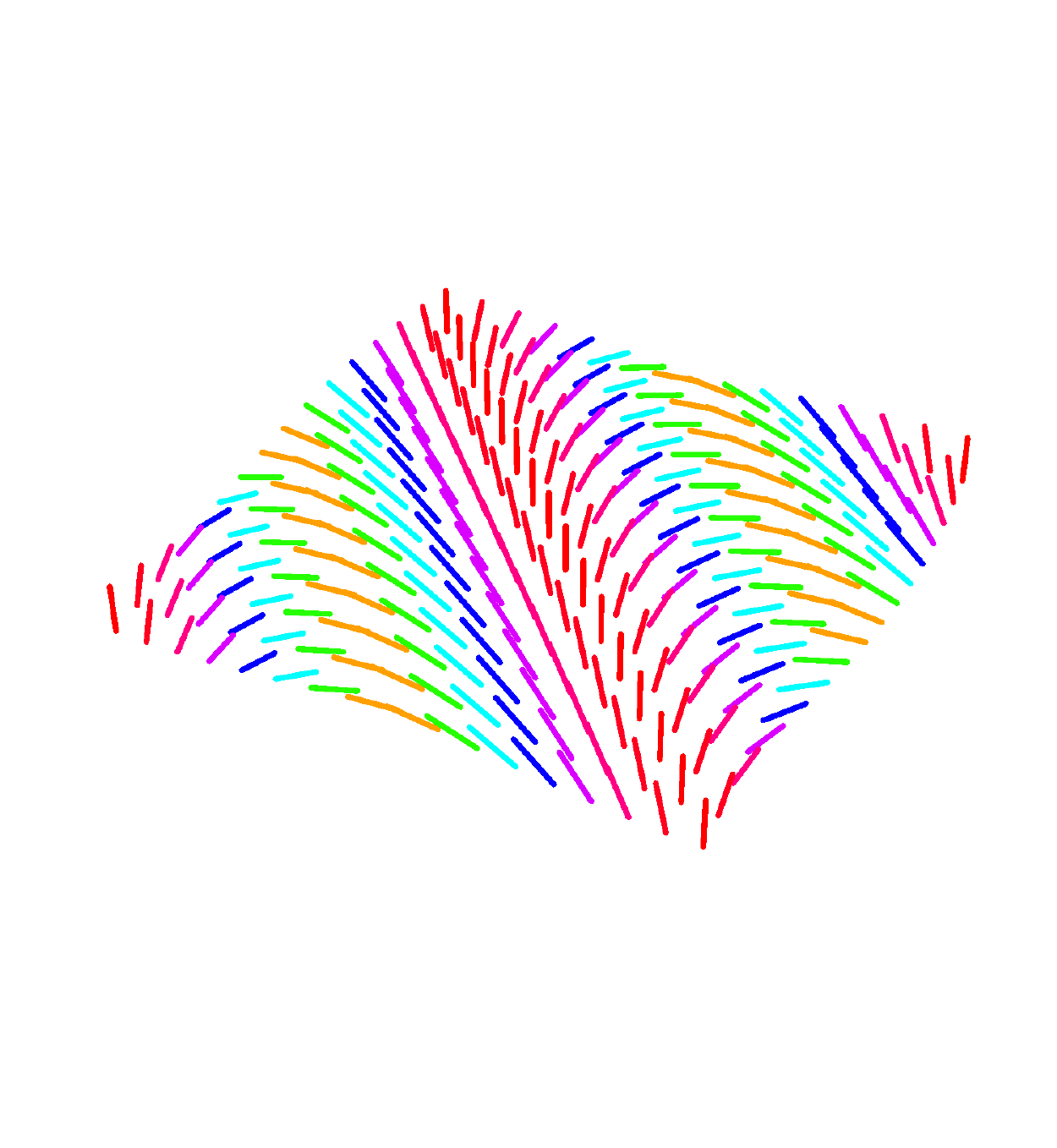}&
\includegraphics[width=0.3\textwidth]{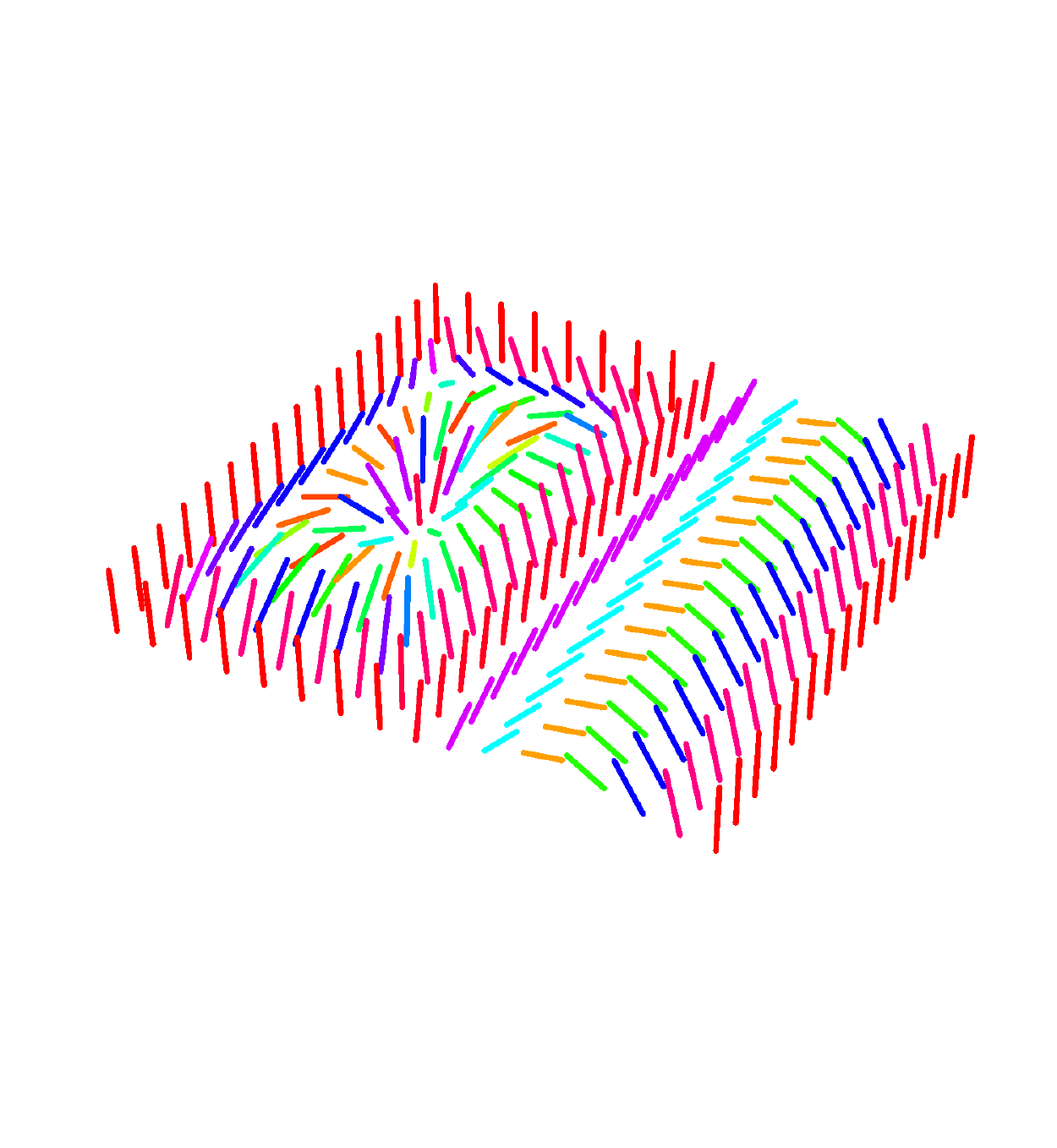}
\end{tabular}
\caption{(Color online) Illustrations of some of the textures $[T^2,\bR\bP^2]$. $T^2$ is shown as a square with opposite sides identified, and $\bR\bP^2$ is visualized as rods in $\bR^3$. Clockwise from top left, the first figure shows the configuration $(0,0,1)$, which can be thought of as the stereographic projection of the standard projection map from $S^2$ to $\bR\bP^2$. The second figure shows two copies of the first figure, ``glued'' together to form the configuration $(0,0,2)$. The third figure shows the configuration $(1,0,0)$, and we can glue to it the $(0,0,1)$ configuration to yield the $(1,0,1)$ configuration, shown in the fourth figure. The fifth figure shows the $(1,1,0)$ configuration; it can be ``diagonally glued'' to the $(0,0,1)$ configuration to yield the $(1,1,1)$ configuration (sixth figure).}
\label{fig.t2rp2fig}
\end{figure}

Since we have obtained representatives of the textures in terms of crossed module homomorphisms, it is not difficult to explicitly construct the textures themselves. The parities of the first two integers $\phi_1(a)$ and $\phi_1(b)$ indicates the one-dimensional texture $\Hom(\pi_1(S^1\vee S^1),\pi_1\bR\bP^2)\simeq\bZ_2\times\bZ_2$ when restricted to the one-skeleton $S^1\vee S^1$ of $T^2$. The third integer $\phi_2(t)_0$ has the following interpretation when both $\phi_1(a)$ and $\phi_1(b)$ are even; then we can consider a representative $(0,0,\phi_2(t)_0)$ which is constant on the 1-skeleton. In that case, we can pinch the 1-skeleton to a point, and the 2-cell becomes a sphere $S^2$. Homotopy classes of maps on the sphere are classified by $[S^2,\bR\bP^2]_0=\pi_2\bR\bP^2\simeq\bZ$; and the $(0,0,\phi_2(t)_0)$ configuration corresponds to the class $\phi_2(t)_0\in\pi_2\bR\bP^2$. In the other cases when at least one of $\phi_1(a)$ or $\phi_1(b)$ is odd; this characterization is not possible, but ``gluing'' the $(0,0,n)$ configuration to the $(\phi_1(a),\phi_1(b),\phi_2(t)_0)$ configuration yields a configuration homotopic to the $(\phi_1(a),\phi_1(b),\phi_2(t)_0+n)$ configuration. Visualizations of the $\bR\bP^2$ textures, thought of as rods in $\bR^3$, on $T^2$ are shown in figure \ref{fig.t2rp2fig}.

A straightforward generalization shows that $\bR\bP^2$ textures on a surface $\Sigma_g$ of genus $g$ is given by
\beq [\Sigma_g,\bR\bP^2] = \bN \cup \bigcup^{2^{2g}-1}\bZ_2, \eeq
where the first factor $\bN$ correspond to textures inducing the trivial map on fundamental groups, while there are $2$ textures inducing each of the other $2^{2g}-1$ homomorphisms in $\Hom(\pi_1\Sigma_g,\pi_1\bR\bP^2)$.

\subsubsection{Textures of the cross cap sigma model} \label{sec.rp2rp2}
In this example we compute the classes $[\bR\bP^2,\bR\bP^2]$, which can be thought of as the topological sectors of a sigma model on a cross cap (or an orientifolded sphere) valued in the target space $\bR\bP^2$ of liquid crystals in uniaxial nematic phase. The fundamental crossed module of $\bR\bP^2$ was computed in example \ref{eg.rp2} as
\beq \p_{\bR\bP^2}:\bZ[\bZ_2]\to\bZ, \eeq
where $\bZ_2=\{e_0,e_1\}$ and the crossed module map sends both $e_0$ and $e_1$ to $2\in\bZ$. The generator $1\in\bZ$ acts on $\bZ[\bZ_2]$ by exchanging $e_0$ and $e_1$. Its automorphisms
\beq \begin{tikzcd}
\bZ[\bZ_2] \ar[r,"\p"] \ar[d,"\phi_2"]& \bZ \ar[d,"\phi_1"] \\
\bZ[\bZ_2] \ar[r,"\p"]& \bZ
\end{tikzcd} \eeq
are completely determined by two integers $\phi_1(1)\in\bZ$ and the component $\phi_2(e_0)_0$ of $\phi_2(e_0)=\phi_2(e_0)_0 e_0+\phi_2(e_0)_1 e_1$. Indeed, the commuting diagram requires that $\phi_2(e_0)_1=\phi_1(1)-\phi_2(e_0)_0$, and equivariance requires that $\phi_2(e_1)=\leftidx{^{\phi_1(1)}}{\phi_2(e_0)}$.

Two automorphisms $\phi,\psi$ are homotopic if there exists a map $\theta:\bZ\to\bZ[\bZ_2]$ satisfying
\begin{align}
\theta(n+m) &= \theta(n)+\leftidx{^{\psi_1(n)}}{\theta(m)}, \\
\p\theta(n) &= \phi_1(n)-\psi_1(n), \\
\theta(\p x) &= \phi_2(x)-\psi_2(x),
\end{align}
for all $m,n\in\bZ$ and $x\in\bZ[\bZ_2]$. A straightforward calculation shows that the homotopy classes of automorphisms are
\beq [\bR\bP^2,\bR\bP^2]_0 = \bZ_2\cup\bZ. \eeq
The first factor $\bZ_2$ corresponds to automorphisms with $\phi_1(1)$ being even, or, equivalently, inducing the trivial map on fundamental groups. This matches with $H^2_0(\bR\bP^2,\bZ)=\bZ_2$. The second factor $\bZ$ corresponds to automorphisms with $\phi_1(1)$ being odd, or, equivalently, inducing the identity map on fundamental groups, which matches with $H^2_{\text{id}}(\bR\bP^2,\bZ)=\bZ$. In the notation $(\phi_1(1),\phi_2(e_0)_0)$ labelling the automorphism $\phi$, representatives are given by
\beq [\bR\bP^2,\bR\bP^2]_0 = \{[0,0],[0,1]\}\cup\{[1,n]:n\in\bZ\}. \eeq

\begin{figure}
\centering
\includegraphics[width=0.3\textwidth]{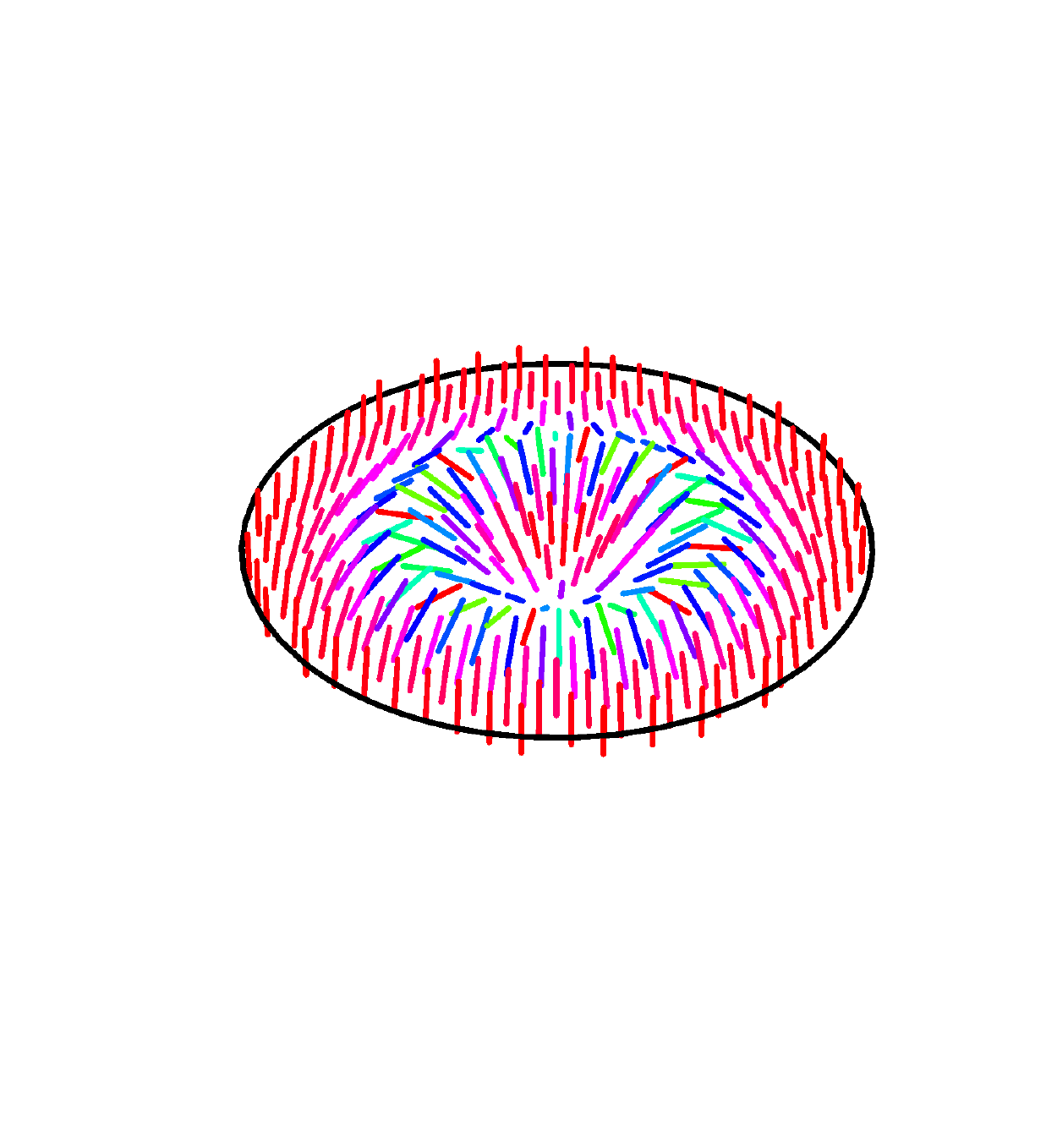}
\includegraphics[width=0.3\textwidth]{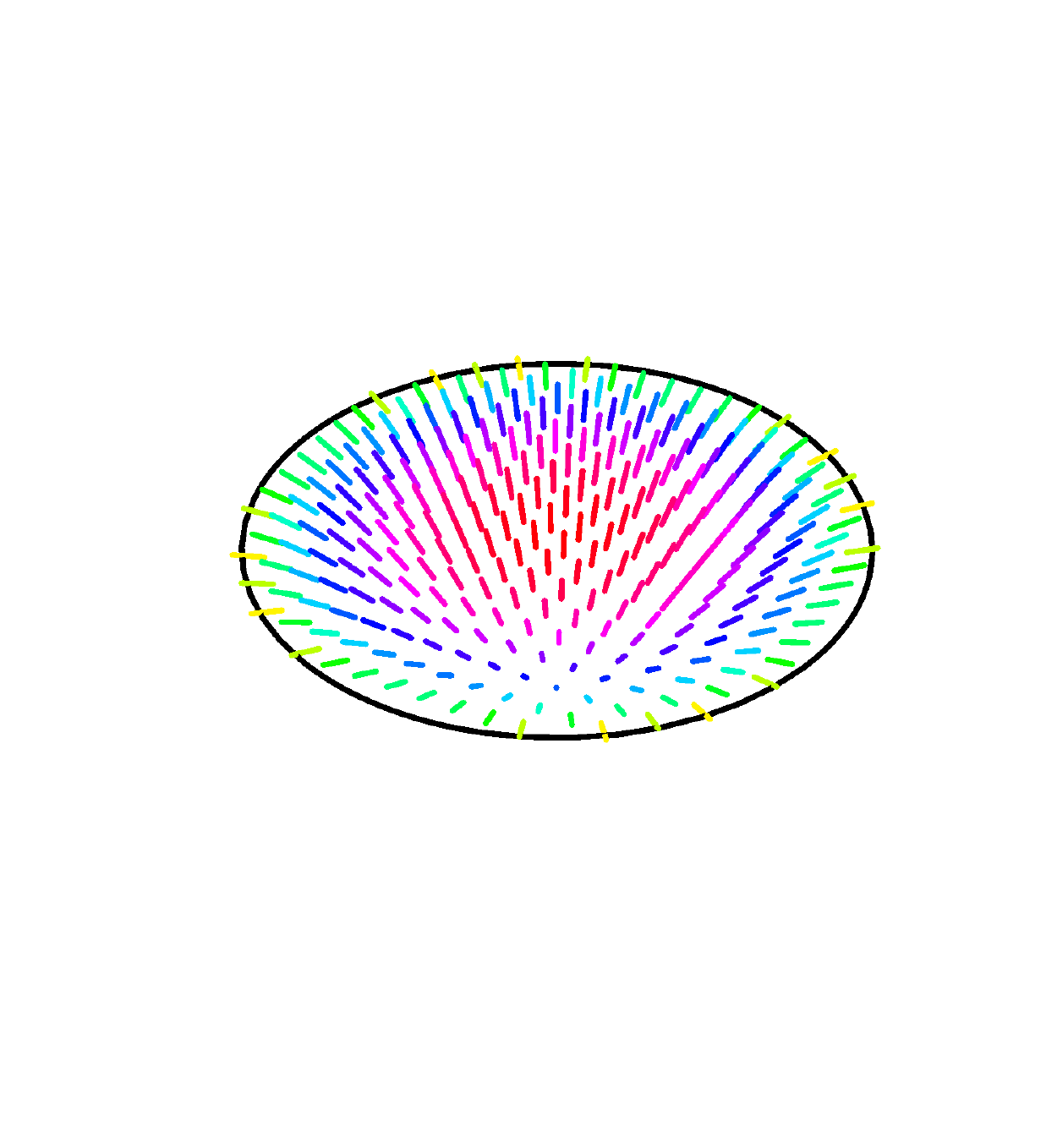}
\includegraphics[width=0.3\textwidth]{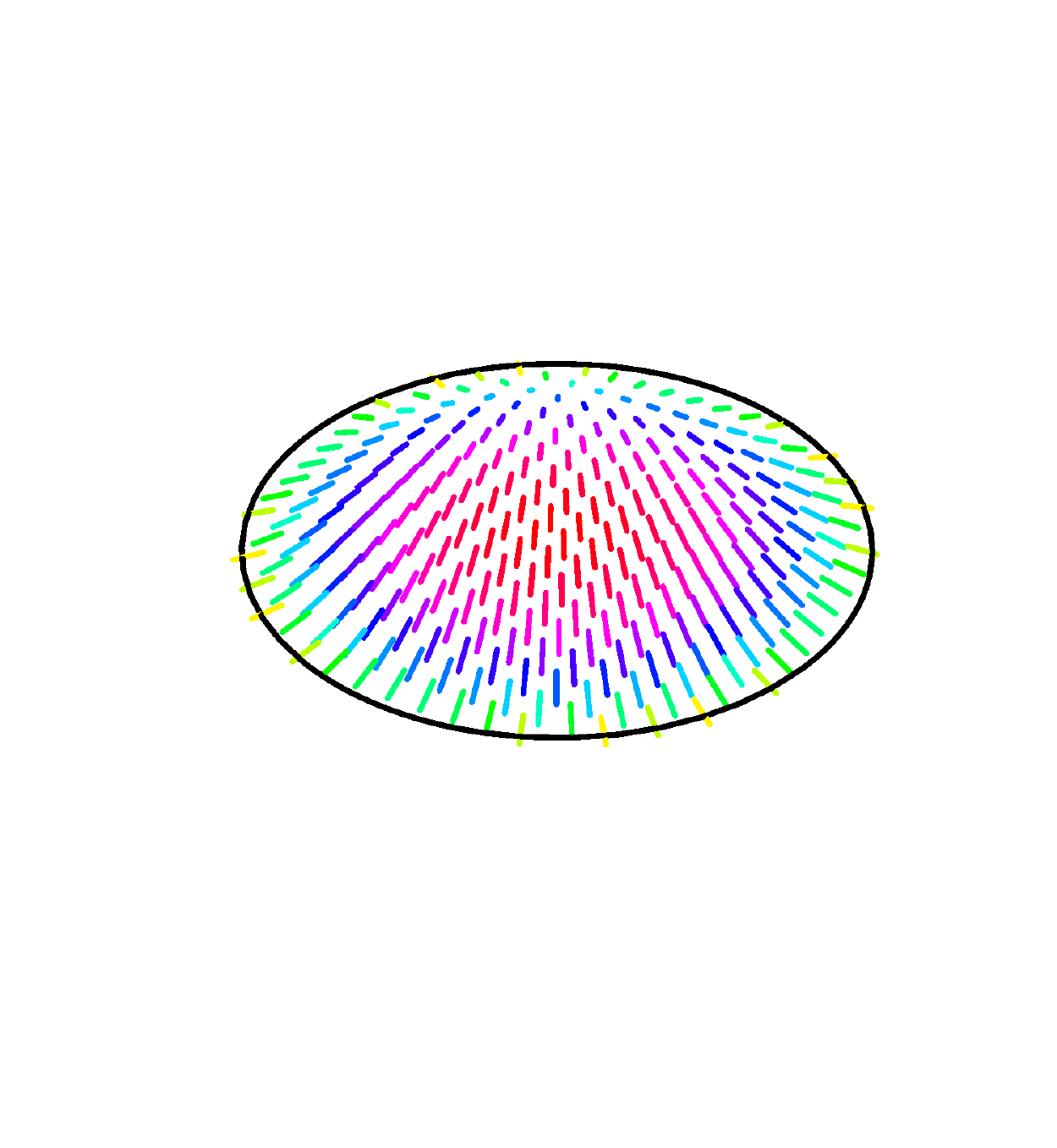}
\caption{Some examples of textures $[\bR\bP^2,\bR\bP^2]$ with $\bR\bP^2$ modeled by the cross cap, which is a disk with antipodes on its boundary identified. The left figure shows the texture $(0,1)$; the center figure shows the texture $(1,1)$, and the right figure shows $(1,0)$. The last two textures are free homotopic but not based homotopic.}
\label{fig.rp2rp2}
\end{figure}

Self-maps $\bR\bP^2\to\bR\bP^2$ realizing these crossed module automorphisms can be described as follows. Model $\bR\bP^2$ as the unit disk with antipodal points on its boundary identified, with the usual polar coordinates $(r,\theta)$, $r\in[0,1],\theta\in\bR/2\pi\bZ$. The base point is the point $(1,0)$. The constant map to the base point clearly realizes the class $[0,0]$. A map realizing the class $[0,1]$ is given by
\beq (r,\theta) \mapsto \begin{cases} (\tan (r\pi/2),\theta)~~~~~ & r\in[0,1/2], \\ \((\tan (r\pi/2))^{-1},\theta+\pi\)~~~~~ & r\in[1/2,1]. \end{cases} \label{eqn.rp2wrap} \eeq
The maps $(r,\theta)\mapsto (r^{\abs{2n-1}},(2n-1)\theta)$ realize the classes $[1,n]$ for $n\in\bZ$ (see figure \ref{fig.rp2rp2} for illustrations).\footnote{Notice that the maps $(r,\theta)\mapsto(r^{2n},2n\theta)$ are homotopic to the constant map.}

The free homotopy classes $[\bR\bP^2,\bR\bP^2]$ are the orbits of the $\pi_1\bR\bP^2=\bZ_2$ action on the based classes $[\bR\bP^2,\bR\bP^2]_0$. The nontrivial element of $\pi_1\bR\bP^2$ exchanges $e_0$ with $e_1$, and hence sends $(\phi_1(1),\phi_2(e_0)_0)$ to $(\phi_1(1),\phi_1(1)-\phi_2(e_0)_0)$. The classes $[0,0]$ and $[0,1]$ are fixed by this action, while $[1,n]$ is sent to $[1,1-n]$. Therefore, the homotopy classes of self-maps of $\bR\bP^2$ are given by
\beq [\bR\bP^2,\bR\bP^2] = \{[0,0],[0,1],[1,\bN]\} \eeq
where the first class is represented by the constant map, the second class is represented by the map in \eqref{eqn.rp2wrap}, and the family of maps indexed by the natural numbers $[1,\bN]$ are represented by $(r,\theta)\mapsto(r^{\abs{2n-1}},(2n-1)\theta)$. Notice that under the $\bZ_2$ action, $2n-1$ is sent to $1-2n$, so this map undergoes an orientation reversal. In figure \ref{fig.rp2rp2}, some examples of textures are illustrated.

\subsubsection{Torus knot defects in nematic liquid crystals}
\begin{figure}
\centering
\begin{tabular}{ccc}
\raisebox{-0.5\height}{\includegraphics[width=0.15\textwidth]{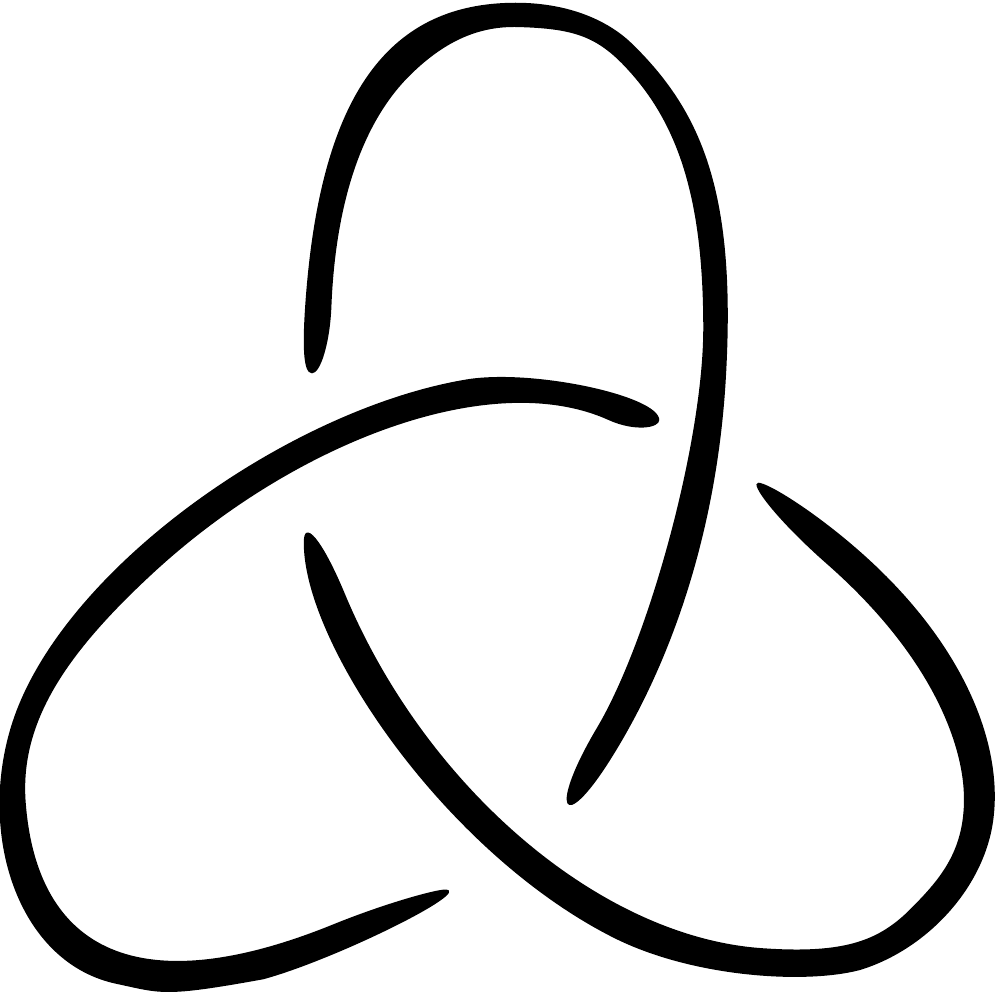}}~~~~~~&
\raisebox{-0.5\height}{\includegraphics[width=0.2\textwidth]{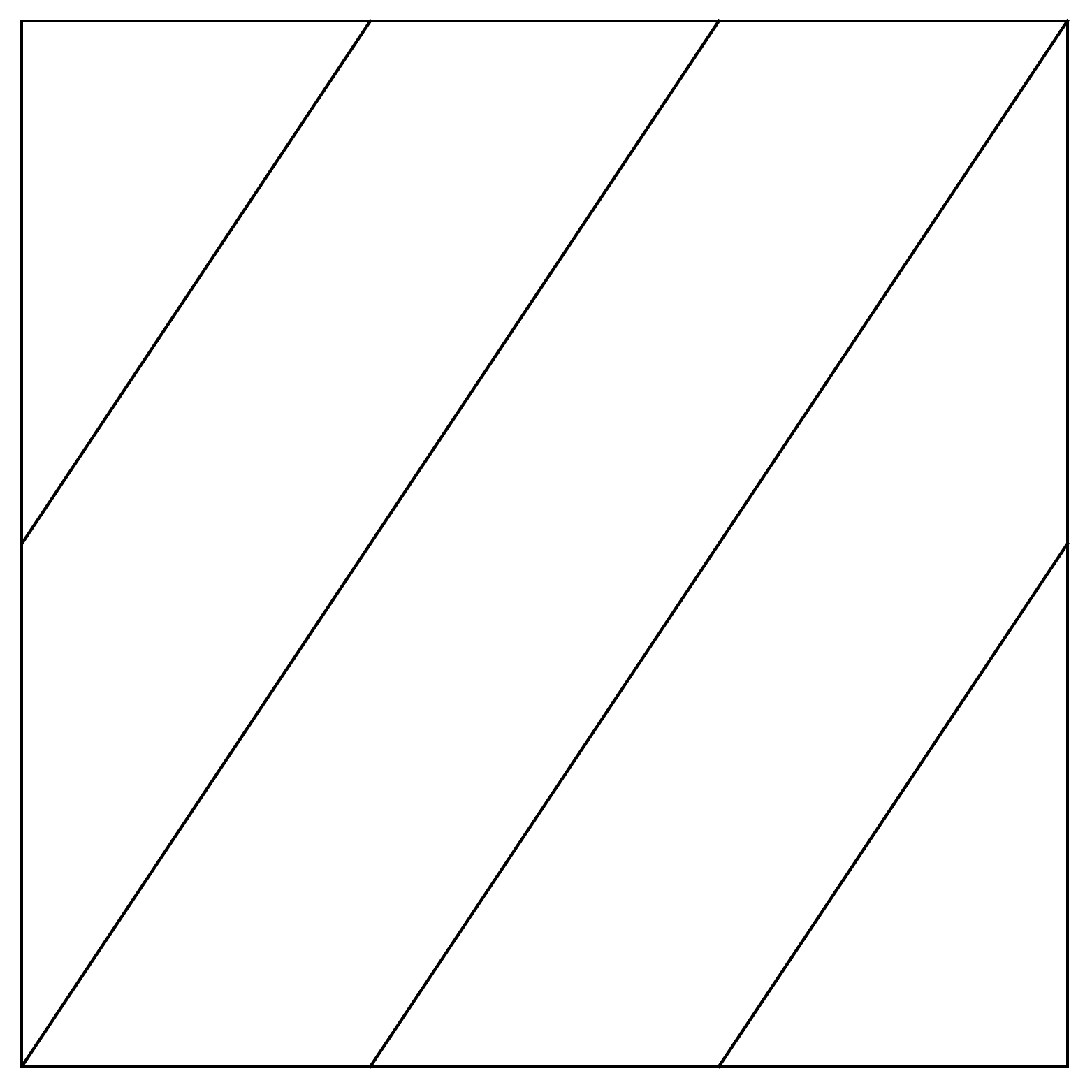}}~~~~~~&
\raisebox{-0.5\height}{\includegraphics[width=0.3\textwidth]{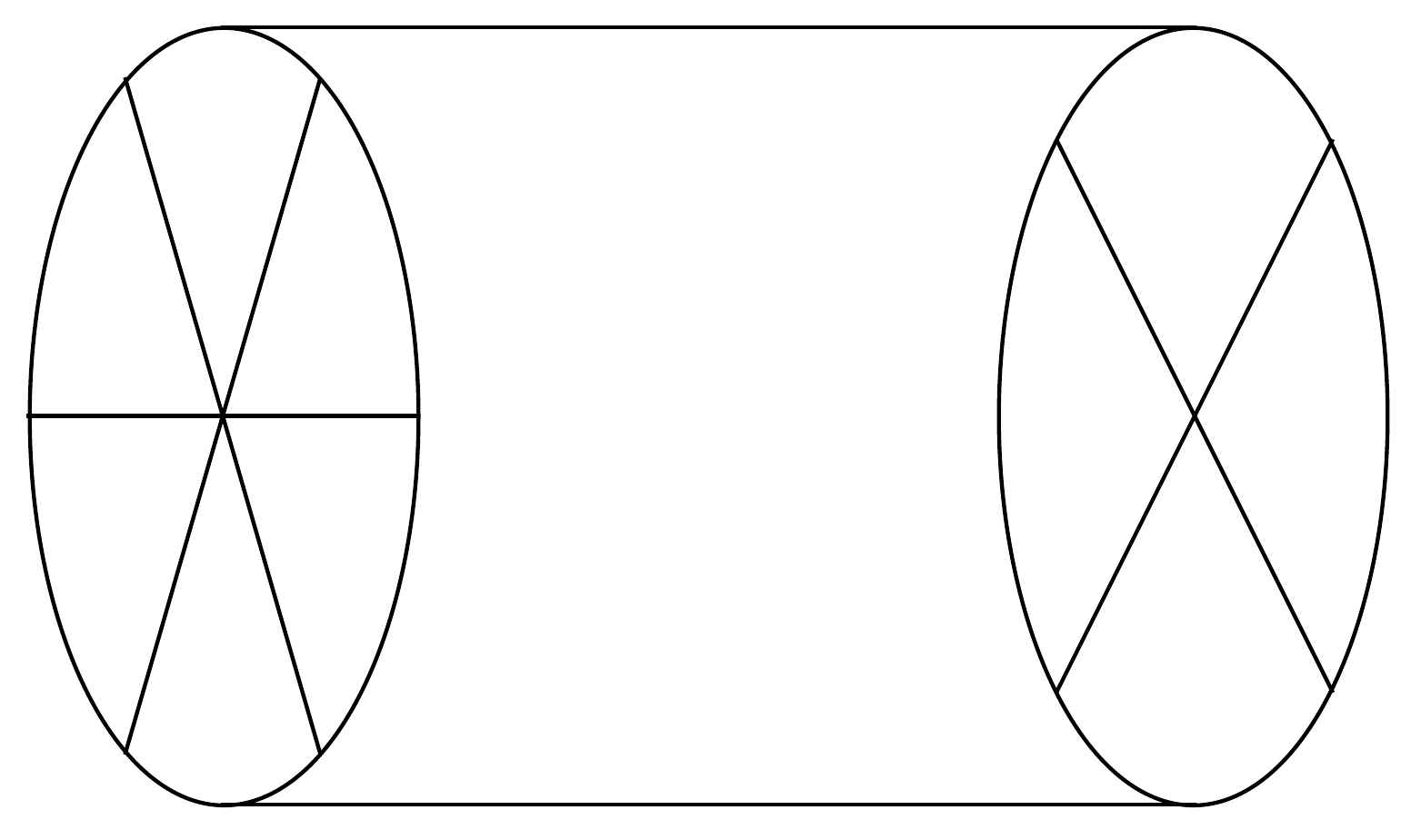}}
\end{tabular}
\caption{The trefoil knot (left) can be inscribed on a torus (middle) wrapping one cycle twice and the other cycle thrice. The trefoil knot complement in $S^3$ retracts onto $M_{2,3}$ (right), which is the cylinder with a cross cap at one end and a three-fold cap at the other.}
\label{fig.knot}
\end{figure}
A torus knot is a knot which can be inscribed on a torus $T^2$ in the 3-sphere $S^3$. Torus knots are labeled by two coprime integers; the knot labeled $(p,q)$ winds round one of the fundamental cycles of the torus $p$ times and the other fundamental cycle $q$ times before joining its ends. The complement of the $(p,q)$ torus knot in $S^3$ retracts onto the quotient $M_{p,q}$ of the cylinder $[0,1]\times S^1$ by the identifications $(0,\theta)\sim(0,\theta+2\pi/p)$ and  $(1,\theta)\sim(1,\theta+2\pi/q)$ (see figure \ref{fig.knot}).\footnote{See, e.g. \cite{lickorish1997} for more information about knots.} This space $M_{p,q}$ admits a CW structure with two 1-cells $\{a,b\}$ and one 2-cell $\{t\}$ glued to $a^pb^{-q}$. In particular, one can read off the knot group
\beq \pi_1M_{p,q} = \moy{a,b|a^p=b^q} \eeq
from the fundamental crossed module
\beq H \hookrightarrow \moy{a,b}, \eeq
where $H$ is the free crossed module on $t\mapsto a^pb^{-q}$ (see appendix \ref{app.freexm}).

In this example, we compute the topological sectors for a line defect along a $(p,q)$-torus knot for the liquid crystal in the uniaxial nematic phase, which has Goldstone modes in the space $\bR\bP^2$. In fact, we shall be slightly more general and not assume that $p$ and $q$ are coprime, although that is the case for the torus knot. The non-coprime spaces are also of physical interest; for example, $M_{2,2}$ is the connected sum $\bR\bP^2\#\bR\bP^2$ of two cross caps, which is the Klein bottle. Indeed, by redefining $x=ab^{-1},y=b$, we see that the CW structure gives the usual fundamental domain of the Klein bottle, because the 2-cell is glued to $a^2b^{-2}=xyxy^{-1}$.

Each crossed module homomorphism
\beq \begin{tikzcd}
H \ar[r,hook] \ar[d,"\phi_2"] & \moy{a,b} \ar[d,"\phi_1"] \\
\bZ[\bZ_2] \ar[r,"\p_{\bR\bP^2}"]& \bZ
\end{tikzcd} \eeq
determines a triple of integers $(\phi_1(a),\phi_1(b),\phi_2(t)_0)\in\bZ^3$, as before. However, this time, not every triple of integers define homomorphisms, as the commutativity of the diagram requires that $p\phi_1(a)-q\phi_1(b)=2(\phi_2(t)_0+\phi_2(t)_1)$. In particular, $p\phi_1(a)-q\phi_1(b)$ has to be even. If $p$ and $q$ are both even, then all triples correspond to homomorphisms; if $p$ is odd and $q$ even, then $\phi_1(a)$ has to be even; if $p$ is even and $q$ odd, then $\phi_1(b)$ has to be even; if $p$ and $q$ are both odd, then $\phi_1(a)$ and $\phi_1(b)$ need to have the same parity.

The condition for homomorphisms $\phi,\psi$ to be based homotopic is the existence of a function $\theta:\moy{a,b}\to\bZ[\bZ_2]$ satisfying
\beq \theta(gg') = \theta(g)\ \leftidx{^{\psi_1(g)}}{\theta(g')}, \label{eqn.knoteqn} \eeq
\beq \phi_1(g)-\psi_1(g) = 2(\theta(g)_0+\theta(g)_1) \eeq
for $g,g'\in\moy{a,b}$, and
\beq \phi_2(t) - \psi_2(t) = \theta(a^pb^{-q}) = \sum_{j=1}^p \leftidx{^{(j-1)\psi_1(a)}}{\theta(a)} -\sum_{k=1}^q \leftidx{^{p\psi_1(a)-k\psi_1(b)}}{\theta(b)}. \eeq
The second equality follows from condition \eqref{eqn.knoteqn}. At this point, we consider separately the four cases corresponding to the parities of $p$ and $q$. Let $r=\gcd(p,q)$. After some straightforward calculation, we can see that the homomorphisms fall into the following homotopy classes:
\begin{itemize}
\item For $p$ and $q$ even:
\beq [M_{p,q},\bR\bP^2]_0 = \bZ_r \cup \bZ_q \cup \bZ_p \cup \bZ \eeq
corresponding respectively to the homomorphisms $(0,0),(1,0),(0,1),(1,1)\in\Hom(\pi_1 M_{p,q},\pi_1\bR\bP^2)$ on fundamental groups. In the notation $(\phi_1(a),\phi_1(b),\phi_2(t)_0)$, representatives of these classes are given by
\beq \bZ_r = \{[0,0,0],[0,0,1],\ldots,[0,0,r-1]\}, \label{eqn.zr} \eeq
\beq \bZ_q = \{[1,0,0],[1,0,1],\ldots,[1,0,q-1]\}, \label{eqn.zq} \eeq
\beq \bZ_p = \{[0,1,0],[0,1,1],\ldots,[0,1,p-1]\}, \label{eqn.zp} \eeq
\beq \bZ = \{[1,1,n],n\in\bZ\}. \eeq
The $\pi_1\bR\bP^2\simeq\bZ_2$ action on the homomorphisms fixes $\phi_1$ and exchanges $\phi_2(t)_0$ with $\phi_2(t)_1=\frac{1}{2}(p\phi_1(a)-q\phi_1(b))-\phi_2(t)_0$. Hence the following classes belong in the same $\pi_1\bR\bP^2$ orbit:
\begin{align} [0,0,x]&\sim[0,0,-x]=[0,0,r-x], \nn \\
[1,0,x]&\sim[1,0,p/2-x], \nn \\
[0,1,x]&\sim[0,1,-q/2-x]=[0,1,p-q/2-x], \nn \\ 
[1,1,x]&\sim[1,1,(p-q)/2-x]. \label{eqn.knotequiv} \end{align}

\item For $p$ odd, $q$ even:
\beq [M_{p,q},\bR\bP^2]_0 = \bZ_r \cup \bZ_p \eeq
corresponding respectively to the homomorphisms $(0,0),(0,1)\in\Hom(\pi_1M_{p,q},\pi_1\bR\bP^2)$ on fundamental groups (recall that $\phi_1(a)$ has to be even in this case -- equivalently, the relation $a^pb^{-q}=1$ requires that $a$ is sent to an even element). A set of representatives is also given by equations \eqref{eqn.zr} and \eqref{eqn.zp}.

Free homotopy groups are obtained by taking the quotient as in \eqref{eqn.knotequiv}.

\item For $p$ even, $q$ odd:
\beq [M_{p,q},\bR\bP^2]_0 = \bZ_r \cup \bZ_q \eeq
corresponding to the homomorphisms $(0,0),(1,0)$ on fundamental groups. A set of representatives is given by equations \eqref{eqn.zr} and \eqref{eqn.zq}. Free homotopy groups are obtained by taking the quotient as in \eqref{eqn.knotequiv}.

\item For $p$ and $q$ odd:
\beq [M_{p,q},\bR\bP^2]_0 = \bZ_r \cup [1,1,0] \eeq
In this case, there is a unique homotopy class inducing the homomorphism $(1,1)$ on fundamental groups. A set of representatives is given by equations \eqref{eqn.zr}. Free homotopy groups are obtained by taking the quotient as in \eqref{eqn.knotequiv}.
\end{itemize}

Hence, the textures on a Klein bottle $M_{2,2}$ on $\bR\bP^2$ are
\beq [M_{2,2},\bR\bP^2] = \{[0,0,0],[0,0,1]\}\cup\{[1,0,0]\}\cup\{[0,1,0]\}\cup\{[1,1,n],n\geq 0,n\in\bZ\}, \eeq
where the four sets correspond respectively to maps inducing the homomorphisms $(0,0),(1,0),(0,1),(1,1)$ on fundamental groups. Notice that the based homotopy classes do indeed correspond to the (twisted) cohomology groups of the Klein bottle $\bR\bP^2\#\bR\bP^2=M_{2,2}$
\beq H^2_{(0,0)}(M_{2,2},\bZ)=\bZ_2,~~~ H^2_{(0,1)}(M_{2,2},\bZ)=\bZ_2,~~~ H^2_{(1,0)}(M_{2,2},\bZ)=\bZ_2,~~~ H^2_{(1,1)}(M_{2,2},\bZ)=\bZ. \eeq

For the $(p,q)$ torus knot defect, $\gcd(p,q)=1$ so there is one texture inducing the trivial map on fundamental groups; the constant map is an example of such a configuration. If one of $p$ or $q$ is even, say, $p$; then there also are $\bZ_q$ based homotopy classes inducing the map $(1,0)$ on fundamental groups. If both $p$ and $q$ are odd, then there is one based homotopy class inducing the map $(1,1)$ on fundamental groups. This is consistent with previous results \cite{lickorish1997,machon2016} that the number of nontrivial based homotopy classes from a knot complement to $\bR\bP^2$ is equal to the knot determinant.\footnote{The determinant of the knot is the absolute value of the Alexander polynomial evaluated at $-1$.} Indeed, the determinant of the $(p,q)$ torus knot is $q$ if $p$ is even, $p$ if $q$ is even, and $1$ if both $p$ and $q$ are odd.

Finally, the defect sectors of a torus knot are obtained by considering the $\pi_1\bR\bP^2$ orbits as in equation \eqref{eqn.knotequiv}. As an example, for the trefoil defect $M_{2,3}$, we see that $[1,0,0]$ and $[1,0,1]$ belong to the same orbit, while $[1,0,2]$ is in its own orbit. This yields the result that there are three topological sectors of a nematic liquid crystal on a trefoil knot defect in $S^3$
\beq [M_{2,3},\bR\bP^2] = \{[0,0,0],[1,0,0],[1,0,2]\}. \eeq

\subsection{Relation to categorical groups and 2-groups} \label{sec.2gsect}
Following Whitehead's discovery that crossed modules were algebraic models of homotopy 2-types, mathematicians began to realize that crossed modules could be thought of as higher categorical generalizations of groups. In 1976, Brown and Spencer proved that crossed modules are equivalent to categorical groups \cite{brown1976}. A year earlier, Grothendieck's student Ho\`ang\footnote{Ho\`ang Xu\^an S\'inh's surname is Ho\`ang and given name is Xu\^an S\'inh. In some of the literature, her surname was wrongly identified as S\'inh.} wrote her thesis \cite{sinh1975} on what she termed ``gr-categories'', but we shall call (weak) 2-groups, which is a generalization of categorical groups where the group axioms need to hold only up to isomorphism. Over the years, it was realized that these higher categorical group-like structures are ubiquitous in mathematics and physics. To give a few examples, the Yetter TQFT \cite{yetter1993} is a generalization of the Dijkgraaf-Witten TQFT to 2-groups. Gerbes \cite{giraud1971,Breen:2006nh} are 2-group (and higher group) principal bundles, whose connection forms are ubiquitous in string and M theory (see e.g. \cite{Sharpe:2015mja}), as well as the Wess-Zumino-Witten conformal field theory. Higher category global symmetries in quantum field theories lead to simplifications and selection rules \cite{Gaiotto:2014kfa} much like ordinary symmetries. They can also be used to classify phases of matter, for example in three dimensional symmetry enriched topological phases \cite{Barkeshli:2014cna}, in four dimensional gauge theories \cite{Gukov:2013zka,Gaiotto:2014kfa} by studying their spontaneous breaking, and by forming phases protected by higher categorical symmetries \cite{Kapustin:2013uxa}. Anomalies of higher group symmetries are similarly preserved under renormalization group flow and can be used to constrain IR behaviors of theories and test conjectured infrared dualities \cite{Benini:2017dus,Benini:2018reh,Cordova:2017kue}. In this section, we would like to elaborate how the classification of defects and textures fits within this framework.

There are several subtly different ways in which an ordinary group can be generalized into a 2-group. We keep the discussion in this section informal, and refer the interested reader to appendix \ref{app.category}, which is a self-contained introduction to higher groups, for more details. One possible definition, drawing on the definition of a group as a category with one object whose morphisms are invertible, is that a 2-group is a 2-category with one object, and whose morphisms and 2-morphisms are invertible. (2-morphisms are maps between morphisms. The following
\beq \begin{tikzcd}
\bG \arrow[r, bend left=50, "\phi"{name=U}]
\arrow[r, bend right=50, "\psi"'{name=D}]
& \mathbb{H}
\arrow[Rightarrow, from=U, to=D, "\theta"]
\end{tikzcd} \eeq
depicts a 2-morphism $\theta$ from a morphism $\phi$ to another morphism $\psi$.) This defines a \emph{strict 2-group}.

It turns out that strict 2-groups are isomorphic to crossed modules. The bijection between them can be described as follows \cite{Baez2003}. Notice that the morphisms $\bG_1$ and the 2-morphisms $\bG_2$ of a strict 2-group $\bG$ both constitute ordinary groups under their respective composition laws. Let $s,t:\bG_2\to\bG_1$ be the (ordinary group) homomorphisms assigning the source and target to each 2-morphism. The crossed module associated to the strict 2-group $\bG$ is
\beq \p_{\bG}=t:\ker s\to \bG_1. \eeq
Conversely, given a crossed module $\p:H\to G$, we associate the strict 2-group with morphisms $G$ and 2-morphisms $H\rtimes G$, with $G$ acting on $H$ via $\p$. A 2-morphism $(h,g)\in H\rtimes G$ has source $g\in G$ and target $g\p(h)\in H$.

Strict 2-groups, and therefore also crossed modules, are the objects of a 2-category $\operatorname{\bf St2G}=\operatorname{\bf XMod}$. Under the above bijection, homomorphisms between crossed modules correspond to homomorphisms between strict 2-groups\footnote{See e.g. \cite{Baez2003} for definitions of homomorphisms and 2-morphisms of strict 2-groups.} and homotopies between homomorphisms of crossed modules correspond to 2-morphisms between homomorphisms of strict 2-groups. Hence, an alternative way of phrasing the content of this section is that we are counting isomorphism classes of 2-group homomorphisms between the fundamental 2-groups of two topological spaces.

In her thesis, Ho\^ang has classified isomorphism classes of strict 2-groups (and therefore crossed modules) in terms of four pieces of algebraic data $(\pi_1,\pi_2,\alpha,[\beta])$, sometimes known as the Ho\^ang data.\footnote{In some references, this has been termed S\'inh data, but as explained in the previous footnote, her surname is actually Ho\^ang, so we shall refer to it as the Ho\^ang data.} Given a crossed module $\p:H\to G$, the associated Ho\^ang data is
\beq \pi_1 = \coker\p,~~~ \pi_2 = \ker\p, \eeq
$\alpha:\pi_1\to\Aut\pi_2$ is the action induced from the action of $G$ on $H$, and $[\beta]\in H^3_\alpha(\pi_1,\pi_2)$ is the $\alpha$-twisted group cohomology class characterizing the extension of $\pi_2$ by $\pi_1$ (see Appendix \ref{app.grpext}). Notice that, when $\p$ is the fundamental crossed module, $\pi_1$ and $\pi_2$ are exactly the first two homotopy groups, $\alpha$ is the standard action of $\pi_1$ on $\pi_2$, and $[\beta]\in H^3_\alpha(\pi_1,\pi_2)=H^3(K(\pi_1,1),\pi_2)$ is the first Postnikov class.\footnote{See e.g. \cite{hatcher2003} for the definition of Postnikov systems.} Conversely, any (ordinary) groups $\pi_1,\pi_2$, with any action $\alpha:\pi_1\to\Aut\pi_2$ and any member of the group cohomology $[\beta]\in H^3_\alpha(\pi_1,\pi_2)$ is the Ho\^ang data for an isomorphism class of strict 2-groups (or crossed modules).\footnote{As mentioned in appendix \ref{app.category}, the Ho\^ang data also classifies isomorphism classes of coherent 2-groups and special 2-groups.}

Recently, the classification of symmetry enriched topological phases in 2+1 dimensions~\cite{Hermele2014SET,Barkeshli:2014cna,Tarantino2016SET} was understood in terms of counting equivalence classes of 2-group homomorphisms from an ordinary group, thought of trivially as a 2-group (i.e. setting $\pi_2=1$), to the automorphism 2-group of a modular tensor category \cite{Barkeshli:2014cna,Benini:2018reh,Cordova:2018cvg}. Equivalence classes of homomorphisms (i.e. homomorphisms taken up to natural isomorphism) between two 2-groups $\bG$ and $\bG'$ inducing a given homomorphism $\phi_1$ at the level of $\pi_1$ exist if and only if there exists some homomorphism $\phi_2$ at the level of $\pi_2$ such that $\phi_1^\ast[\beta']=\phi_{2\ast}[\beta]\in H^3(\pi_1,\pi_2')$. This is sometimes known as the ``$H^3$ obstruction to symmetry localization''~\cite{Barkeshli:2014cna,Fidkowski2017H3obstruction} in the literature. If this condition is satisfied, then the equivalence classes of 2-group homomorphisms form a torsor for $H^2_{\phi_1}(\pi_1,\pi_2')$. This is sometimes known as the $H^2$ ways in which symmetry can fractionalize. We note that our result on the classification of topological sectors fits within this paradigm (see proposition \ref{thm.2dcoh}), with the caveat that the existence of maps is never obstructed (essentially because a two dimensional $M$ is unable to probe the third cohomology of the target space $X$). In fact, the discussion in this section shows that the mathematical structure underpinning both of these problems (classification of symmetry enriched topological phases and classification of two dimensional textures and defects) is identical -- that of counting classes of homomorphisms between 2-groups.

\section{Dimension three} \label{sec.three}
\subsection{A special case}
Whitehead's results are actually more general than what we had discussed until now; they work for any $d$-dimensional $M$ as long as $X$ satisfies the further condition that $\pi_i=0$ for all $i=2,3,\ldots,d-1$ \cite{whitehead1949b,ellis1988}.
\begin{prop}
Let $M$ be a reduced CW complex of dimension $d$, and $X$ be a reduced CW complex such that $\pi_i=0$ for $i=2,3,\ldots,d-1$. Then
\beq [M,X]_0 \simeq \bigcup_{\phi_1} H_{\phi_1}^d(M,\pi_dX), \eeq
where the union is over homomorphisms of fundamental groups $\phi_1\in\Hom(\pi_1M,\pi_1X)$, and the cohomology groups have local coefficients via $\phi_1$ and the action of $\pi_1X$.
\end{prop}
This shows, for instance, that $[T^d,S^d]=[T^d,S^d]_0\simeq H^d(T^d,\bZ)=\bZ$ for all $d\geq 2$. In three (and higher odd dimensions), this generalizes to Lens spaces $L_{p,q}$, which are free quotients of $S^3$ with $\pi_1L_{p,q}=\bZ_p$, $\pi_2L_{p,q}=0$ and $\pi_3L_{p,q}=\bZ$. In that case we obtain
\beq [T^3,L_{p,q}]_0 = \bigcup^{p^3} \bZ = (\bZ_p)^3\times\bZ, \eeq
corresponding to a copy of $H^3(T^3,\bZ)\simeq\bZ$ for each homomorphism of $\pi_1T^3$ to $\pi_1L_{p,q}=\bZ_p$. Since the Lens spaces are orientable, $\pi_1L_{p,q}$ acts trivially on $\pi_3L_{p,q}$, and so
\beq [T^3,L_{p,q}] = (\bZ_p)^3\times\bZ. \eeq
\subsubsection{Example: Spiral phase of antiferromagnets}
The spiral phase of antiferromagnets spontaneously breaks all of the rotational symmetry, and is this described at low energies by an order parameter taking values in $SO(3)$ \cite{bhattacharjee2011}. The topological space underlying the Lie group $SO(3)$ is $\bR\bP^3=L_{2,1}$. The above remark shows that the the three dimensional textures of antiferromagnets in the spiral phase with periodic boundary conditions are classified by $8$ copies of the integers
\beq [T^3,SO(3)] = (\bZ_2)^3 \times \bZ. \eeq

\subsection{The general case}
Without the condition that the intermediate homotopy groups vanish, the problem becomes significantly more complicated. In this section, we examine the case when $M$ is a three dimensional CW complex, and do not assume that $\pi_2X=0$. Following Whitehead's success at algebraically modeling a homotopy 2-type with a crossed module, attempts were made to generalize these methods to higher dimensions. The groundwork was laid by Brown and Loday who proved van Kampen-like theorems for higher dimensional groups \cite{brown1987,brown1987b}. From this, several subtly different algebraic models of homotopy 3-types were introduced, such as crossed squares, introduced in \cite{guinwalery1981}, 2-crossed modules of \cite{conduche1984}, and quadratic modules of \cite{baues1991}. In this paper, we shall use the crossed squares model, as we feel, from our extremely limited exposure to this vast topic, that it has the easiest generalization to higher dimensions (see for example \cite{ellis1987}). We will mainly follow \cite{Ellis1993}. For an overview of the different models for homotopy $n$-types, see \cite{brown1990}.

The homotopy classification result in three dimensions is similar to the two dimensional case, but with the fundamental crossed module replaced by the \emph{fundamental crossed square}. We shall state the theorem before defining and explaining the algebraic concepts.
\begin{thm}[\cite{Ellis1993}] \label{thm.3d}
Let $M$ be a reduced CW complex of dimension 3, and $X$ be a reduced CW complex. Then
\beq [M,X]_0 \simeq [\Pi_{\leq 3}M,\Pi_{\leq 3}X]_0, \eeq
and
\beq [M,X] \simeq [\Pi_{\leq 3}M,\Pi_{\leq 3}X], \eeq
where $\Pi_{\leq 3}M$ is the fundamental crossed square of $M$, and $[\Pi_{\leq 3}M,\Pi_{\leq 3}X]_0$ (resp. $[\Pi_{\leq 3}M,\Pi_{\leq 3}X]$) denotes the based (resp. free) homotopy classes of $\Hom(\Pi_{\leq 3}M,\Pi_{\leq 3}X)$.
\end{thm}

\subsection{Crossed squares}
\begin{defn}
A crossed square \cite{guinwalery1981} is a commuting square of groups
\beq \begin{tikzcd}
L \ar[r,"\kappa"] \ar[d,"\eta"] &K \ar[d,"\nu"] \\
H \ar[r,"\mu"] &G
\end{tikzcd} \eeq
together with actions $G\to\Aut L,G\to\Aut K,G\to\Aut H$ of $G$ on $L$, $K$ and $H$ satisfying
\begin{enumerate}
\item 
each of the five homomorphisms $\kappa,\eta,\nu,\mu,\nu\kappa=\mu\eta$ are crossed modules, where the action of $K$ and $H$ on $L$ is given via the map to $G$,
\item 
$\kappa$ and $\eta$ are $G$-equivariant (so both $(\mu,\kappa)$ and $(\nu,\eta)$ are homomorphisms of crossed modules),
\end{enumerate}
and a function $[\ ,\ ]:H\times K\to L$, satisfying
\begin{enumerate}[resume]
\item 
$[hh',k] = [\leftidx{^h}{h'},\leftidx{^h}{k}][h,k]$ and $[h,kk']=[h,k][\leftidx{^k}{h},\leftidx{^k}{k'}]$,
\item 
$\eta[h,k] = h\leftidx{^k}{h^{-1}}$ and $\kappa[h,k]=\leftidx{^h}{k}k^{-1}$,
\item 
$[\eta l,k] = l\leftidx{^k}{l^{-1}}$ and $[h,\kappa l]=\leftidx{^h}{l}l^{-1}$,
\item 
$[\leftidx{^g}{h},\leftidx{^g}{k}] = \leftidx{^g}{[h,k]}$.
\end{enumerate}
\end{defn}
$K$ and $H$ act on $L,K$ and $H$ via their images in $G$, and $G$ understood to act on itself by conjugation.

\begin{figure}
\centering
\parbox{0.3\textwidth}{\includegraphics[width=0.3\textwidth]{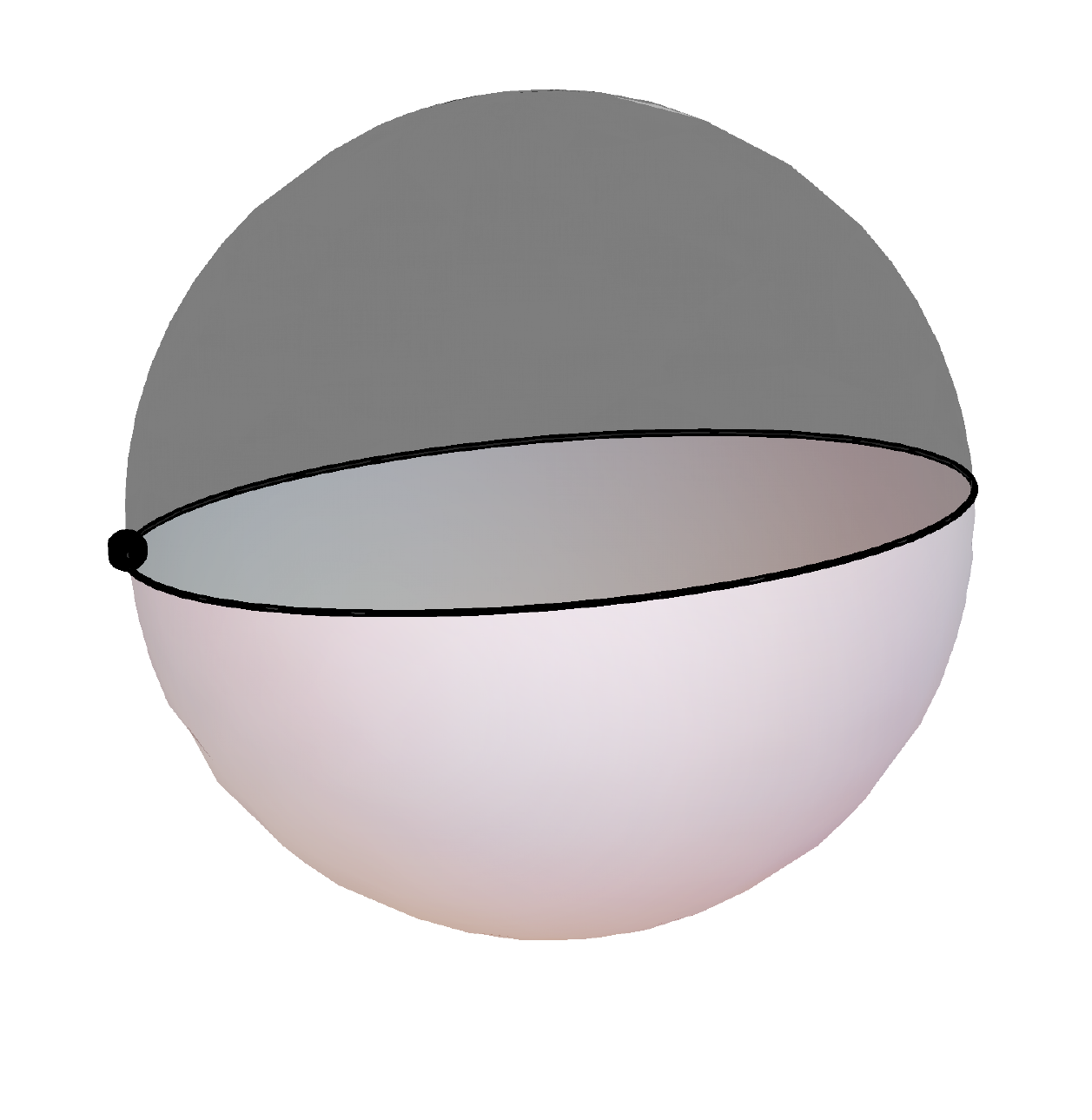}}
\parbox{0.1\textwidth}{\includegraphics[width=0.1\textwidth]{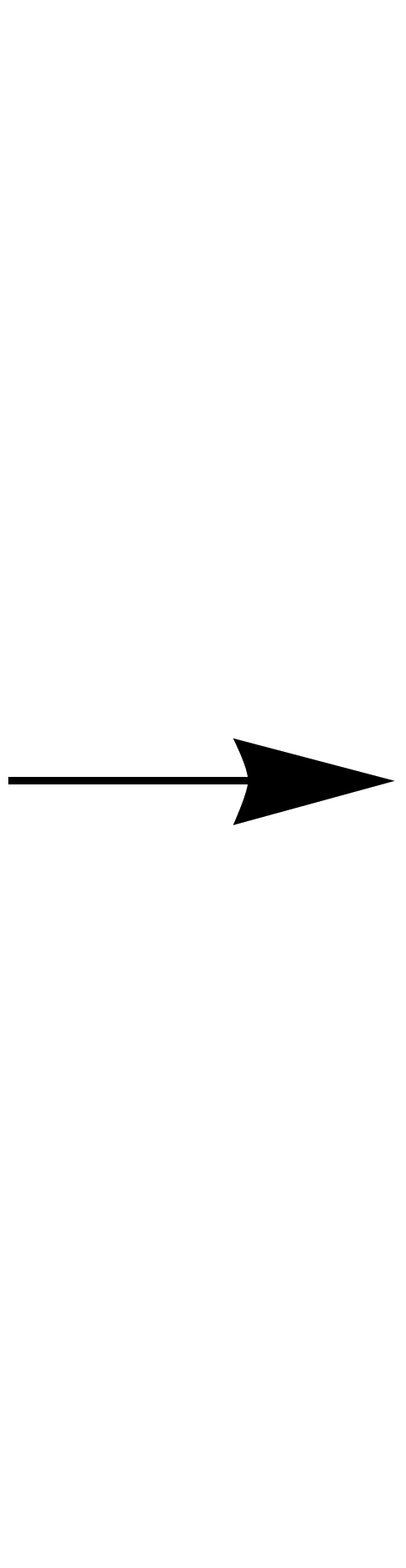}}
\parbox{0.4\textwidth}{\includegraphics[width=0.4\textwidth]{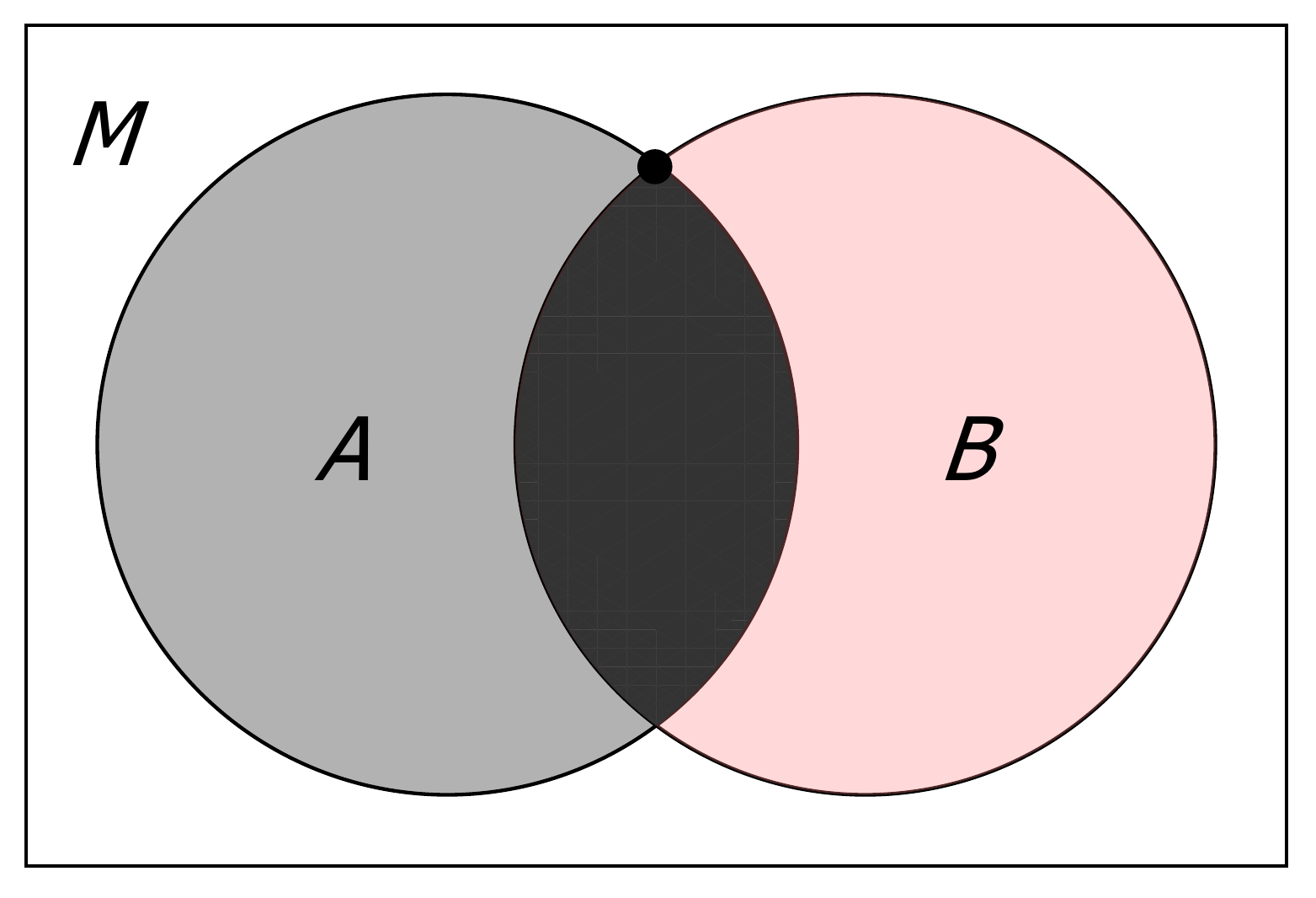}}
\caption{$\pi_3(M;A,B,\ast)$ consists of homotopy classes of triad maps $(B^3;B^2_+,B^2_-,p)\to (M;A,B,\ast)$.}
\label{fig.triadhom}
\end{figure}

\begin{eg}[Triad fundamental groups] \label{eg.triad}
The prototypical example of a crossed square is the crossed square associated to a pointed triad of spaces $(M;A,B,\ast)$. A pointed triad consists of two subspaces $A\subset M$, $B\subset M$ of a space $M$, with the base point contained in both subspaces $\ast\in A\cap B$. The third homotopy group $\pi_3(M;A,B,\ast)$ of the triad is defined to consist of homotopy classes of maps\footnote{The notation in \eqref{eqn.homtriad} represents a map $B^3\to M$ which maps $B^2_+\subset B^3$ into $A\subset M$, $B^2_-\subset B^3$ into $B\subset M$, and $p$ to $\ast$. Homotopy classes of such maps are taken with respect to homotopies along which $B^2_+,B^2_-,p$ are always mapped into $A,B,\ast$ respectively.}
\beq (B^3;B^2_+,B^2_-,p) \to (M;A,B,\ast) \label{eqn.homtriad} \eeq
where $B^3$ is the 3-dimensional unit disk, and $B^2_\pm$ are the upper and lower hemispheres of its surface \cite{blakers1951,blakers1952,blakers1953} (see figure \ref{fig.triadhom}). Then the commuting square\footnote{To avoid clutter, we once again suppress the base point dependence. The triad, relative and usual homotopy groups we consider are always pointed.}
\beq \begin{tikzcd}
\pi_3(M;A,B) \ar[r,"\p_-"] \ar[d,"\p_+"] &\pi_2(B,A\cap B) \ar[d,"\p"] \\
\pi_2(A,A\cap B) \ar[r,"\p"] &\pi_1(A\cap B)
\end{tikzcd} \eeq
together with the generalized Whitehead product \cite{blakers1953b} $\pi_2(A,A\cap B)\times\pi_2(B,A\cap B)\to\pi_3(A\cup B;A,B)\subset \pi_3(M;A,B)$ is a crossed square. In the diagram, $\p_\pm$ are the obvious projections of the boundary homomorphism from $\pi_3(M;A,B)$ to the relative homotopy groups (since the equator $B^2_+\cap B^2_-$ is mapped to $A\cap B$).
\end{eg}

\begin{defn}
A homomorphism of crossed squares
\beq \phi:\begin{bmatrix}L&K\\H&G\end{bmatrix}\to\begin{bmatrix}L'&K'\\H'&G'\end{bmatrix} \eeq
consists of group homomorphisms $\phi_L:L\to L',\phi_K:K\to K',\phi_H:H\to H',\phi_G:G\to G'$ such that the resulting cube
\beq \begin{tikzcd}[row sep=scriptsize, column sep=scriptsize]
& L \arrow[dl] \arrow[rr] \arrow[dd,"\phi_L",near start] & & K \arrow[dl] \arrow[dd,"\phi_K"] \\
H \arrow[rr, crossing over] \arrow[dd,"\phi_H"] & & G \\
& L' \arrow[dl] \arrow[rr] & & K' \arrow[dl] \\
H' \arrow[rr] & & G' \arrow[from=uu, crossing over,"\phi_G",near start]\\
\end{tikzcd} \eeq
of group homomorphisms commutes, each of $\phi_L,\phi_K$ and $\phi_H$ is $\phi_G$-equivariant, and $\phi_L[h,k]=[\phi_Hh,\phi_Kk]$.
\end{defn}

\subsection{Fundamental crossed square} \label{sec.fundxs}
\begin{figure}
\centering
\begin{subfigure}{0.4\textwidth}
\centering
\includegraphics[width=\textwidth]{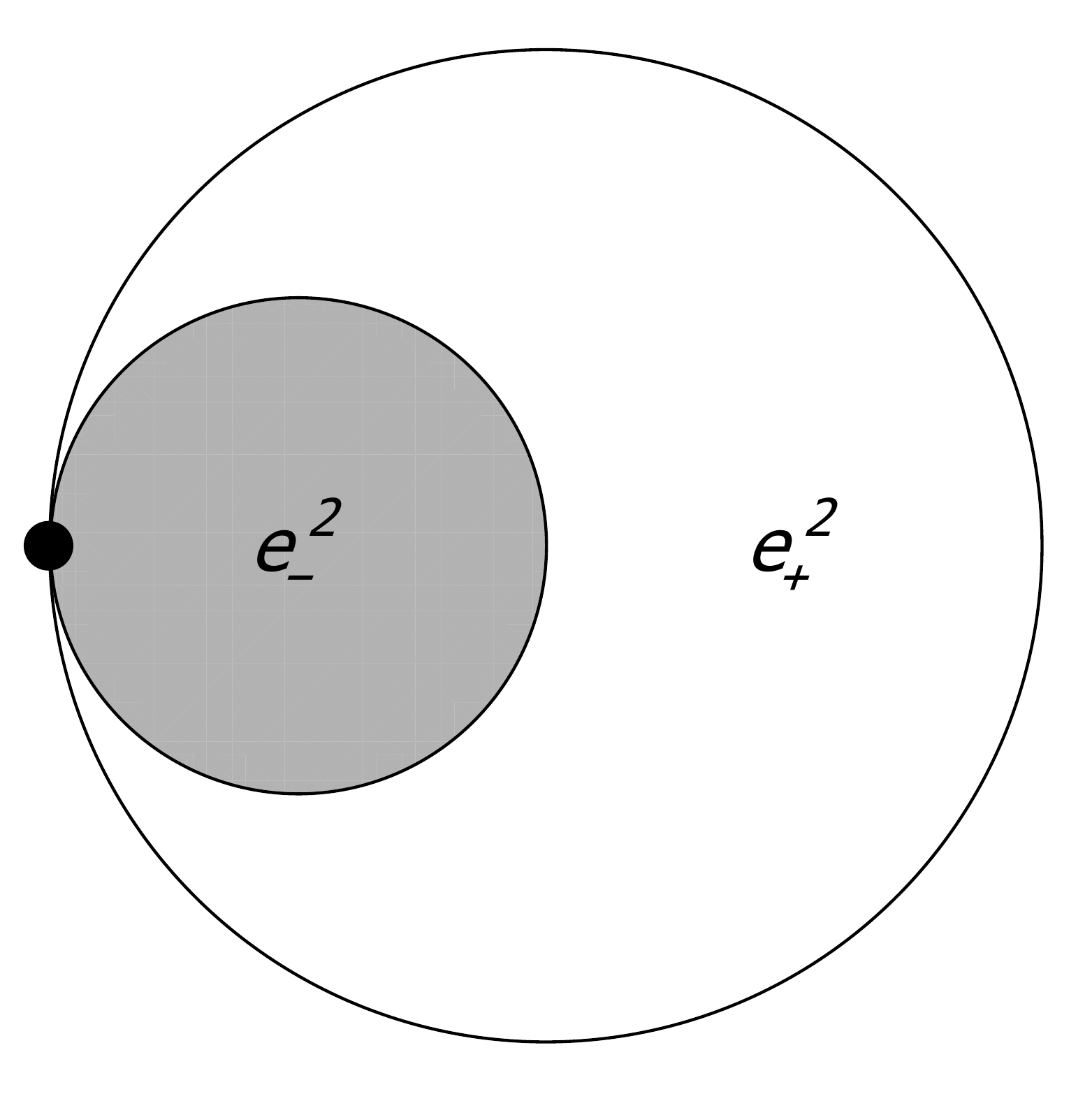}
\caption{Triad structure on each 2-cell.}
\label{fig.2cell}
\end{subfigure}
\begin{subfigure}{0.4\textwidth}
\centering
\includegraphics[width=\textwidth]{fig_triadhom2.pdf}
\caption{Standard triad structure on $S^2$.}
\label{fig.s2}
\end{subfigure}
\caption{Examples of triad structures.}
\end{figure}

In order to define the fundamental crossed square of a CW complex, we require some additional structure on the complex. Suppose that $M$ is a reduced CW complex, so that $M^1=\bigvee_{\Sigma_1} S^1$ is a wedge of circles. Each 2-cell of $M$ is attached to $M^1$ via a pointed map. Furthermore, define the following triad structure $(e^2,e^2_+,e^2_-)$ (see example \ref{eg.triad}) on each 2-cell $B^2$: take $e^2_-$ to be a smaller disk inside the 2-cell, touching the boundary $\p e^2$ only at the base point; and take $e^2_+$ to be the closure of the complement of $e^2_-$ in $e^2$. See figure \ref{fig.2cell}. The triad structure on each 2-cell induces an obvious triad structure $(M^2;M^2_+,M^2_-)$ on the 2-skeleton, by taking $M^2_\pm$ to be the subspace\footnote{but not necessarily CW-subspace of $M^2$.} of $M^2$ composed of $M^1$ and the union of all the $e^2_\pm$. Notice that if a 2-cell is attached trivially to the base point, then the triad structure corresponds to the northern and southern hemispheres of the sphere $(S^2;S^2_+,S^2_-)$ (see figure \ref{fig.s2}). Finally, we assume that the attaching maps of the 3-cells are in fact triad maps $(S^2;S^2_+,S^2_-)\to(M^2;M^2_+,M^2_-)$. A CW complex carrying such a triad structure and satisfying the above conditions is known as a \emph{canonical} CW complex. We shall assume that we are dealing with canonical CW complexes, and since every reduced CW complex is homotopy equivalent to a canonical one (with the same number of cells) \cite{Ellis1993}, there is no loss of generality in doing so.

Note the following properties of $M^2_\pm$: first, since $e^2_+$ retracts to the boundary of the 2-cell, and $e^2_-$ is contractible, both $M^2_+$ and $M^2_-$ retract to $M^1$. Second, $M^2_+$ and $M^2_-$ form a cover for the 2-skeleton: $M^2_+\cup M^2_-=M^2$. Third, the intersection $M^2_+\cap M^2_-$ is a wedge sum of circles, one for each 1-cell \emph{and} each 2-cell.

We now define a functor $\Pi_{\leq 3}$, known as the \emph{fundamental crossed square}, from canonical CW complexes to crossed squares, by taking $\Pi_{\leq 3}M$ to be the crossed square associated to the triad $(M;M^2_+,M^2_-)$ (see Example \ref{eg.triad}). In other words,
\beq \label{eqn.fundxs} \Pi_{\leq 3}M =
\begin{tikzcd}
\pi_3(M;M^2_+,M^2_-) \ar[r] \ar[d] &\pi_2(M^2_-,M^2_+\cap M^2_-) \ar[d] \\
\pi_2(M^2_+,M^2_+\cap M^2_-) \ar[r] &\pi_1(M^2_+\cap M^2_-)
\end{tikzcd}. \eeq

\subsection{A construction of the fundamental crossed square}\label{sec.summfundxs}
As the computation of triad homotopy groups is somewhat esoteric, we include in this section an explicit algorithm for computing the fundamental crossed square of a three dimensional CW complex. For a more detailed discussion, including definitions of several concepts used below, consult appendix \ref{app.fundxs}.

The fundamental crossed square of a three dimensional canonical CW complex
\beq \Pi_{\leq 3}M =
\begin{tikzcd}
\pi_3(M;M^2_+,M^2_-) \ar[r,"\kappa"] \ar[d,"\eta"] &\pi_2(M^2_-,M_+^2\cap M^2_-) \ar[d,"\nu"] \\
\pi_2(M^2_-,M^2_+\cap M^2_-) \ar[r,"\mu"] &\pi_1(M^2_+\cap M^2_-)
\end{tikzcd}
=
\begin{tikzcd}
L \ar[r,"\kappa"] \ar[d,"\eta"] &\bar H \ar[d,"\nu"] \\
H \ar[r,"\mu"] &G
\end{tikzcd} \eeq
has the following description:
\begin{enumerate}
\item 
Let $F=\pi_1M^1=\moy{\Sigma_1}$ and $\sigma_2:\Sigma_2\to F$ be the gluing maps of the 2-cells. Then $H$ is the free pre-crossed module on $\sigma_2$. In other words,
\beq H = \moy{F\times\Sigma_2} \eeq
with $\p:H\to F$ defined by $\p(f,t)=f\sigma_2(t)f^{-1}$ and $F$ acts on $H$ freely via $\leftidx{^f}{(f',t)}=(ff',t)$, where $f,f'\in F$ and $t\in\Sigma_2$.
\item 
$G$ is isomorphic to the semi-direct product $F\ltimes H$ (which is also isomorphic to the free group $\moy{\Sigma_1\cup\Sigma_2}$). The crossed module $\mu:H\to G$ is the obvious normal inclusion of $H$ into $F\ltimes H$.
\item 
$\bar H$ is isomorphic to $H$, and the crossed module $\nu:\bar H\to G$ is the normal inclusion given by
\beq \bar h \mapsto (\p h,h^{-1}) \in F\ltimes H. \eeq
\item 
Since the gluing maps of the 3-cells are triad maps, the restriction of the gluing maps to $e^2_+\cap e^2_-\approx S^1$ maps into $M^2_+\cap M^2_-$ and therefore defines a map $\sigma_3:\Sigma_3\to G$. Since both $e^2_+$ and $e^2_-$ retracts to $e^2_+\cap e^2_-$, $\sigma_3$ in fact maps into both $H$ and $\bar H$. Let $\p:C\to G$ be the free crossed module on $\sigma_3$.
\item 
$L$ is isomorphic to the coproduct $(H\otimes\bar H)\circ C$ modulo the relations
\beq \begin{cases}
i(\p c\otimes\bar h) = j(c)j(\leftidx{^{\bar h}}{c^{-1}}), \\
i(h\otimes\p c) = j(\leftidx{^h}{c})j(c^{-1}),
\end{cases} \eeq
where $h\in H$, $\bar h\in\bar H$ and $c\in C$. See definition \ref{def.tensor} for the nonabelian tensor product, and definition \ref{def.coproduct} for the coproduct $\circ$. The crossed module maps $\eta:L\to H$ and $\kappa:L\to\bar H$ are induced from the homomorphisms $\p:C\to H\cap\bar H$, $\eta':H\otimes\bar H\to H$ and $\kappa':H\otimes\bar H\to\bar H$, with the latter two defined by
\beq \eta'(h\otimes\bar h)=h\leftidx{^{\bar h}}{h^{-1}},~~~~~ \kappa'(h\otimes\bar h)=\leftidx{^h}{\bar h}{\bar h}^{-1}. \eeq
\end{enumerate}

Finally, we remark that the first three homotopy groups of $M$ can be recovered from its fundamental crossed square as the homology groups of the complex
\beq L \biglongto^{\eta\times\kappa} H\rtimes\bar H \biglongto^{\mu\nu} G, \label{eqn.homgrpxs} \eeq
or equivalently, the homology groups of the complex
\beq L \bigto^\eta H \bigto^\p F \eeq
where $\p:H\to F$ is the pre-crossed module map (see 1 above). In other words,
\beq \pi_1M = \coker \p = F/\p(H), ~~~~~ \pi_2M=\ker\p/\eta(L), ~~~~~ \pi_3M = \ker\eta. \eeq

\begin{eg}[$\Pi_{\leq 3}S^2$] \label{eg.s2xs}
The standard CW structure on $S^2$ with one 2-cell attached trivially admits a triad structure given by the northern and southern hemispheres (see figure \ref{fig.s2}). Therefore $H=\bar H=G=\bZ$, the free group on the singleton set $\Sigma_2$, with the crossed module maps $\mu$ and $\nu$ being identity maps. $L$ is the tensor product $\bZ\otimes\bZ$, which reduces in this case to the standard abelian tensor product over $\bZ$, so $L\simeq\bZ$. From point 5 above, the crossed module maps $\eta$ and $\kappa$ are trivial maps. Therefore,
\beq \Pi_{\leq 3}S^2 = \begin{tikzcd}
\bZ \ar[r,"0"] \ar[d,"0"] &\bZ \ar[d,"\id"] \\
\bZ \ar[r,"\id"] &\bZ
\end{tikzcd}. \eeq
\end{eg}
Notice that from the above remark, we can immediately read off $\pi_3(S^2)=\bZ$.

\begin{figure}
\centering
\includegraphics[width=0.3\textwidth]{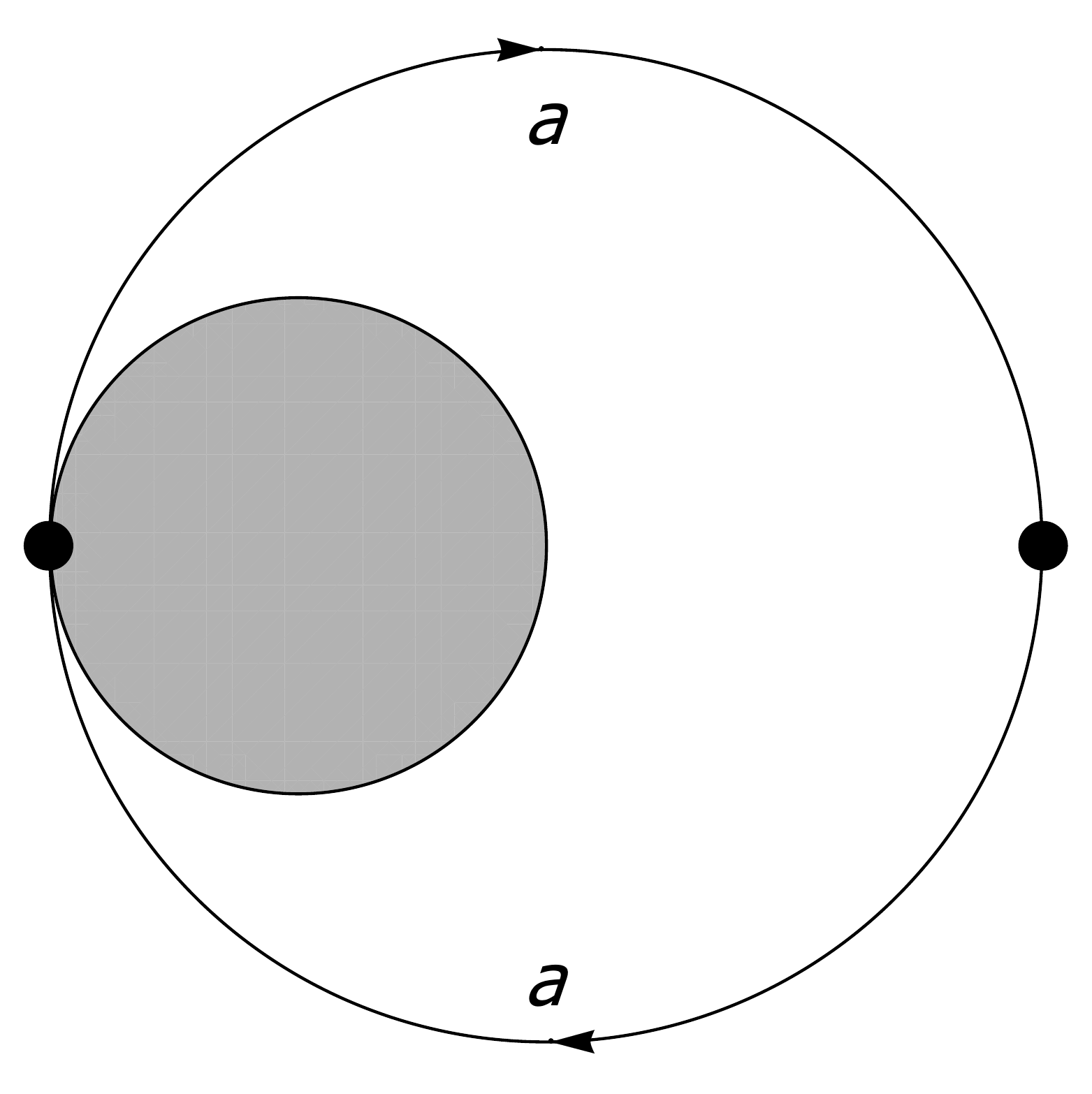}
\caption{Triad structure on $\bR\bP^2$ taken as the cross cap.}
\label{fig.rp2}
\end{figure}

\begin{eg}[$\bR\bP^2$ and $\bR\bP^3$]
The standard CW structure on $\bR\bP^3$ consists of one 1-cell $\Sigma_1=\{a\}$, one 2-cell $\Sigma_2=\{t\}$ and one 3-cell $\Sigma_3=\{x\}$, and its 2-skeleton is $\bR\bP^2$. Therefore, we have $F=\moy{a}\simeq\bZ$. The 2-cell is glued to two copies of the 1-cell, $\sigma_2(t)=a^2$. $H=\moy{F\times\Sigma_2}$ is generated by elements of the form $\leftidx{^n}{t}:=\leftidx{^{a^n}}{t}$ with $n\in\bZ$, and the pre-crossed module homomorphism sends $\p(\leftidx{^n}{t})=a^2$. Note that $G=F\ltimes H$ is isomorphic to the free group $\moy{a,t}$ under the isomorphism defined on generators by
\beq \(a^{n_0},\leftidx{^{n_1}}{t}\ \leftidx{^{n_2}}{t}\ \ldots\leftidx{^{n_k}}{t}\) \mapsto a^{n_1}ta^{-n_1}a^{n_2}ta^{-n_2}\ldots a^{n_k}ta^{-n_k}a^{n_0}. \eeq

A set of representatives for the free crossed module corresponding to $H$ comprises elements of the form $t^{n_0}\ \leftidx{^1}{t}^{n_1}$, where $(n_0,n_1)\in\bZ^2$, so $\pi_2(\bR\bP^2,S^1)=\bZ^2$, recovering the fundamental crossed module of $\bR\bP^2$ in example \ref{eg.rp2}, where we have written it as $\bZ[\bZ_2]$. The triad structure on $\bR\bP^2$ is shown in figure \ref{fig.rp2}. The 3-cell attaches to $\bR\bP^2$ killing $\pi_2\bR\bP^2$, which is isomorphic to the kernel of $\pi_2(\bR\bP^2,S^1)\to\pi_1S^1$. Clearly, $t\ \leftidx{^1}{t}^{-1}$ is a representative of the generator of the kernel, so we can take the triad gluing map to be $\sigma_3(x)=t\ \leftidx{^1}{t}^{-1}$.

The group $L$ is complicated, and we do not know of a simpler way of presenting it than as the coproduct $L=(H\otimes\bar H)\circ C$, with $C$ the free crossed module on $\sigma_3$. We note that the kernel $\pi_3\bR\bP^3=\ker\eta$ is generated by the element in $L$ represented by
\beq \(t\otimes\overline{\leftidx{^2}{t}{}}\)^{-1} \circ x\ \leftidx{^{(a,1)}}{x}, \eeq
where representatives of elements in the free crossed module $C$ are generated by elements of the form $\leftidx{^{(f,h)}}{x}$, $f\in F,h\in H$.
\end{eg}

\begin{figure}
\centering
\includegraphics[width=0.6\textwidth]{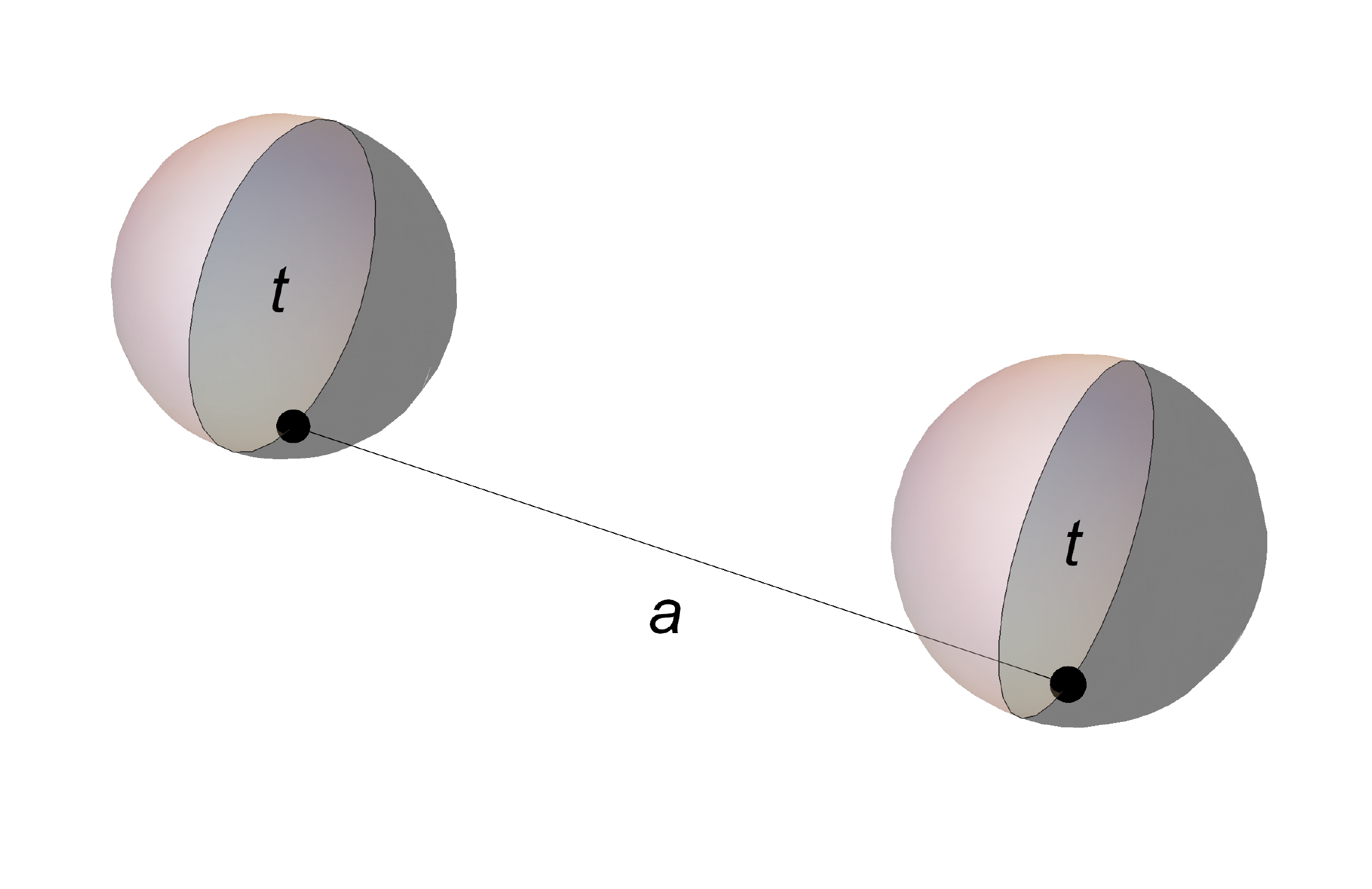}
\caption{Triad structure on the 2-skeleton $S^1\vee S^2$ of $S^1\times S^2$. The spheres on either end of the line are identified.}
\label{fig.s1s2}
\end{figure}

\begin{eg}[$\Pi_{\leq 3}(S^1\times S^2)$] \label{eg.s1s2xs}
The standard CW structure on $S^1\times S^2$ consists of one 1-cell $\Sigma_1=\{a\}$, one 2-cell $\Sigma_2=\{t\}$ and one 3-cell $\Sigma_3=\{x\}$. The 2-cell is attached via the trivial map $\sigma_2(t)=1$ to $F=\moy{a}\simeq\bZ$. Therefore, $H=\moy{F\times\Sigma_2}$ is generated by elements of the form $(a^n,t)$, and the pre-crossed module homomorphism $\p:H\to F$ is the trivial homomorphism. Note that the free crossed module on $\sigma_2$ is the abelianization $H^{\mathrm{ab}}$, which is isomorphic to $\pi_2(S^1\vee S^2)=\bZ[\bZ]$, in agreement with example \ref{eg.ringr3}.

Notice that $\bar H=H$ as subgroups of $F\ltimes H$, since $H\to F$ is the trivial homomorphism. This also implies that the homomorphism $\eta':H\otimes\bar H\to H$ is the commutator map. The 3-cell is attached via the triad map $\sigma_3(x)=t\ \leftidx{^a}{t}^{-1}$ (see figure \ref{fig.s1s2}), and the free crossed module $C\to F\ltimes H$ on $\sigma_3$ is injective. The kernel $\eta:(H\otimes\bar H)\circ C\to H$, which is isomorphic to $\pi_3(S^1\times S^2)=\bZ$, is generated by $\(t\otimes t\)\circ 1$.
\end{eg}

\begin{figure}
\centering
\includegraphics[width=0.4\textwidth]{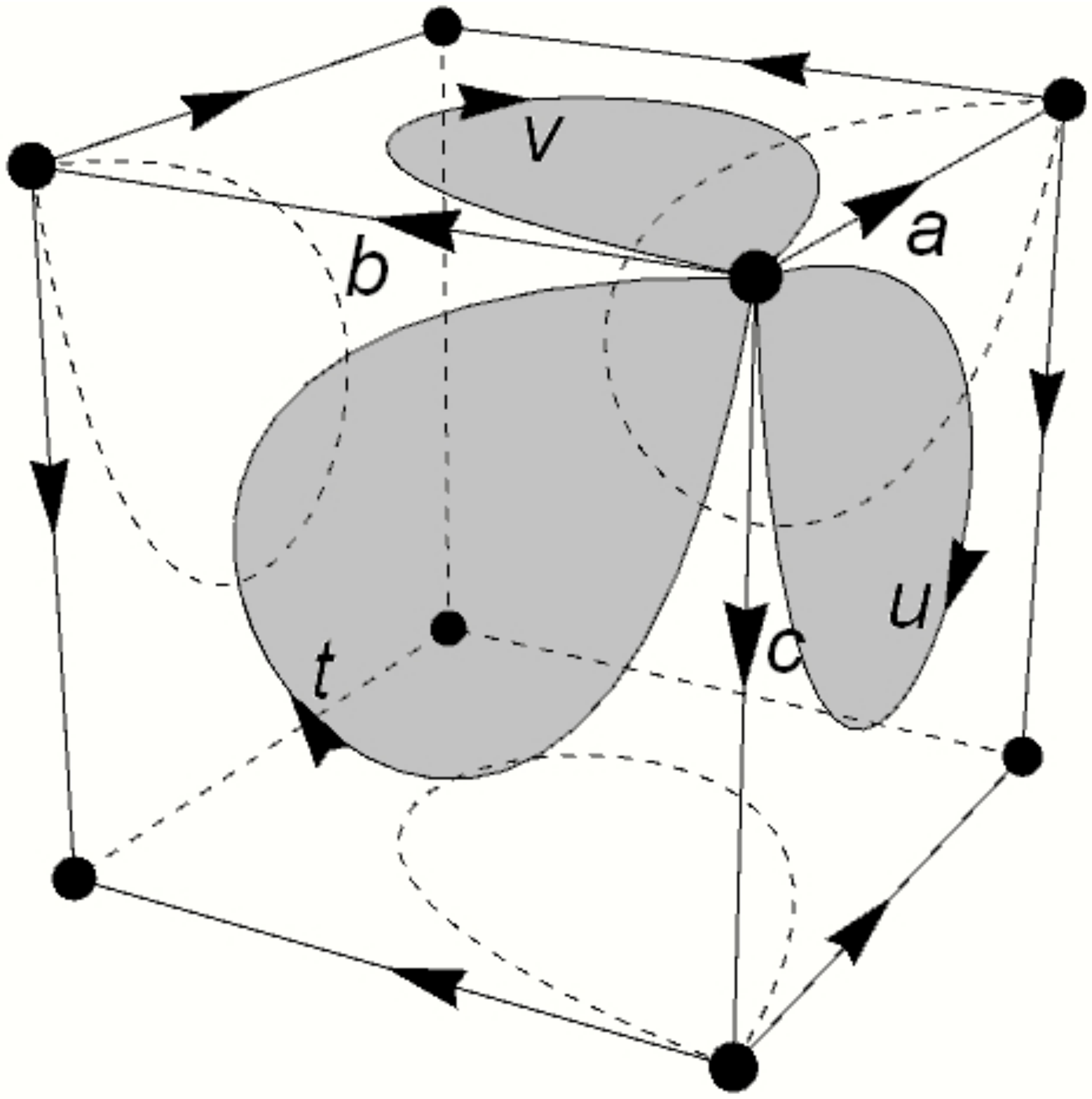}
\caption{Triad structure on the 2-skeleton of $T^3$. The opposite faces of the cube are identified.}
\label{fig.t3}
\end{figure}

\begin{eg}[$\Pi_{\leq 3}T^3$] \label{eg.t3}
The standard CW structure on $T^3=S^1\times S^1\times S^1$ consists of three 1-cells $\Sigma_1=\{a,b,c\}$, three 2-cells $\Sigma_2=\{t,u,v\}$ and one 3-cell $\Sigma_3=\{x\}$. The 2-cells are attached via commutators
\beq \sigma_2(t)=bcb^{-1}c^{-1},~~~~ \sigma_2(u)=cac^{-1}a^{-1},~~~~ \sigma_2(v)=aba^{-1}b^{-1}, \eeq
which defines the pre-crossed module $\p:H=\moy{\moy{a,b,c}\times\{t,u,v\}}\to\moy{a,b,c}$.

The 3-cell is attached via the triad map (see figure \ref{fig.t3})
\beq \sigma_3(x) = t\ \leftidx{^c}{v}^{-1}\ u\ \leftidx{^a}{t}^{-1}\ v\ \leftidx{^b}{u}^{-1} \in H\cap\bar H. \eeq
\end{eg}

\subsection{Homomorphisms of free crossed squares}
When counting homomorphisms of free crossed squares, the following remark from \cite{Ellis1993} is useful.
\begin{prop}[\cite{Ellis1993}] \label{thm.hom}
Homomorphisms
\beq \begin{tikzcd}[row sep=scriptsize, column sep=scriptsize]
& L \arrow[dl] \arrow[rr] \arrow[dd,"\phi_L",near start] & & K \arrow[dl] \arrow[dd,"\phi_K"] \\
H \arrow[rr, crossing over] \arrow[dd,"\phi_H"] & & G \\
& L' \arrow[dl] \arrow[rr] & & K' \arrow[dl] \\
H' \arrow[rr] & & G' \arrow[from=uu, crossing over,"\phi_G",near start]\\
\end{tikzcd} \eeq
from the free crossed square on the attaching maps $\sigma_2,\sigma_3$ to another free crossed square are completely determined by their restriction to the cells
\beq \phi_1:\Sigma_1\to F',~~~~~ \phi_2:\Sigma_2\to H',~~~~~ \phi_3:\Sigma_3\to L'. \eeq
Conversely, any three functions $\phi_1,\phi_2,\phi_3$ as defined above, for which the following diagrams
\beq \begin{tikzcd}
\Sigma_2 \ar[r,"\phi_2"] \ar[d,"\p"] &H' \ar[d,"\p'"] \\
F \ar[r,"\phi_1^\ast"] &F'
\end{tikzcd}~~~~~\text{{and}}~~~~~
\begin{tikzcd}
\Sigma_3 \ar[r,"\phi_3"] \ar[d,"\eta"] &L' \ar[d,"\eta'"] \\
H \ar[r,"\phi_2^\ast"] &H'
\end{tikzcd} \eeq
commute, extend uniquely to a homomorphism of crossed squares. Here, $\phi_1^\ast$ is the unique homomorphism extending $\phi_1$ and $\phi_2^\ast$ is the unique $\phi_1^\ast$-equivariant homomorphism extending $\phi_2$ and satisfying the commutative diagram
\beq \begin{tikzcd}
H \ar[r,"\phi_2^\ast"] \ar[d,"\p"] &H' \ar[d,"\p'"] \\
F \ar[r,"\phi_1^\ast"]& F'
\end{tikzcd}. \eeq
\end{prop}

\subsection{Homotopies between homomorphisms of free crossed squares}
Taking inspiration from section \ref{sec.2dmap}, we could define a homotopy between homomorphisms $\phi,\psi$ of fundamental crossed squares from $\Pi_{\leq 3}M\to\Pi_{\leq 3}X$ to be a homomorphism from $\Pi_{\leq 3}I_0M$ (or $\Pi_{\leq 3}IM$ for free homotopies) to $\Pi_{\leq 3}X$ restricting to $\phi$ and $\psi$ on the subcomplexes $0\times M$ and $1\times M$ (see section \ref{sec.2dmap} for definitions of $IM$ and $I_0M$). However, we have not actually given an algorithmic description of the triad homotopy group $\pi_3(IM;IM^2_+,IM^2_-)$ of a four dimensional CW complex. In reality, it is much simpler to realize $IM$ (or $I_0M$) as another type of algebraic object, the fundamental squared complex \cite{Ellis1993}, which is the generalization of the fundamental crossed complex \cite{whitehead1949b} to incorporate crossed squares.
\begin{defn}
A squared complex is a collection of groups and group homomorphisms
\beq \begin{tikzcd}
&&& K \ar[rd,"\nu"]& \\
\to C_5 \ar[r,"\p_5"] &C_4 \ar[r,"\p_4"] &L \ar[ur,"\kappa"] \ar[dr,"\eta"] && G \\
&&& H \ar[ru,"\mu"]&
\end{tikzcd} \eeq
together with actions of $G$ on $L,K,H,C_4,C_5,\ldots$ and a $G$-equivariant function $[,]:H\times K\to L$, satisfying the properties
\begin{enumerate}
\item $L,H,K,G$ is a crossed square,
\item $C_j$, $j\geq 4$ are abelian groups,
\item $\p_j\p_{j+1}=0$ for $j\geq 4$, and $\kappa\p_4=\eta\p_4=0$,
\item $\nu(K)$ and $\mu(H)$ act trivially on $C_j$, $j\geq 4$. This and 2 together imply that each $C_j$ is a $\pi_1=G/\mu(H)\nu(K)$-module.
\item Each $\p_j$, $j\geq 4$, is $\pi_1$-equivariant.
\end{enumerate}
\end{defn}
The fundamental squared complex $\tilde\Pi_{\leq 3}M$ of a canonical CW complex $M$ of arbitrary dimension is
\beq \begin{tikzcd}
&&&\pi_2(M^2_-,M^2_+\cap M^2_-) \ar[dr,"\nu"]& \\
\ldots\ar[r,"\p_5"] &\pi_4(M^4,M^3) \ar[r,"\p_4"] &\pi_3(M^3;M^2_+,M^2_-) \ar[ur,"\bar\eta"] \ar[dr,"\eta"] &&\pi_1(M^2_+\cap M^2_-) \\
&&&\pi_2(M^2_+,M^2_+\cap M^2_-) \ar[ur,"\mu"]&
\end{tikzcd}. \eeq
The Hurewicz isomorphisms (see e.g. \cite{hatcher2003}) tell us that the relative homotopy groups $\pi_j(M^j,M^{j-1})$, $j\geq 4$, are free $\pi_1M$-modules over the set $\Sigma_j$ of of $j$-cells. A free squared complex is a squared complex whose crossed square part is free and groups $C_j$, $j\geq 4$ are free as $\pi_1$-modules, so the fundamental squared complex of a complex is free.

Homomorphisms of squared complexes can be defined in an obvious way, generalizing homomorphisms of crossed squares. As in the case of crossed squares, a map between canonical CW complexes induces a homomorphism between fundamental squared complexes. In dimension at most four, the converse also applies: a homomorphism of fundamental squared complexes can always be realized by some map between the CW complexes \cite{Ellis1993,baues2006}. Therefore, we can define the notion of homotopy between homomorphisms of crossed squares as a homomorphism of squared complexes.
\begin{defn}[Homotopy of crossed square homomorphisms]
Let $M$ be a three dimensional canonical CW complex and $X$ be a canonical CW complex of arbitrary dimension. Two homomorphisms $\phi,\psi:\Pi_{\leq 3}M\to\Pi_{\leq 3}X$ of fundamental crossed squares are said to be based homotopic (resp. free homotopic) if there exists a homomorphism $\Phi:\tilde\Pi_{\leq 3}I_0M\to\Pi_{\leq 3}X$ (resp. $\Phi:\tilde\Pi_{\leq 3}IM\to\Pi_{\leq 3}X$) of fundamental squared complexes extending $\phi$ and $\psi$ (that is, such that $\Phi|_{0\times M}=\phi$ and $\Phi|_{1\times M}=\psi$).
\end{defn}

We end this section by remarking that the notion of homotopy groups of crossed squares (see the discussion around equation \eqref{eqn.homgrpxs}) can be generalized to squared complexes by defining $\pi_j$ to be the homology groups of the chain complex
\beq \ldots\biglongto^{\p_{j+1}} C_j \biglongto^{\p_j} C_{j-1} \to\ldots\biglongto^{\p_5} C^4 \biglongto^{\p_4} L \biglongto^{\eta\times\kappa} H\rtimes K \biglongto^{\mu\nu} G. \eeq
When applied to the fundamental squared complex, these homotopy groups coincide with the topological homotopy groups.

\subsection{Examples of three dimensional sigma models and their topological sectors}
\subsubsection{Heisenberg spins in three dimensions}
We compute the topological sectors of a two spatial dimensional thermal theory of Heisenberg spins asymptoting to a constant at spatial infinity. Hence, $M=S^1\times S^2$ and $X=S^2$. Alternatively, this also describes a codimension three spatial defect in a theory of spins at finite temperature. The result which we will derive is
\beq [S^1\times S^2,S^2] = \bigcup_{q\in\bZ} \bZ_{\abs{2q}} \eeq
with the provision that $\bZ_0=\bZ$. When $q=0$, it is possible to choose a representative configuration which is constant on the 2-skeleton, and the integer characterizing the homotopically inequivalent maps reducing to the constant map on the 2-skeleton corresponds to the Hopf invariant in $\pi_3S^2=\bZ$. This can be interpreted as the sector describing the pair production of skyrmions of opposite unit charge, braided a number of times equal to the Hopf invariant, before annihilation \cite{wilczek1983}. When $q$ is nonzero, the picture is more complicated and can be thought of as a generalization of the situation in \cite{wilczek1983}. $q$ can be thought of as the wrapping number $q\in\pi_2S^2$ of any $S^2$ cross section. For each $q$, there are $2q$ homotopically distinct classes of maps which restrict on the 2-skeleton to the configuration with wrapping number $q$ (see example \ref{eg.ringr3} for an analysis of the topological sectors on the 2-skeleton).

We explain in some detail how the result is obtained. The fundamental crossed squares of $S^1\times S^2$ and $S^2$ are (see examples \ref{eg.s2xs} and \ref{eg.s1s2xs})
\beq \Pi_{\leq 3}(S^1\times S^2) =
\begin{tikzcd}
L \ar[r,"\bar\eta"] \ar[d,"\eta"] &\overline{\moy{\bZ\times\{t\}}} \ar[d,hook] \\
\moy{\bZ\times\{t\}} \ar[r,hook] &\moy{a}\ltimes\moy{\bZ\times\{t\}}
\end{tikzcd},~~~~
\Pi_{\leq 3}S^2 =
\begin{tikzcd}
\bZ \ar[r,"0"] \ar[d,"0"] & \bZ \ar[d,"\id"] \\
\bZ \ar[r,"\id"] & \bZ
\end{tikzcd}. \eeq
Following proposition \ref{thm.hom}, the homomorphisms from $\Pi_{\leq 3}(S^1\times S^2)$ to $\Pi_{\leq 3}S^2$ are determined by two integers $\phi_2(t)\in\bZ$ and $\phi_3(x)\in\bZ$. One can check that the diagram
\beq \begin{tikzcd}
\{x\} \ar[r,"\phi_3"] \ar[d,"\sigma_3"] &\bZ \ar[d,"0"] \\
\moy{\bZ\times\{t\}} \ar[r,"\phi_2^\ast"] &\bZ
\end{tikzcd} \eeq
commutes for all values of $\phi_2(t)$ and $\phi_3(x)$, so conversely every pair of integers determines a homomorphism of crossed squares.

Two homomorphisms $\phi,\psi$ are homotopic (note that since $S^2$ is simply connected, based homotopy and free homotopy classes coincide) if $\phi$ and $\psi$ extends to a homomorphism of squared complexes from $\tilde\Pi_{\leq 3}I_0(S^1\times S^2)$ to $\tilde\Pi_{\leq 3}S^2$. The CW structure of $I_0(S^1\times S^2)$ consists of one 0-cell, two 1-cells $\{a_0,a_1\}$, three 2-cells $\{t_0,t_1,a_I\}$, three 3-cells $\{x_0,x_1,t_I\}$ and one 4-cell $\{x_I\}$, where e.g. $a_0$ denotes the copy of the cell $a$ along $0\times(S^1\times S^2)$, while $a_I$ denotes the 2-cell $[0,1]\times a$. The attaching maps for $t_0,t_1,x_0,x_1$ are the same as those in $\Pi_{\leq 3}(S^1\times S^2)$, and it is straightforward to see that the attaching maps for $a_I$ and $t_I$ can be chosen as
\beq \sigma_2(a_I) = a_1a_0^{-1}, \eeq
\beq \sigma_3(t_I) = t_1t_0^{-1}. \eeq
The boundary map $\p_4(x_I)\in\pi_3(I_0M^3;I_0M^2_+,I_0M^2_-)$ is more difficult to construct, in no part because
\beq L=\pi_3(I_0M^3;I_0M^2_+,I_0M^2_-)=\pi_3(I_0M^2;I_0M^2_+,I_0M^2_-)\circ C=(H\otimes\bar H)\circ C \eeq
is complicated to describe. First, note that $x_I$ is bounded by four 3-cells: $x_0,x_1$ and $t_I$ twice. Since the boundary homomorphism $\pi_4(I_0M^4;I_0M^3)\to\pi_3(I_0M^3)$ to sends $x_I$ to $x_0-x_1$, we conclude that the $C$ portion of $\p_4(x_I)$ can be chosen to be of the form
\beq \p_4(x_I)_C = \leftidx{^{(f_1,h_1)}}{t_I}\ \leftidx{^{(f_2,h_2)}}{x_1}\ \leftidx{^{(f_3,h_3)}}{t_I}^{-1}\ \leftidx{^{(f_4,h_4)}}{x_0}^{-1} \eeq
for some $f_i\in F$ and $h_i\in H$, which can be solved for by noting that $\eta\p_4(x_I)=1$. Therefore, $\eta(\p_4(x_I)_C)$ should take the form $\eta'(l')$ for some $l'\in H\otimes\bar H$. After some work, we find
\beq \p_4(x_I)_C = \leftidx{^{t_0^{-1}}}{}\(t_I\ x_1\ t_I^{-1}\ \leftidx{^{a_I}}{x_0}^{-1}\), \eeq
from which we find
\beq \eta(\p_4(x_I)_C) = \leftidx{^{a_1}}{t_0}^{-1}\ a_I^{-1}\ \leftidx{^{a_0}}t_0\ t_0^{-1}\ a_I\ t_0 = \eta'\left[ (\leftidx{^{a_1}}{t_0}^{-1}\otimes a_I)(a_I^{-1}\otimes t_0)\right] \eeq
so finally
\beq \p_4(x_I) = (a_I^{-1}\otimes t_0)^{-1}(\leftidx{^{a_1}}{t_0}^{-1}\otimes a_I)^{-1} \circ \leftidx{^{t_0^{-1}}}{}\(t_I\ x_1\ t_I^{-1}\ \leftidx{^{a_I}}{x_0}^{-1}\). \eeq
Note that $(t_0\otimes t_0)\circ 1$ and $(t_1\otimes t_1)\circ 1$ also lie in $\ker\eta$, but these correspond to generators of $\pi_3(S^2)$, to which $x_I$ does not attach to, so $\p_4(x_I)$ should not contain terms of this form.

A homomorphism $\Phi$ of squared complexes $\tilde\Pi_{\leq 3}I_0(S^1\times S^2)\to\tilde\Pi_{\leq 3}S^2$ extending $\phi$ and $\psi$ exists if the cells $a_I$, $t_I$ and $x_I$ can be mapped by $\Phi$ such that the diagrams\footnote{The targets of $a_0,t_0,x_0$ and $a_1,t_1,x_1$ are already determined by $\phi$ and $\psi$ respectively.}
\beq \begin{tikzcd}
\{a_I\} \ar[r,"\Phi_2"] \ar[d,"\p"] &\bZ \ar[d] \\
\moy{a_0,a_1} \ar[r,"\Phi_1^\ast"] &1
\end{tikzcd},~~~~~~
\begin{tikzcd}
\{t_I\} \ar[r,"\Phi_3"] \ar[d,"\sigma_3"] &\bZ \ar[d,"0"] \\
H \ar[r,"\Phi_2^\ast"] &\bZ
\end{tikzcd},~~~~~~
\begin{tikzcd}
\{x_I\} \ar[r] \ar[d,"\p_4"] &0 \ar[d] \\
L \ar[r,"\Phi_3^\ast"] & \bZ
\end{tikzcd}
\eeq
commute. The first diagram is trivially satisfied; the second implies
\beq \Phi_2(t_1t_0^{-1}) = \psi(t)-\phi(t) = 0, \eeq
and the third implies
\begin{align}
0=\Phi_3^\ast\p_4(x_I) &= \Phi_3^\ast\left[(a_I^{-1}\otimes t_0)^{-1}(\leftidx{^{a_1}}{t_0}^{-1}\otimes a_I)^{-1} \circ \leftidx{^{t_0^{-1}}}{}\(t_I\ x_1\ t_I^{-1}\ \leftidx{^{a_I}}{x_0}^{-1}\)\right] \nn \\
&= -\Phi_2^\ast(a_I^{-1})\Phi_2^\ast(t_0)-\Phi_2^\ast(\leftidx{^{a_1}}{t_0}^{-1})\Phi_2^\ast(a_I)+\Phi_3^\ast\left[\leftidx{^{t_0^{-1}}}{}\(t_I\ x_1\ t_I^{-1}\ \leftidx{^{a_I}}{x_0}^{-1}\)\right] \nn \\
&= 2\phi(t)\Phi_2(a_I) + \psi(x)-\phi(x).
\end{align}
Therefore, the homotopy classes are given by
\beq [S^1\times S^2,S^2] = \bigcup_{q\in\bZ}\bZ_{2q}. \eeq
The integer $q$ can be viewed as the winding number $q\in[S^1\vee S^2,S^2]\simeq\pi_2(S^2)\simeq\bZ$ of the map restricted to the 2-skeleton $S^1\vee S^2$ (see section \ref{eg.ringr3}); for each $q$, there are $2q$ homotopically inequivalent ways of extending the map to all of $S^1\times S^2$.

This is consistent with a classic result of Pontrjagin \cite{Pontrjagin1941} classifying homotopy classes of maps from a three dimensional complex $M$ to $S^2$. We briefly summarize his result in the following. Given a map $\phi:M\to S^2$, let $\phi|_2$ be the restriction to a 2-skeleton of $M$. Then the pullback of the volume form on $S^2$ defines an element in the cohomology group $\phi|_2^\ast \omega_{S^2}\in H^2(M,\bZ)$. Each element of $H^2(M,\bZ)$ arises in this way as the pullback of some map. Fix an element $\alpha\in H^2(M,\bZ)$; then the number of homotopy classes of maps giving rise in the above manner to $\alpha$ is equal to the index of the sublattice $2\alpha\cup H^1(M,\bZ)$ in $H^3(M,\bZ)$. In other words,
\beq [M,S^2] \simeq \bigcup_{\alpha\in H^2(M,\bZ)} H^3(M,\bZ)/(2\alpha\cup H^1(M,\bZ)). \eeq

\subsubsection{Textures of Heisenberg spins in three dimensions with periodic boundary conditions} \label{sec.t3s2}
We compute the textures of a three dimensional model of spins with periodic boundary conditions, so $M=S^1\times S^1\times S^1=T^3$ and $X=S^2$. The result is
\beq \label{eqn.t3s2} [T^3,S^2] = \bigcup_{(q_1,q_2,q_3)\in\bZ^3}\bZ_{2q} \eeq
where $q=\gcd(\abs{q_1},\abs{q_2},\abs{q_3})$, and $\bZ_0=\bZ$. The triplet of integers $(q_1,q_2,q_3)$ can be understood as the skyrmion numbers (i.e. the integers in $H^2(T^2,\pi_2(S^2))\simeq\bZ$ characterizing the texture) on each face of the fundamental domain of the torus (the 3 faces shown in the figure \ref{fig.t3}). For each triplet of integers $(q_1,q_2,q_3)$, there are $\bZ_{2q}$ homotopy classes of maps which restrict to the above mentioned three-skyrmion configuration labeled by $(q_1,q_2,q_3)$ on the 2-skeleton.

Most of the calculations in this section are similar to but more complicated than the previous example, so we shall assume familiarity with the previous example and be more sparing with the details.

The fundamental crossed square of $T^3$ has been computed in example \ref{eg.t3} above; it is the free crossed square on $\sigma_2:\Sigma_2=\{t,u,v\}\to\moy{\Sigma_1}=\moy{a,b,c}$ and $\sigma_3:\Sigma_3=\{x\}\to\moy{\moy{a,b,c}\times\{t,u,v\}}$ where
\beq \sigma_2(t)=bcb^{-1}c^{-1},~~~~~ \sigma_2(u)=cac^{-1}a^{-1},~~~~~ \sigma_2(v)=aba^{-1}b^{-1}, \eeq
\beq \sigma_3(x)=t\ \leftidx{^c}{v}^{-1}\ u\ \leftidx{^a}{t}^{-1}\ v\ \leftidx{^b}{u}^{-1}. \eeq
Homomorphisms $\phi$ from $\Pi_{\leq 3}T^3$ to $\Pi_{\leq 3}S^2$ correspond bijectively to quadruplets of integers \\ \mbox{$(\phi(t),\phi(u),\phi(v),\phi(x))\in\bZ^4$} (one should check that the diagram
\beq \begin{tikzcd}
\{x\} \ar[r,"\phi_3"] \ar[d,"\sigma_3"] &\bZ \ar[d,"0"] \\
M \ar[r,"\phi_2^\ast"] &\bZ
\end{tikzcd} \eeq
commutes for all values of $(\phi(t),\phi(u),\phi(v),\phi(x))$~).

The CW structure of $I_0T^3$ has one 0-cell, six 1-cells $\{a_0,b_0,c_0,a_1,b_1,c_1\}$, nine 2-cells \\ $\{t_0,u_0,v_0,t_1,u_1,v_1,a_I,b_I,c_I\}$, five 3-cells $\{x_0,x_1,t_I,u_I,v_I\}$ and one 4-cell $\{x_I\}$; with attaching maps
\beq \sigma_2(a_I) = a_1a_0^{-1} \eeq
(and similarly for $b_I$ and $c_I$),
\beq \sigma_3(t_I) = t_1\ \leftidx{^c}{b_I}\ c_I\ t_0^{-1}\ b_I^{-1}\ \leftidx{^b}{c_I}^{-1} \eeq
(and cyclically with $(t,a)\to(u,b)\to(v,c)\to(t,a)$).
The attaching map $\p_4:\{x_I\}\to\pi_3(I_0M^3;I_0M^2_+,I_0M^2_-)$ maps $x_I$ to a generator of $\ker\eta=\pi_3(I_0M^3)$ (since $\pi_3(I_0T^3)=0$). $x_I$ is bounded by eight 3-cells, and is sent to $x_1-x_0$ under the boundary homomorphism $\pi_4(I_0M^4,I_0M^3)\to\pi_3(I_0M^3)$, so the $C$ portion of $\p_4(x_I)$ can be chosen to take the form
\beq \p_4(x_I)_C = \leftidx{^{(f_1,h_1)}}{x_1}\ \leftidx{^{(f_2,h_2)}}{t_I}^{-1}\ \leftidx{^{(f_3,h_3)}}{v_I}\ \leftidx{^{(f_4,h_4)}}{u_I}^{-1}\ \leftidx{^{(f_5,h_5)}}{x_0}^{-1}\ \leftidx{^{(f_6,h_6)}}{t_I}\ \leftidx{^{(f_7,h_7)}}{v_I}^{-1}\ \leftidx{^{(f_8,h_8)}}{u_I} \eeq
for some $f_i\in F$ and $h_i\in H$, which can be obtained by solving $\eta\p_4(x_I)=1$, or, equivalently, writing $\eta(\p_4(x_I)_C)$ as $\eta'(l')$ for some $l'\in H\otimes\bar H$.

A homotopy $\Phi$ between $\phi$ and $\psi$ exists if and only if $\Phi$ can be defined on the cells $\{a_I,b_I,c_I,t_I,u_I,v_I,x_I\}$ in such a way that the diagrams
\beq \begin{tikzcd}
\{a_I,b_I,c_I\} \ar[r,"\Phi_2"] \ar[d,"\sigma_2"] & \bZ \ar[d] \\
F \ar[r] & 1 \end{tikzcd},~~~~~
\begin{tikzcd}
\{t_I,u_I,v_I\} \ar[r,"\Phi_3"] \ar[d,"\sigma_3"] &\bZ \ar[d,"0"] \\
H \ar[r,"\Phi_2^\ast"] &\bZ
\end{tikzcd},~~~~~
\begin{tikzcd}
\{x_I\} \ar[r,"\Phi_4"] \ar[d,"\p_4"] & 0 \ar[d] \\
L \ar[r,"\Phi_3^\ast"] & \bZ
\end{tikzcd} \eeq
commute. The first diagram places no requirement on the target of $a_I,b_I,c_I$, while the second requires $\phi(t)=\psi(t)$, $\phi(u)=\psi(u)$ and $\phi(v)=\psi(v)$. After some considerable calculation (see appendix \ref{app.calc} for more details), one can show that the third diagram becomes
\beq 0=\Phi_3^\ast\p_4(x_I)=\psi(x)-\phi(x)-2\phi(t)\Phi(a_I)-2\phi(u)\Phi(b_I)-2\phi(v)\Phi(c_I). \label{eqn.t3s2hom} \eeq
Therefore, $\phi$ and $\psi$ are homotopic if and only if $(\phi(t),\phi(u),\phi(v))=(\psi(t),\psi(u),\psi(v))$ and $\phi(x)-\psi(x)$ is twice of some integer linear combination of $\phi(t),\phi(u),\phi(v)$. Thus we arrive at \eqref{eqn.t3s2}, which is consistent with Pontrjagin's result \cite{Pontrjagin1941}.

\section{Conclusion and further discussion}\label{sec.conc}
In this paper, a systematic approach to classifying the topological sectors of non-linear sigma models was presented. The approach generalizes the statement that based homotopy classes of maps from a one dimensional space to an arbitrary space is classified by homomorphisms between the respective fundamental groups. In $d$ dimensions, the fundamental groups are generalized to fundamental crossed $(d-1)$-cubes, and the homotopy classification amounts to counting equivalence classes of homomorphisms between these algebraic objects. The situation in $d=2$ and $3$ was discussed in detail. Explicit constructions of the algebraic objects -- fundamental crossed modules and fundamental crossed squares -- were provided, in principle reducing the problem of enumerating topological sectors in two and three dimensions to an algorithmic combinatorial exercise. However, the complexity of the algebraic objects increases dramatically as we go up in dimension, as exhibited by the three dimensional case we considered in section \ref{sec.three}. In fact, computing the three dimensional textures of nematics (with target space $\bR\bP^2$) was extremely daunting  despite being algorithmic, and would have to be revisited in further work, probably with the aid of machine computation.

The mathematical procedure demonstrated by replacing fundamental groups with fundamental crossed cubes is that of \emph{categorification}. Recently, and rather belatedly, methods from higher category theory were applied to physics, particularly in the study of symmetries and classification of phases. This paper can be viewed as a continuation of the theme, applying methods of higher categories to the classification of topological phases of sigma models. In two dimensions, the situation is fairly well understood, as crossed modules are equivalent to strict 2-groups, which have been classified by Ho\^ang. In three dimensions, however, there are many subtly different flavors of 3-group or cat$^2$ group one could define (as discussed in appendix \ref{app.category}), and it is unclear exactly how our formulation relates to others. Further work, studying exactly how this work fits in the language of higher categories, particularly with regards to fundamental $\infty$-groupoids, is in the pipeline.

\section{Acknowledgements}
We would like to thank Alexander Abanov, Alexander Kirillov, Martin  Ro\v cek, Tobias Shin, Dennis Sullivan and Ying Hong Tham for numerous helpful discussions and comments. AP would like to thank the Kavli Institute for Theoretical Physics for hospitality during the preparation of this manuscript. JPA is supported in part by NSF grant PHY1620628. AP was supported in part by NSF grants PHY1333903, PHY1314748, and PHY1620252 and is currently supported by the Simons Foundation via the ICTS-Simons postdoctoral fellowship.

\appendix
\section{Categorical groups, group objects in categories, etc.} \label{app.category}
In this appendix, we state the content of this paper in a category theoretical language. Due to our background as physicists, we will keep the discussion self-contained and accessible, beginning with an informal introduction to higher categories. We highly recommend the paper \cite{Baez2003} which introduces 2-groups in an accessible manner.

We begin with some mathematical terminology. A $0$-category is a set of objects. A $1$-category, or simply category, is a collection of objects with morphisms between objects. An $n$-category $\rC$ is a collection of objects, such that associated to any two objects $x,y\in\rC$ is an $n-1$-category $\rC(x,y)$. The objects of $\rC(x,y)$ are 1-morphisms from $x$ to $y$; the 1-morphisms of $\rC(x,y)$ are 2-morphisms of $\rC$, and so on. For instance, the collection $\Cat$ of all (small) categories is a 2-category, with 1-morphisms being functors and 2-morphisms being natural transformations between functors.

An ordinary group can be defined in a number of ways:
\begin{enumerate}
\item
A category $G$ with one object and invertible morphisms.
\item
An object $G$ in a category with morphisms $m:G\times G\to G$, $\inv:G\to G$ and $\id:I\to G$, which describe multiplication, inverse and identity, satisfying certain axioms. Such an object is known as a group object, so we could succintly say that $G$ is a group object in some category.
\item
An object in the category $\Grp$ of groups.
\end{enumerate}

\subsection{2-groups}
Depending on which definition one works with, and the choices one can make in the process, different higher categorical generalizations of groups are possible. For example, working with definition 1 yields probably the most economical version of a 2-group, a 2-category with one object and invertible morphisms and 2-morphisms. This is known as a strict 2-group. We could have instead chosen a slightly different generalization: a 2-category with one object, \emph{weakly} invertible\footnote{A weakly invertible morphism is a morphism $\phi$ for which there exist morphisms $\alpha$ and $\beta$ such that $\alpha\phi$ and $\phi\beta$ are both isomorphic to the identity.} morphisms and invertible 2-morphisms. This is known as a weak 2-group. If one appends to the definition of weak 2-group fixed choices of weak inverses of morphisms and 2-isomorphisms to the identity morphism, one gets the definition of a coherent 2-group.

Definitions 2 and 3 seem to be more complicated that definition 1, but they have an advantage in that they apply to groups with extra structures, such as topological groups or Lie groups, via a process known as internalization. For example, in definition 2, a group object in the category $\operatorname{\bf Diff}$ of smooth manifolds is a Lie group. In definition 3, let the category $\operatorname{\bf DiffGrp}$ be the category of group objects in $\operatorname{\bf Diff}$ and homomorphisms in $\operatorname{\bf Diff}$; then tautologically, an object in $\operatorname{\bf DiffGrp}$ is a Lie group. Definitions 2 and 3 generalize to 2-groups as follows
\begin{enumerate}
\item[2'.]
A strict 2-group is a group object in some 2-category.
\item[3'.]
A categorical group is an object in the 2-category $\Grp\!\Cat$.
\end{enumerate}
Given any category $\rC$, the \emph{internal category} $\rC\!\Cat$ is a 2-category whose objects are categories in $\rC$, 1-morphisms are functors in $\rC$ and 2-morphisms are natural transformations in $\rC$. A category in $\rC$, say, $X$, consists of two objects $X_0,X_1\in\rC_0$, known as the objects of $X$ and morphisms of $X$ respectively, and three morphisms $s_X,t_X\in\rC(X_1,X_0)$ and $i_X\in\rC(X_0,X_1)$ known as the source-assigning, target-assigning and identity-assigning morphisms of $X$, satisfying the usual axioms of a category. Functors in $\rC$ and natural transformations in $\rC$ are the expected structures generalizing functors and natural transformations in the prototypical 2-category $\Cat$.

2-groups in the category $\rC$ can be defined by taking in definition 2 the 2-category $\rC\!\Cat$, and categorical groups in the category $\rC$ can be defined by taking in definition 3 the 2-category $\rC\!\Grp\!\Cat=(\rC\!\Grp)\Cat$, the 2-category of internal categories in $\rC\!\Grp$. So for example a strict Lie 2-group is a group object in $\operatorname{\bf Diff}\!\Cat$, and a categorical Lie group is an object in $\operatorname{\bf Diff}\!\Grp\!\Cat$.

A remarkable result of category theory, the commutativity of internalization \cite{maclane1971}, states that $\rC\!\Cat\!\Grp=\rC\!\Grp\!\Cat$, i.e. a strict 2-group is equivalent to a categorical group, so definitions 2 and 3 are equivalent. Moreover, in section \ref{sec.2gsect} we showed that a strict 2-group was equivalent to a crossed module, so crossed modules are algebraic realizations of both strict 2-groups and categorical groups.

We could also have generalized definition 2 in a weaker sense by defining a coherent (or weak, or special) 2-group to be a coherent 2-group object in some 2-category, leading to flavors of 2-group for which the group operation is not associative, but rather associative only up to natural isomorphism.
\begin{enumerate}
\item[2''.]
A (weak, coherent, special, etc.) 2-group is a (weak, coherent, special etc.) 2-group object in some 2-category.
\end{enumerate}
The collection of (weak, coherent, special, strict etc.) 2-groups form 2-categories $\operatorname{\bf W2G}$, $\operatorname{\bf C2G}$, $\operatorname{\bf S2G}$, $\operatorname{\bf St2G}$, etc. with 2-group homomorphisms as morphisms and 2-group natural transformations as 2-morphisms \cite{Baez2003}. The content of section \ref{sec.two} can be understood as counting natural isomorphism classes of morphisms in the 2-category $\operatorname{\bf S2G}\simeq\operatorname{\bf XMod}$.

Coherent, special and strict 2-groups have the same isomorphism classes \cite{Baez2003}, so the classification results of Ho\^ang similarly apply: isomorphism classes of coherent, special and strict 2-groups are classified by four pieces of data $(\pi_1,\pi_2,\alpha,[\beta])$ where $\pi_1$ and $\pi_2$ are groups, $\alpha:\pi_1\to\Aut\pi_2$ is an action and $[\beta]\in H_\alpha^3(\pi_1,\pi_2)$. We discussed in section \ref{sec.2gsect} that $[\beta]$ could be interpreted as the crossed module extension class in the case of the 2-category $\operatorname{\bf XMod}$. For special 2-groups, $[\beta]$ has a different interpretation; this time as the cohomology class of the associator natural isomorphism.

Finally, we remark that the homotopy 2-type of a space can also be modeled by 2-groups of a different flavor. For example, the fundamental 2-group $\tilde\Pi_{\leq 2}(X,\ast)$ is a coherent 2-group, described as follows. The objects of the category are based loops in $X$; and the morphisms between two loops are the homotopies between them, modulo homotopies-of-homotopies. The group structure is given by the concatenation of loops and their homotopies.

The skeletal version of the fundamental 2-group has objects equal to the isomorphism classes of the objects of $\tilde\Pi_{\leq 2}(X,\ast)$, which is exactly $\pi_1(X,\ast)$. The class $[\beta]$ then has the interpretation as the homotopy-of-homotopy class of the homotopy implementing the associator isomorphism.

\subsection{3-groups}
One could generalize definitions 2 and 3 to $n$-groups:
\begin{enumerate}
\item[2.]
A strict $n$-group is a group object in some $n$-category. A strict $n$-group in category $\rC$ is a group object in the $n$-category $\rC\!\Cat^{n-1}:=\rC\!\Cat\!\Cat\ldots\Cat$.
\item[3.]
A cat$^{n-1}$ group is an object in the $n$-category $\Grp\!\Cat^{n-1}:=\Grp\!\Cat\!\Cat\ldots\Cat$. A cat$^{n-1}$ group in category $\rC$ is an object in $\rC\!\Grp\!\Cat^{n-1}$.
\end{enumerate}
For $n>2$, definitions 2 and 3 no longer define equivalent objects, so a $3$-group is not the same as a cat$^2$ group. Crossed $n$-cubes (crossed squares when $n=2$) are algebraic models for cat$^n$ groups \cite{ellis1987}, but we are unclear on exactly how they relate to strict $(n+1)$-groups, and indeed, to other flavors of $(n+1)$-groups.

The content of section \ref{sec.three} can be interpreted as counting 2-isomorphism classes of morphisms between objects in the 3-category $\Grp\!\Cat^2$.

\section{The fundamental crossed module of a two dimensional complex}
\label{app.freexm}
The fundamental crossed module $\Pi_{\leq 2}M$ of a reduced CW complex $M$ is free, in the following sense. The 1-skeleton $M^1$ is a wedge sum of circles, and so $\pi_1M^1$ is the free group $\moy{\Sigma_1}$ on the set $\Sigma_1$ of 1-cells. Let $\Sigma_2$ be the set of 2-cells, and $\sigma_2:\Sigma_2\to\pi_1M^1$ be the gluing map of the 2-cells. Then consider the free group $\moy{\pi_1M^1\times\Sigma_2}$ on the product of $\pi_1M^1$ with the 2-cells, and the homomorphism
\beq \tilde\p:\moy{\pi_1M^1\times\Sigma_2}\to \pi_1M^1,~~~~~\tilde\p(g,t)=g\sigma_2(t)g^{-1}, \eeq 
where $g\in\pi_1M^1$ and $t\in\Sigma_2$. Together with the free action of $\pi_1M^1$ on $\moy{\pi_1M^1\times\Sigma_2}$ by $\leftidx{^g}{(g',t)}=(gg',t)$, this defines a pre-crossed module (see definition \ref{def.xm}). In fact, both $\pi_1M^1$ and $\moy{\pi_1M^1\times\Sigma_2}$ are free groups, and the pre-crossed module is fully determined by $\Sigma_1,\Sigma_2$ and the attaching map $\sigma_2$, so we shall call such a structure the \emph{free} pre-crossed module on $\sigma_2:\Sigma_2\to\moy{\Sigma_1}$.

We can ``force'' condition 2 of definition \ref{def.xm} on a free pre-crossed module by taking the quotient by the normal subgroup $N$ generated by all elements of the form $(\leftidx{^{\tilde\p(\bar h)}}{\bar h'})\bar h{\bar h'}{}^{-1}\bar h^{-1}$. This yields a crossed module
\beq \p:\moy{\pi_1M^1\times\Sigma_2}/N \to \pi_1M^1 \eeq
which is known as the \emph{free crossed module} on $\sigma_2$. In fact, this is isomorphic to the relative homotopy group
\beq \pi_2(M^2,M^1) \simeq \moy{\pi_1M^1\times\Sigma_2}/N, \eeq
and $\tilde\p$ descends to give the boundary homomorphism $\p:\pi_2(M^2,M^1)\to\pi_1M^1$.

\section{Double extensions of groups by crossed modules}
\label{app.grpext}
\subsection{Group extensions are classified by $H^2_\alpha(G,N)$}
We first recall the classification of group extensions. If the sequence
\beq 1\to N\to E\to G\to 1 \eeq
is exact, we say that $E$ is a group extension of $G$ by $N$. If $N$ is abelian, then the action of $E$ on itself by conjugation descends to an action $G\to \Aut N$. We have the following result
\begin{prop}
Let $G$ be a group, $N$ be an abelian group, and $\alpha:G\to\Aut N$ be a given action of $G$ on $N$. Then group extensions of $G$ by $N$ inducing the action $\alpha$ are classified by the second $\alpha$-twisted group cohomology $H^2_\alpha(G,N)$.
\end{prop}
Given an extension $1\to N\to E\to G\to 1$, we can get a representative of the corresponding group cohomology class as follows. Choose a lift $s:G\to E$; this will in general fail to be a homomorphism (in fact, if $s$ is a homomorphism, the extension is split and corresponds to the trivial cohomology class). Let $\omega(g,h)=s(g)s(h)s(gh)^{-1}$ measure its failure to be a homomorphism. Since $\omega(g,h)$ projects to the trivial element in $G$, it is in fact valued in $N$. Furthermore, associativity of $G$ enforces the cocycle condition on $\omega$, so in fact $\omega(g,h)$ is a 2-cocycle representing the class corresponding to the extension $E$. Different choices of lifts yield different representatives of the cohomology class.

\subsection{Crossed module extensions are classified by $H^3_\alpha(\pi_1,\pi_2)$}
If the sequence
\beq 1\to\pi_2\to H\bigto^\p G\to\pi_1\to 1 \eeq
is exact, and $\p:H\to G$ is a crossed module, then $\p:H\to G$ is an crossed module extension (or double extension) of $\pi_1$ by $\pi_2$. (Note that this automatically implies that $\pi_2$ is abelian.) The crossed module action of $G$ on $H$ descends to an action of $\pi_1$ on $\pi_2$. There is the following, analogous result (see e.g. \cite{brown1982})
\begin{prop}
Let $\pi_1$ be a group, $\pi_2$ be an abelian group, and $\alpha:\pi_1\to\Aut\pi_2$ be a given action of $\pi_1$ on $\pi_2$. Then crossed module extensions of $\pi_1$ by $\pi_2$ inducing the action $\alpha$ are classified up to equivalence by the third $\alpha$-twisted group cohomology $H^3_\alpha(\pi_1,\pi_2)$.
\end{prop}
Two crossed module extensions $\p:H\to G$ and $\p':H'\to G'$ are equivalent if there exists a crossed module homomorphism $\phi$ between them such that the following diagram commutes
\beq \begin{tikzcd}
&&H'\ar[r,"\p"] \ar[dd,"\phi_2"] &G'\ar[rd] \ar[dd,"\phi_1"]&& \\
1\ar[r] &\pi_2 \ar[ru] \ar[rd] &&& \pi_1 \ar[r]& 1\ . \\
&&H\ar[r,"\p"] &G \ar[ru]&&
\end{tikzcd} \eeq

Given an extension $\ds1\to\pi_2\to H\bigto^\p G\to\pi_1\to 1$, we can get a representative of the corresponding group cohomology class as follows. Choose a lift $s:\pi_1\to G$, and as before, measure its failure to be a homomorphism with $\omega(a,b)=s(a)s(b)s(ab)^{-1}$. Since $\omega(a,b)$ projects to the identity, it lies in the image $\p H$, and so we can choose a lift $\tilde\omega:\pi_1\times\pi_1\to H$ of $\omega$. Now, measure the failure of $\tilde\omega$ to be associative with
\beq \beta(a,b,c) = \delta\tilde\omega(a,b,c) = \leftidx{^{\alpha(a)}}{\tilde\omega(b,c)}\tilde\omega(a,bc)\tilde\omega(ab,c)^{-1}\tilde\omega(a,b)^{-1}. \eeq
Since $\pi_1$ is associative, the image $\p\beta=\delta\omega=1$ is trivial, so $\beta$ in fact is valued in $\pi_2$. From the definition and the fact that $\delta^2=0$, it is clear that $\beta$ is a 3-cocycle. In fact, $\beta$ represents the cohomology class corresponding to the equivalence class of crossed module extension. Different choices of lifts made in the definition of $\beta$ yield different representatives of the cohomology class. This class $[\beta]$ is the class referred to in the main body of the text.

\section{Explicit construction of the fundamental crossed square}\label{app.fundxs}
In this appendix, we give an explicit construction of the fundamental crossed square, following \cite{Ellis1993}. This is an elaboration of the outline given in section \ref{sec.summfundxs}.

The fundamental crossed square of a canonical CW complex is free on the cells of the complex and their gluing maps \cite{Ellis1993}. We shall elaborate what this means. Let $\Sigma_i$ be the set of $i$-cells of $M$ and $F=\pi_1M^1=\moy{S_1}$. The relative homotopy group $H=\pi_2(M^2_+,M^2_+\cap M^2_-)$ can be constructed as the free pre-crossed module (see definition \ref{def.xm}) on the attaching maps $\sigma_2:\Sigma_2\to F$ of the 2-cells
\beq \label{eqn.prexm1} H = \pi_2(M^2_+,M^2_+\cap M^2_-) = \moy{F\times\Sigma_2}. \eeq
Now consider the crossed module corresponding to the pair $(M^2_+,M^2_+\cap M^2_-)$. Since $M^2_+$ retracts onto $M^1\subset M^2_+\cap M^2_-$, the crossed module map is injective, yielding the exact sequence
\beq 0\to \pi_2(M^2_+,M^2_+\cap M^2_-) \bigto^{\tilde\p} \pi_1(M^2_+\cap M^2_-) \to \pi_1M^2_+\simeq \pi_1M^1 \to 1 \eeq
which splits via the homomorphism induced by the inclusion $M^1\hookrightarrow M^2_+\cap M^2_-$. This yields our desired description of $G$ as
\beq G = \pi_1(M^2_+\cap M^2_-) \simeq \pi_1M^1\ltimes\pi_2(M^2_+,M^2_+\cap M^2_-) = F\ltimes H \label{eqn.g} \eeq
where the action of $F$ on $H$ is as in the pre-crossed module \eqref{eqn.prexm1}.

Recall that $M^2_+\cap M^2_-$ is a wedge sum of circles, one for each 1-cell and each 2-cell, so $G=\moy{\Sigma_1\cup\Sigma_2}$. One may explicitly check that this is consistent with \eqref{eqn.g}.

The crossed module $H\to G=F\ltimes H$ making up the bottom of the fundamental crossed square \eqref{eqn.fundxs} is the inclusion $h\mapsto (1,h)$. There is a second copy of $H$ in $G$, which we denote by
\beq \bar H = \{(\p h,h^{-1}):h\in H\} \subset F\ltimes H \eeq
which we identify with $K=\pi_2(M^2_-,M^2_+\cap M^2_-)$. The crossed module $K=\bar H\to G$ making up the right side of the fundamental crossed square \eqref{eqn.fundxs} is the inclusion $h\mapsto \bar h:=(\p h,h^{-1})$.

We are left with the triad homotopy group $L=\pi_3(M;M^2_+,M^2_-)$, which is more complicated to describe. We will make use of the notions of tensor product and coproduct of crossed modules.
\begin{defn} \label{def.tensor}
The nonabelian tensor product of two crossed modules $\mu:H\to G$ and $\nu:K\to G$ is a crossed module $H\otimes K\to G$ where $H\otimes K$ is generated by $h\otimes k$, with $h\in H$ and $k\in K$, modulo the relations
\beq \begin{cases} hh'\otimes k=(\leftidx{^h}{h'}\otimes \leftidx{h}{k})(h\otimes k), \\ 
h\otimes kk'=(h\otimes k)(\leftidx{^k}{h}\otimes\leftidx{^k}k'), \end{cases} \eeq
where $H$ and $K$ act on one another via $G$. $G$ acts on $H\otimes K$ via $\leftidx{^g}{(h\otimes k)}=\leftidx{^g}{h}\otimes\leftidx{^g}{k}$.
\end{defn}
The tensor product of two crossed modules fits in a crossed square \cite{brown1987}
\beq \begin{tikzcd}
H\otimes K \ar[r,"\kappa"] \ar[d,"\eta"] &K \ar[d,"\nu"] \\
H \ar[r,"\mu"] &G
\end{tikzcd} \eeq
where the homomorphisms $\eta$ and $\kappa$ are defined by $\eta(h\otimes k)=h\leftidx{^k}h^{-1}$ and $\kappa(h\otimes k)=\leftidx{^h}{k}k^{-1}$. The product is defined by $[h,\bar h]=h\otimes\bar h$.

\begin{defn} \label{def.coproduct}
The coproduct of two crossed modules $\beta:B\to G$ and $\gamma:C\to G$ is the crossed module $\p:B\circ C\to G$ associated\footnote{Recall from the discussion towards the end of \ref{sec.fundxm} that the crossed module $\p:H\to G$ associated to a pre-crossed module $\tilde\p:\tilde H\to G$ is obtained by taking $H$ to be the quotient of $\tilde H$ by the normal subgroup generated by elements of the form $hh'h^{-1}\leftidx{^{\tilde\p h}}{h'{}^{-1}}$. The homomorphism $\tilde\p$ descends to a crossed module homomorphism $\p$ on the quotient.} to the pre-crossed module $\tilde\p:B\rtimes C\to G$.\footnote{The pre-crossed module homomorphism is $\tilde\p(b,c)=\beta(b)\gamma(c)$, and one can check that it is a homomorphism with respect to the group law $(b_1,c_1)(b_2,c_2)=(b_1\leftidx{^{c_1}}{b_2},c_1c_2)$.}
\end{defn}
The obvious inclusions $B\hookrightarrow B\rtimes C$ and $C\hookrightarrow B\rtimes C$ induce homomorphisms $i:B\to B\circ C$ and $j:C\to B\circ C$.

The higher dimensional van Kampen theorem of Brown and Loday \cite{brown1987b} shows that the triad homotopy group of the 2-skeleton is the nonabelian tensor product of the relative homotopy groups considered earlier
\beq \pi_3(M^2;M^2_+,M^2_-) \simeq \pi_2(M^2_+,M^2_+\cap M^2_-)\otimes\pi_2(M^2_-,M^2_+\cap M^2_-). \eeq

The gluing maps of the 3-cells define a map $\sigma_3:\Sigma_3\to H\cap\bar H\subset G$ (recall that the gluing maps are by assumption triad maps, so they map into both $H$ and $\bar H$). Let $\p:C\to G$ be the free crossed module on $\sigma_3$. Then the triad homotopy group $L=\pi_3(M;M^2_+,M^2_-)$ is isomorphic to the coproduct $\pi_3(M^2;M^2_+,M^2_-) \circ C$ of the triad homotopy group of the 2-skeleton with $C$, modulo the relations \cite{brown1987b,Ellis1993}
\beq \begin{cases} i(\p c\otimes \bar h) = j(c) j(\leftidx{^{\bar h}}{c^{-1}}), \\
 i(h\otimes\p c) = j(\leftidx{^h}{c})j(c^{-1}), \end{cases} \eeq
where $c\in C,h\in H$ and $\bar h\in\bar H$. The crossed module homomorphisms $\eta:L\to H$ and $\kappa:L\to\bar H$ are induced by the homomorphisms $\eta:H\otimes\bar H\to H$, $\kappa:H\otimes\bar H\to\bar H$ and $\p:C\to H\cap\bar H$. Finally, the product map which completes the description of the fundamental crossed square is $[h,\bar h]=i(h\otimes\bar h)$.

The fundamental crossed square is completely defined by the cells $\Sigma_1,\Sigma_2,\Sigma_3$, the gluing map $\sigma_2:\Sigma_2\to\moy{\Sigma_1}$ and the triad gluing map $\sigma_3:\Sigma_3\to H\cap\bar H$. It is free, in the sense that any other crossed square of the form
\beq \begin{tikzcd}
\tilde L \ar[r,"\tilde\kappa"] \ar[d,"\tilde\eta"] &\bar H \ar[d,"\nu"] \\
H \ar[r,"\mu"] & G
\end{tikzcd} \eeq
with a function $\alpha:\Sigma_3\to\tilde L$ such that $\tilde\eta\circ\alpha=\eta$, necessarily admits an extension of $\alpha$ to morphism $\phi$ of crossed squares
\beq \phi:
\begin{tikzcd}
L \ar[r,"\kappa"] \ar[d,"\eta"] &\bar H \ar[d,"\nu"] \\
H \ar[r,"\mu"] & G
\end{tikzcd}
\to
\begin{tikzcd}
\tilde L \ar[r,"\tilde\kappa"] \ar[d,"\tilde\eta"] &\bar H \ar[d,"\nu"] \\
H \ar[r,"\mu"] & G
\end{tikzcd} \eeq
such that $\phi_L$ extends $\alpha$, and $\phi_H,\phi_{\bar H}$ and $\phi_G$ are identity homomorphisms.

\section{Calculation of $\Phi\p_4(x_I)$ in example \ref{sec.t3s2}} \label{app.calc}
As mentioned in the text, the squared complex homomorphism $\p_4$ sends $x_I$ to the generator of $\pi_3(I_0M^3)$, and we know that the $C$ component can be chosen to take the form
\beq \p_4(x_I)_C = \leftidx{^{(f_1,h_1)}}{x_1}\ \leftidx{^{(f_2,h_2)}}{t_I}^{-1}\ \leftidx{^{(f_3,h_3)}}{v_I}\ \leftidx{^{(f_4,h_4)}}{u_I}^{-1}\ \leftidx{^{(f_5,h_5)}}{x_0}^{-1}\ \leftidx{^{(f_6,h_6)}}{t_I}\ \leftidx{^{(f_7,h_7)}}{v_I}^{-1}\ \leftidx{^{(f_8,h_8)}}{u_I}, \eeq
with the $f_i\in F$ and $h_i\in H$ determined by the condition that $\eta\p_4(x_I)=1$, or, equivalently, that $\p\p_4(x_I)_C$ can be written in the form $\eta'l'$ for some $l'\in H\otimes\bar H$. An inspection of the attaching maps
\beq \sigma_3(x_j) = t_j\ \leftidx{^{c_j}}{v_j^{-1}}\ u_j\ \leftidx{^{a_j}}{t_j^{-1}}\ v_j\ \leftidx{^{b_j}}{u_j^{-1}},~~~~~ j=0,1 \eeq
\beq \sigma_3(t_I) = t_1\ \leftidx{^{c_1}}{b_I}\ c_I\ t_0^{-1}\ b_I^{-1}\ \leftidx{^{b_1}}{c_I^{-1}} \eeq
\beq \sigma_3(u_I) = u_1\ \leftidx{^{a_1}}{c_I}\ a_I\ u_0^{-1}\ c_I^{-1}\ \leftidx{^{c_1}}{a_I^{-1}} \eeq
\beq \sigma_3(v_I) = v_1\ \leftidx{^{b_1}}{a_I}\ b_I\ v_0^{-1}\ a_I^{-1}\ \leftidx{^{a_1}}{b_I^{-1}} \eeq
a convenient choice of $f_j$s, such that the terms in $\sigma_3(x_1)$ can be cancelled, is given by
\beq f_1=1,~~~ f_2=1,~~~ f_3=c_1,~~~ f_4=1,~~~ f_5=1,~~~ f_6=a_1,~~~ f_7=1,~~~ f_8=b_1. \eeq
The $h_j$s can be chosen so that the $t_I$s, $u_I$s and $v_I$s are ``slotted into the right place''. Since we have manufactured the cancellation of twelve terms, this leaves a mess of 36 terms in the expression for $\p\p_4(x_I)_C$. One has to massage this into a product of terms of the form $\eta'(h\otimes\bar k)=hk^{-1}\leftidx{^{\p k}}{h}^{-1}k$.

For the purposes of computing homotopy classes of $[T^3,S^2]$, we need only compute the image $\Phi\p_4(x_I)$ in $\bZ$ rather than $\p_4(x_I)$ itself. This affords us some simplifications; for example, elements of the form
\beq \eta'(h\otimes\bar k)\otimes \bar k'=(hk^{-1}\leftidx{^{\p k}}h^{-1}k)\otimes\bar k' \eeq
are sent by $\Phi$ to $(\Phi(h)-\Phi(k)-\Phi(h)+\Phi(k))\Phi(k')=0$. With these simplifications, it is a straightforward but long calculation to arrive at equation \eqref{eqn.t3s2hom}.


\bibliography{defect}
\bibliographystyle{jhep}

\end{document}